Eva Ponick, Gabriele Wieczorek

# Künstliche Intelligenz in Governance, Risk und Compliance

Ergebnisse einer Studie zu Anwendungspotentialen von Künstlicher Intelligenz (KI) in Governance, Risk und Compliance (GRC)

2021

**Kontaktdaten:**

Hochschule Hamm-Lippstadt
Marker Allee 76-78
59063 Hamm
Germany

Prof. Dr. Eva Ponick      Prof. Dr. Gabriele Wieczorek
eva.ponick@hshl.de        gabriele.wieczorek@hshl.de

**Hinweis:** Die Datei beinhaltet zunächst die Version auf Deutsch und im Anschluss die Version auf Englisch.
**Notice:** The file contains first the version in German and afterwards the version in English.

# Zusammenfassung

Die digitale Transformation führt zu einem tiefgreifenden Wandel der Organisationsstrukturen in Unternehmen. Um neue Technologien nicht nur punktuell anwenden zu können, müssen Prozesse im Unternehmen überdacht und Funktionen ganzheitlich, gerade auch im Hinblick auf Schnittstellen, betrachtet werden.

So erfolgen zielgerichtete Managemententscheidungen in Unternehmen unter anderem auf Grundlage von Risikomanagement und Compliance Management sowie des internen Kontrollsystems als Governance-Funktionen. Die Effektivität und die Effizienz dieser Funktionen ist entscheidend für die Einhaltung von Richtlinien und regulatorischen Anforderungen sowie für die Bewertung möglicher Handlungsalternativen im Rahmen der unternehmerischen Tätigkeiten. Unter GRC (Governance, Risk und Compliance) wird ein integrierter Governance-Ansatz verstanden, bei dem die genannten Governance-Funktionen ineinandergreifen und nicht einzeln für sich stehen.

Eine wichtige Technologie der digitalen Transformation stellen die Methoden der Künstlichen Intelligenz dar. Diese Technologie, die ein breites Spektrum von Methoden wie beispielsweise Maschinelles Lernen, künstliche neuronale Netze, Natural Language Processing oder Deep Learning aufweist, bietet Anwendungsmöglichkeiten in unterschiedlichsten Unternehmensbereichen vom Einkauf über die Produktion oder Kundenbetreuung. Und auch in GRC kann Künstliche Intelligenz eingesetzt werden, beispielsweise zur Aufbereitung und Analyse unstrukturierter Daten.

Die vorliegende Studie führt die Ergebnisse einer Umfrage aus dem Jahr 2021 zur Identifikation und Analyse von Anwendungspotentialen Künstlicher Intelligenz in GRC auf.

# Abstract

The digital transformation leads to fundamental change in organizational structures. To be able to apply new technologies not only selectively, processes in companies must be revised and functional units must be viewed holistically, especially with regard to interfaces.

Target-oriented management decisions are made, among other things, on the basis of risk management and compliance in combination with the internal control system as governance functions. The effectiveness and efficiency of these functions is decisive to follow guidelines and regulatory requirements as well as for the evaluation of alternative options for acting with regard to activities of companies. GRC (Governance, Risk and Compliance) means an integrated governance-approach, in which the mentioned governance functions are interlinked and not separated from each other.

Methods of artificial intelligence represents an important technology of digital transformation. This technology, which offers a broad range of methods such as machine learning, artificial neural networks, natural language processing or deep learning, offers a lot of possible applications in many business areas from purchasing to production or customer service. Artificial intelligence is also being used in GRC, for example for processing and analysis of unstructured data sets.

This study contains the results of a survey conducted in 2021 to identify and analyze the potential applications of artificial intelligence in GRC.

# Inhaltsverzeichnis

# 1  Einleitung

Künstliche Intelligenz (KI) stellt einen wichtigen Faktor dar, um im Rahmen der digitalen Transformation Prozesse zu automatisieren. Mittlerweile stehen viele erprobte Verfahrung zur Verfügung, die in unterschiedlichen Bereichen und Prozessen im Unternehmen eingesetzt werden können. Doch auch, wenn der Einsatz Künstlicher Intelligenz häufig als Mehrwert betrachtet wird, hinken Umsetzung und Einbindung der Technologie in die Prozessstrukturen der Unternehmen deutlich hinterher.[1] Denn der Einsatz von Künstlicher Intelligenz in Unternehmen ist kein Selbstläufer, sondern beruht darauf, dass gewisse Rahmenbedingungen erfüllt werden. Tarafdar et al., 2019 untersuchen beispielsweise den Einsatz Künstlicher Intelligenz zur Optimierung von Geschäftsprozessen und ermitteln fünf Fähigkeiten, die Unternehmen besitzen sollten, um Künstliche Intelligenz erfolgreich in Prozesse zu integrieren. Hierzu gehört die „Data Science Competence", der sichere Umgang mit Daten und die Bedeutung von Daten für die Methoden der Künstlichen Intelligenz und die „Business Domain Proficiency" und somit die genaue Kenntnis über die Unternehmensbereiche und deren Geschäftsprozesse und eine Vorstellung, wie der Einsatz von Künstlicher Intelligenz hier zu Verbesserungen führen kann. Des Weiteren beinhaltet dies die „Enterprise Architecture Expertise", das Wissen um das Zusammenspiel zwischen der Unternehmensarchitektur und der Modellierung von Geschäftsprozessen, den „Operational IT Backbone", d.h. die bestehenden IT-Strukturen und -Technologien, sowie „Digital Inquisitiveness", also eine Wissbegierde, sich mit neuen Technologien auseinanderzusetzen.

In der vorliegenden Studie stehen nun ganz konkret die Funktionen Risikomanagement, Compliance und das interne Kontrollsystem und ihre Verzahnung im Fokus. Gerade der funktions- und bereichsübergreifende Umgang mit Chancen und Risiken vor dem Hintergrund der großen Unsicherheit über zukünftige Entwicklungen, denen unternehmerische Entscheidungen ausgesetzt sind, macht das integrierte GRC-Konzept (GRC steht dabei für Governance, Risk und Compliance) interessant für Anwendungen der Künstlichen Intelligenz, da die Erwartung besteht, dass eine höhere Integrationsreife des GRC-

---

[1] Vgl. Tarafdar et al., 2019 und Ransbotham et al., 2017.

Konzepts mit einem erhöhten Grad der Automatisierung und des Datenaustausches einher geht. Die Quantität und Qualität strukturierter und unstrukturierter Daten sind eine wichtige Voraussetzung für den erfolgreichen Einsatz Künstlicher Intelligenz. Neben dem Einsatz in GRC muss aber auch die unternehmensweite Risiko- oder GRC-Strategie sicherstellen, dass die Verwendung von Künstlicher Intelligenz in anderen Prozessen des Unternehmens, wie z.B. der Produktion oder der Kundenbetreuung, regelkonform abläuft und risikotechnisch analysiert wird.[2]

Aus diesem Blickwinkel heraus wurde 2021 die Studie durchgeführt, um das Zusammenspiel zwischen GRC und KI genauer herauszuarbeiten. Mit dem Online-Fragebogentool Sosci Survey[3] wurde ein Fragebogen erstellt und auf [www.soscisurvey.de](www.soscisurvey.de) zur Verfügung gestellt. Die Datenerhebung erfolgte anonym. Die Umfrage richtete sich speziell an Personen mit GRC-Bezug. Der Link zur Umfrage wurde über unterschiedliche Quellen verteilt (Allianz für Sicherheit in der Wirtschaft e.V. (ASW), das Deutsche Institut für Compliance (DICO), das Deutsche Institut für Interne Revision e.V. (DIIR), den Internationalen Controller Verein e.V. (ICV), die ISACA Germany Chapter E.V. (ISACA) und die Risk Management & Rating Association e.V. (RMA) sowie verschiedene (über)regionale KI-Plattformen und Wirtschaftsförderungen). Die Daten wurden im Zeitraum von September 2021 bis Dezember 2021 erhoben. Insgesamt wurden 140 Datensätze erhoben, von denen 67 abgeschlossen waren und als Basis für die Ergebnisse verwendet werden konnten. Auf Grund der geringen Anzahl von Interviews und der Ansprache von Personen über einschlägige Verbände kann das Ergebnis als eingeschränkt repräsentativ erachtet werden. Dennoch ergeben sich interessante Aussagen, die als Grundlage für weitere Forschung zu Anwendungsfeldern von KI in GRC angesehen werden können.

Konkret soll die Studie Antworten auf die Fragen geben

- Wie sieht der aktuelle Stand in Bezug auf den Einsatz von KI im Bereich GRC aus?

---

[2] Für eine Diskussion rechtlicher Herausforderungen beim Einsatz von KI für die Versicherungswirtschaft vgl. beispielsweise Armbrüster, 2022.
[3] Vgl. Leiner, 2019.

- Sind die notwendigen Rahmenbedingungen gegeben, um KI in GRC einzusetzen?
- Welche konkreten Potentiale und Anwendungsmöglichkeiten für KI in GRC bestehen bereits im Unternehmen?
- Welche konkreten Potentiale und Anwendungsmöglichkeiten für KI sehen die Befragten allgemein in Bezug auf GRC?

Die vorliegende Arbeit bietet einen ersten Überblick über die Ergebnisse der Studie zu den Anwendungspotentialen von KI in GRC. Im folgenden Verlauf sind die Fragen, die im Rahmen der Studie gestellt wurden, sowie die Umfrageergebnisse zur jeweils angegebenen Frage aufgeführt. Die vorliegenden Daten dienen dabei als Grundlage weiterer, komplexerer Analysen und zukünftiger Forschungsfragen.

In Kapitel 2 werden die Fragen der Studie und die Ergebnisse in der verwendeten Reihenfolge behandelt. Der Fragebogen war in die drei Abschnitte

- Status Quo zu GRC und KI
- KI-Potentiale in GRC
- Allgemeine Angaben zum Unternehmen

eingeteilt. In Kapitel 3 werden die wichtigsten Ergebnisse noch einmal aufgeführt. Im Anschluss werden in Kapitel 4 die Ergebnisse zusammengeführt. Die Studie schließt mit einer Zusammenfassung in Kapitel 5 ab.

## 2 Grundsätzliche Ergebnisse

### 2.1 Teil 1: Status Quo zu GRC und KI

Teil 1 der Studie soll grundsätzliche Informationen darüber geben, ob KI in GRC bereits eingesetzt wird. Darüber hinaus liefern die Fragen wichtige Informationen zur Integration der GRC-Kernfunktionen und somit auch zu den Rahmenbedingungen eines Einsatzes von KI in GRC.

### *1. Setzen Sie im Unternehmen bereits Künstliche Intelligenz ein?*

**Frage 1**: Setzen Sie im Unternehmen bereits Künstliche Intelligenz ein?

Es wurde zur Frage folgender Hinweis gegeben:

„KI kann z.B. eingesetzt werden für Sprach- oder Textverstehen, Bild- oder Tonerkennung, wissensbasierte Systeme, Mensch-Maschine-Interaktion bzw. Assistenzsysteme, autonome Systeme, Sensorik und Robotik, Virtual Reality und Augmented Reality."

Mögliche Antworten waren: ja, nein, nicht bekannt.

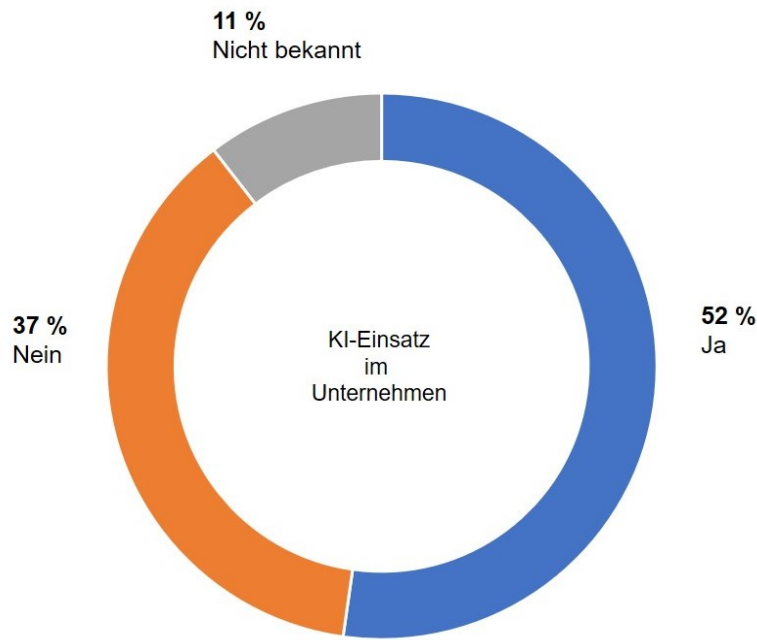

*Abbildung 1: Einsatz von KI im Unternehmen (n=67)*

Die Angabe bezieht sich auf die abgeschlossenen und ausgewerteten Fragebögen (n=67). Die Mehrheit (52 %) gab an, dass KI im Unternehmen bereits eingesetzt wird.

## *2. In welchen Einsatzbereichen im Unternehmen werden KI-Technologien geplant bzw. bereits angewandt?*

**Frage 2**: In welchen Einsatzbereichen im Unternehmen werden KI-Technologien geplant bzw. bereits angewandt?

Es sollten die Einsatzbereiche

- Virtual Reality und Augmented Reality
- Autonome Systeme, Sensorik und Robotik

- Mensch-Maschine-Interaktion bzw. Assistenzsysteme
- Wissensbasierte Systeme
- Bild- oder Tonerkennung
- Sprach- oder Textverstehen

bewertet werden.

Bei mehreren Projekten im Unternehmen sollte der höchste Entwicklungsstand angegeben werden. Mögliche Antworten waren: kein Projekt geplant, Projekt zur Einführung geplant, Projekt zur Anwendung gestartet, Projekt bereits umgesetzt, nicht bekannt.

Die Ergebnisse zu dieser Frage sind in Abhängigkeit der Beantwortung von Frage 1 aufgeführt.

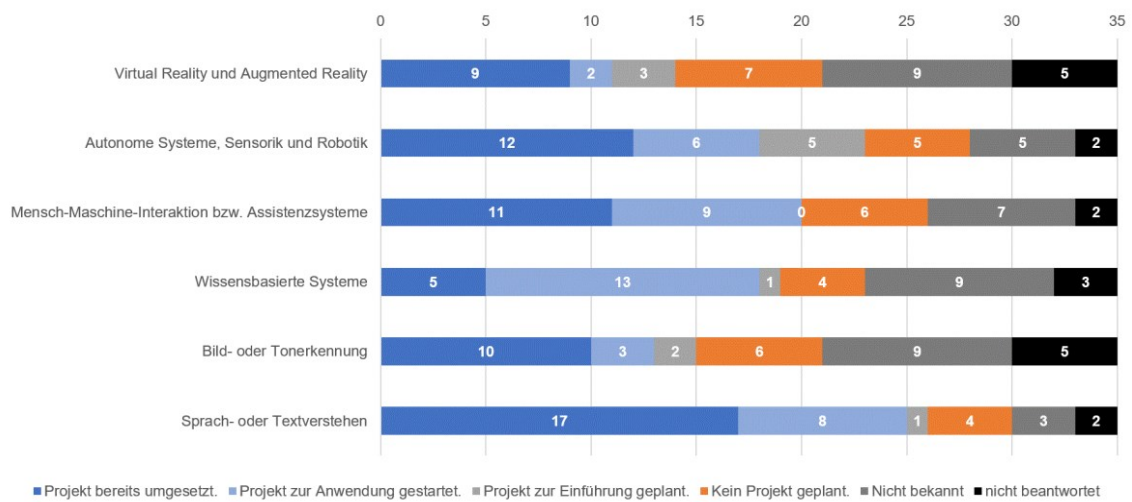

*Abbildung 2: Einsatzbereiche mit geplanten oder bereits umgesetzten KI-Projekten bei Unternehmen, wenn die Frage 1 (KI-Einsatz im Unternehmen) mit „Ja" beantwortet wurde (n=35)*

Bei den Unternehmen, die KI bereits einsetzen (n=35) sticht, in Abbildung 2 der Bereich Sprach- oder Textverstehen deutlich hervor.

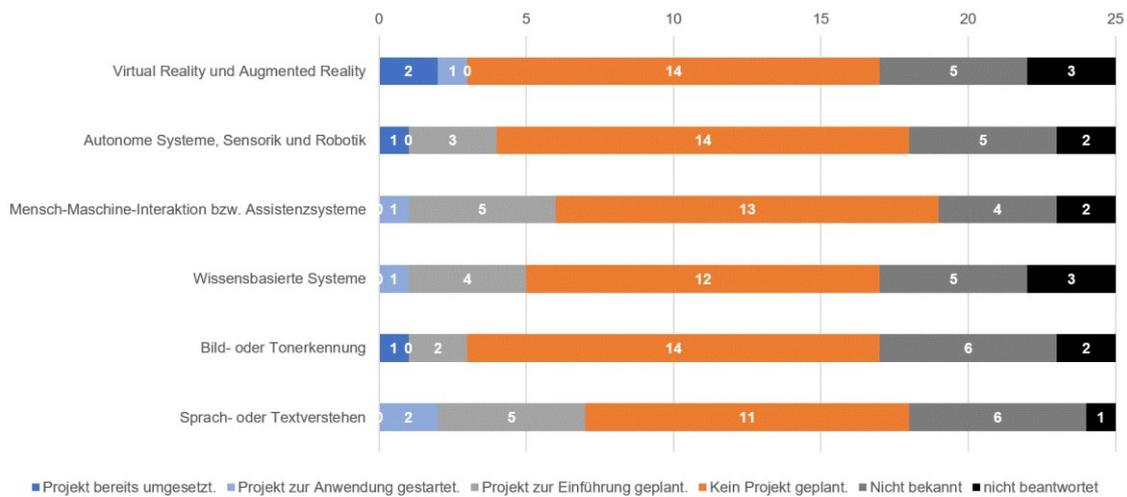

*Abbildung 3: Einsatzbereiche mit geplanten oder bereits umgesetzten KI-Projekten bei Unternehmen, wenn die Frage 1 (KI-Einsatz im Unternehmen) mit „Nein" beantwortet wurde (n=25)*

Abbildung 3 zeigt, dass in den Fällen, in denen keine KI-Projekte im Unternehmen umgesetzt wurden (n=25), meist auch keine Projekte dieser Art geplant sind. Die Aussagen zu bereits umgesetzten Projekten in dieser Gruppe fallen ins Auge, da sie nicht konsistent zu der ursprünglichen Aussage sind, dass keine KI eingesetzt wird. Eine Erklärung für diese Angaben kann den vorliegenden Daten nicht entnommen werden.

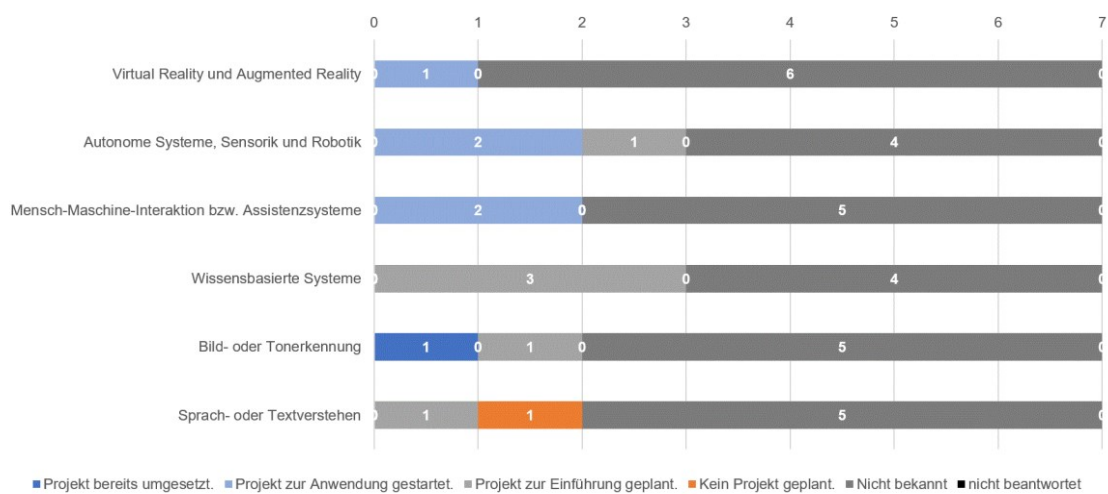

*Abbildung 4: Einsatzbereiche mit geplanten oder bereits umgesetzten KI-Projekten bei Unternehmen, die Frage 1 (KI-Einsatz im Unternehmen) mit „Nicht bekannt" beantwortet haben (n=7)*

Abbildung 4 führt die letzte der drei Gruppen aus Frage 1 auf, welche die Frage zum Einsatz von KI im Unternehmen mit „nicht bekannt" beantwortet hat (n=7). Entsprechend zu dieser Aussage wurde mehrheitlich angegeben, dass auch nicht bekannt ist, ob in den angegebenen Einsatzbereichen Projekte geplant oder umgesetzt werden. Auch hier fiel für den Bereich Bild- oder Tonerkennung ein Eintrag auf, der angibt, dass ein Projekt bereits umgesetzt sei. Zusammen mit den entsprechenden Aussagen aus der Gruppe derjenigen, die angaben, dass kein Einsatz von KI im Unternehmen vorliegt, wäre hier von Interesse, wie die Frage von den Studienteilnehmerinnen und -teilnehmern tatsächlich interpretiert wurde.

### *3. Sind die GRC-Bereiche Risikomanagement, Compliance und IKS in Ihrem Unternehmen bereits in irgendeiner Form miteinander verzahnt, d.h. es besteht zumindest ein informatorischer Austausch zwischen zwei (oder mehr) GRC-Bereichen?*

**Frage 3**: Sind die GRC-Bereiche Risikomanagement, Compliance und IKS in Ihrem Unternehmen bereits in irgendeiner Form miteinander verzahnt, d.h. es besteht zumindest ein informatorischer Austausch zwischen zwei (oder mehr) GRC-Bereichen?

Um ein einheitliches Verständnis zu erzeugen wurde bei dieser Frage eine Beschreibung des verwendeten integrierten GRC-Ansatzes vorangestellt:

„Im Folgenden wird unter GRC, oder besser einem integrierten GRC-Ansatz, die Verzahnung der Governance-Funktionen Risikomanagement, Compliance und Internes Kontrollsystem (IKS) verstanden, die auf die Hebung von Synergien im Umgang mit Risiken abzielt. Dagegen kommt der Internen Revision, als Funktion mit GRC-Bezug, eine übergeordnete und unabhängige Rolle im integrierten GRC-Management zu."

Zudem war die Frage mit dem Hinweis versehen, dass die Beantwortung für den weiteren Verlauf der Befragung essentiell ist. Mögliche Antworten waren: ja, nein, keine Angabe.

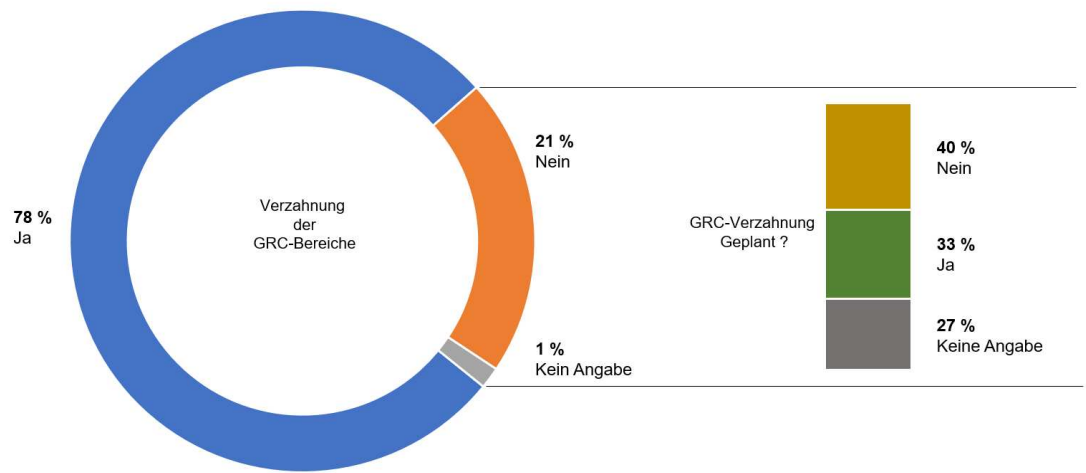

*Abbildung 5: Verzahnung der GRC-Bereiche Risikomanagement, Compliance und IKS (n=67)*

Diejenigen, welche diese Frage mit „Nein" (n=14) oder „Keine Angabe" (n=1) beantwortet haben, wurden im Folgenden gefragt, ob eine Verzahnung von GRC-Bereichen geplant ist. Die Angaben sind dem rechten Bereich der Abbildung 5 zu entnehmen und beziehen sich somit auf n=15 Personen. Eine klare Tendenz bezüglich geplanter Verzahnung ist den Daten nicht zu entnehmen.

## *4. Ist die Verzahnung von GRC-Bereichen geplant?*

**Frage 4**: Ist die Verzahnung von GRC-Bereichen geplant?

Diese Frage wurde nur angezeigt, wenn Frage 3 mit „Nein" oder „Keine Angabe" beantwortet wurde. Mögliche Antworten waren: ja, nein, keine Angabe.

Die Ergebnisse zu dieser Frage können Abbildung 5, Frage 3 entnommen werden.

## *5. Welche Bereiche sind miteinander verzahnt?*

**Frage 5**: Welche Bereiche sind miteinander verzahnt?

Diese Frage wurde *nicht* angezeigt, wenn Frage 3 mit „Nein" oder „Keine Angabe" beantwortet wurde. Mögliche Antworten waren:

- Risikomanagement und Compliance
- Risikomanagement und IKS

- Compliance und IKS
- Risikomanagement, Compliance und IKS.

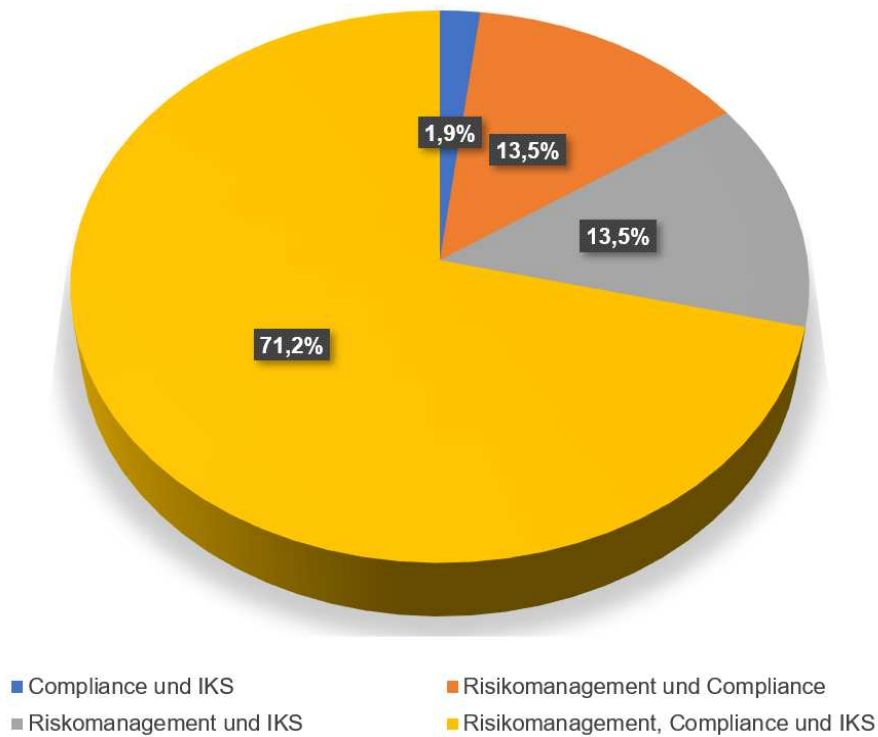

*Abbildung 6: Überblick der miteinander verzahnten Bereiche Risikomanagement, Compliance und/oder IKS (n=52)*

Frage 3 wurde in n=52 Fällen mit „Ja" beantwortet. Auf diese Fragebögen bezieht sich nun die vorliegende Frage 5. Den anderen n=15 Teilnehmerinnen und Teilnehmern lag diese Frage nicht vor. Die Aufteilung bei den Antworten auf die Zusammensetzung der verzahnten Bereiche ist Abbildung 6 zu entnehmen. In der überwiegenden Mehrheit (71 %) lag eine Dreifachverzahnung aller drei Bereiche Risikomanagement, Compliance und IKS vor.

## *6. Wie beurteilen Sie die operative Verzahnung der GRC-Bereiche in Ihrem Unternehmen?*

**Frage 6**: Wie beurteilen Sie die operative Verzahnung der GRC-Bereiche in Ihrem Unternehmen?

Die Frage diente der Beurteilung der vorliegenden operativen Verzahnung. Die folgenden Aussagen wurden zur Bewertung vorgegeben:

- Die GRC-Bereiche haben ein einheitliches Grundverständnis zu Begriffen und Definitionen im Risikokontext (z.B. Risiko, Risikoarten, …).
- Über ein einheitliches Grundverständnis hinaus stimmen sich die GRC-Bereiche inhaltlich von der Risikoanalyse bis zur Risikosteuerung ab.
- Die GRC-Bereihe nutzen abgestimmte, einheitliche Methoden im Risk Assessment.
- Die Zusammenarbeit zwischen den GRC-Bereichen ist inhaltlich und zeitlich harmonisiert.
- Die Erfassung und Analyse von Informationen zwischen den GRC-Bereichen ist redundanzfrei.

Diese Frage wurde nicht angezeigt, wenn Frage 3 nach der Verzahnung der Bereiche mit „Nein" oder „Keine Angabe" beantwortet wurde. Die aufgeführten Aussagen konnten mit einer 5-stufigen Skala (stimme gar nicht zu, stimme eher nicht zu, teils, teils, stimme eher zu, stimme voll zu) bewertet werden.

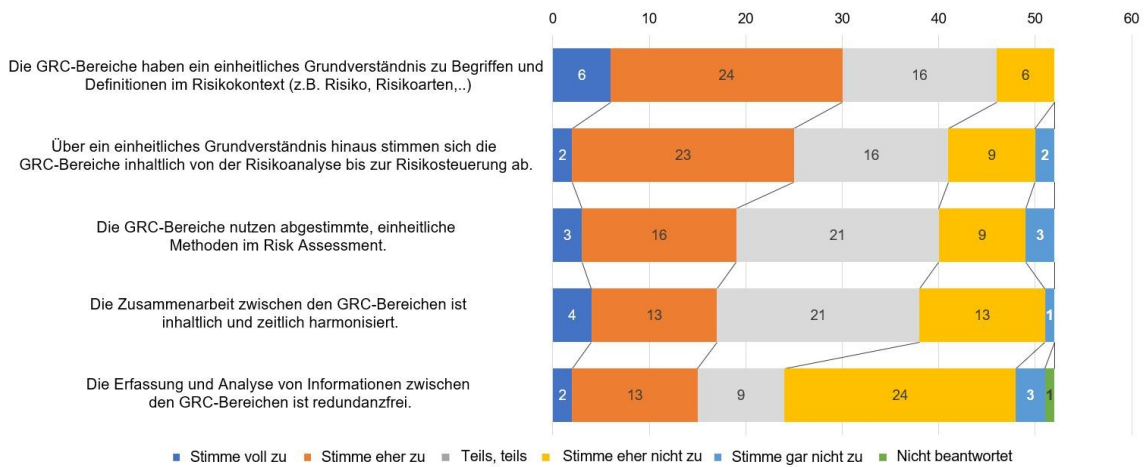

*Abbildung 7: Beurteilung der operativen Verzahnung der GRC-Bereiche (n=52)*

Wurde Frage 3, ob die GRC-Bereiche Risikomanagement, Compliance und IKS in irgendeiner Form verzahnt sind, mit „Ja" beantwortet, sollte Frage 6 Auskunft über die Struktur der Verzahnung geben. Wie in Abbildung 7 dargestellt, stimmte eine Mehrheit von 58 % der 52 Befragten, die diese Frage beantwortet haben, eher oder voll der Aussage zu, dass zumindest ein einheitliches Grundverständnis zu Begriffen und Definitionen im Risikokontext vorliegt.

Der Aussage einer redundanzfreien Erfassung und Analyse von Informationen zwischen den GRC-Bereichen und damit einer Rahmenbedingung für den erfolgreichen Einsatz von KI in GRC stimmten 29 % eher oder voll zu. Jedoch gaben 52 % an, dass sie dieser Aussage eher nicht oder gar nicht zustimmen.

### *7. Wählen Sie bitte aus, welche übergeordneten Elemente bzw. Instrumente bereits in Ihrem Unternehmen zur Integration der GRC-Bereiche existieren oder geplant sind!*

**Frage 7**: Wählen Sie bitte aus, welche übergeordneten Elemente bzw. Instrumente bereits in Ihrem Unternehmen zur Integration der GRC-Bereiche existieren oder geplant sind.

Die folgenden Elemente und Instrumente sollten einer Bewertung unterzogen werden:

- GRC-Komitee
- Integriertes GRC-Reporting
- Integrierte GRC-Strategie
- GRC-Prozess
- GRC als integrierte Funktion
- GRC als integrierter Bereich
- GRC-Handbuch
- Integrierte GRC-Softwarelösung.

Diese Frage wurde nicht angezeigt, wenn Frage 3 nach der Verzahnung der Bereiche mit „Nein" oder „Keine Angabe" beantwortet wurde. Damit flossen insgesamt n=52 Fragebögen in die Bewertung ein.

Für die einzelnen Punkte waren die folgenden Angaben möglich: nicht geplant, in Evaluation, geplant oder in Umsetzung, existiert bereits, nicht bekannt.

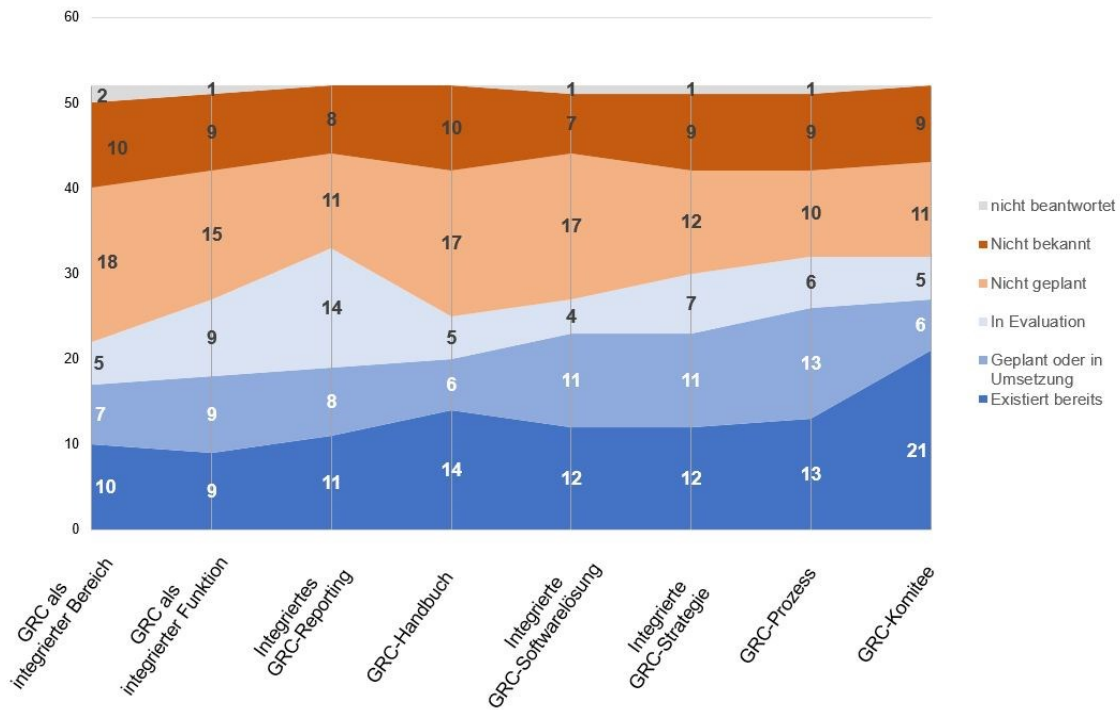

*Abbildung 8: Instrumente und Elemente zur Integration der GRC-Bereiche (n=52)*

Betrachtet man bereits existierende Elemente bzw. Instrumente, die geplant sind oder sich in Umsetzung befinden, so sticht GRC als integrierter Bereich besonders hervor. Hier war der Anteil derjenigen, die angaben, dass dieses Instrument bereits existiert, geplant oder in Umsetzung ist, mit 33 % am geringsten. Insgesamt 35 % gaben sogar an, dass dieses Instrument nicht geplant ist, was den höchsten Wert dieser Kategorie darstellt.

Beim integrierten GRC-Reporting geben 35 % an, dass dies bereits existiert, geplant oder in Umsetzung ist. Es gaben nur 21 % an, dass dieses Instrument nicht geplant ist. Mit 27 % ist hier jedoch der Anteil derjenigen, die angaben, dass sich dieses Instrument in Evaluation befindet, am höchsten. Dies wirft die interessante Frage nach dem Grund dieser intensiven Evaluierung auf.

Die Existenz, Planung oder Umsetzung eines GRC-Komitees wurden von 52 % angegeben, was den höchsten Wert darstellt. Auch hier gaben lediglich 21 % an, dass dieses Instrument nicht geplant ist und bei nur 10 % wurde angegeben, dass der Einsatz evaluiert wird.

Bei dem GRC-Handbuch und der integrierten GRC-Softwarelösung gaben immerhin 33 % an, dass diese Elemente nicht geplant sind.

## 8. Für welche GRC-Bereiche existieren IT-Anwendungen?

**Frage 8**: Für welche GRC-Bereiche existieren IT-Anwendungen?

Diese Frage wurde unabhängig von Antworten anderer Fragen stets gestellt und bezieht sich somit auf n=67 Fragebögen.

Mögliche Antwortmöglichkeiten waren:

- Risikomanagement
- Compliance
- IKS
- Interne Revision
- Keiner der Bereiche.

Eine Mehrfachauswahl war möglich.

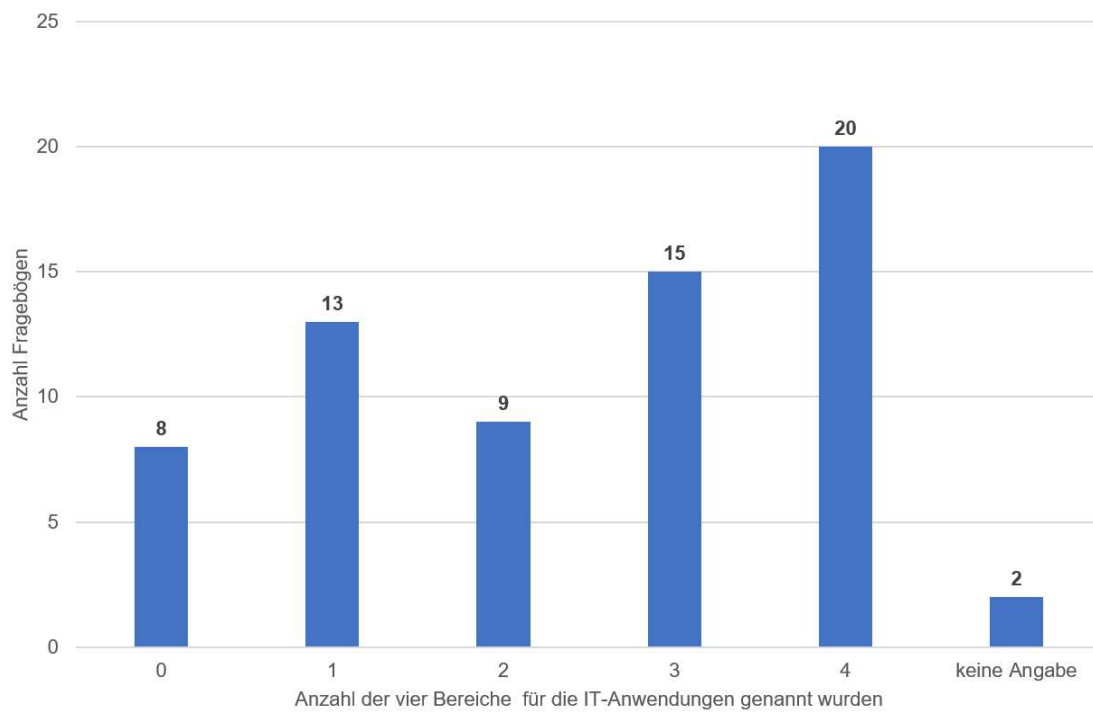

*Abbildung 9: Anzahl der GRC-Bereiche, für die angegeben wurde, dass IT-Anwendungen existieren (n=67)*

Abbildung 9 zeigt einen ersten Überblick, ob insgesamt für keinen, einen, zwei, drei oder alle vier GRC-Bereiche angegeben wurde, dass IT-Anwendungen existieren. Die meisten Angaben (30 %) erfolgten dabei für alle vier Bereiche. Keiner der Bereiche (Angabe 0 in Abbildung 9) wurde von 12 % angegeben und 3 % machten zu dieser Frage keine Angaben.

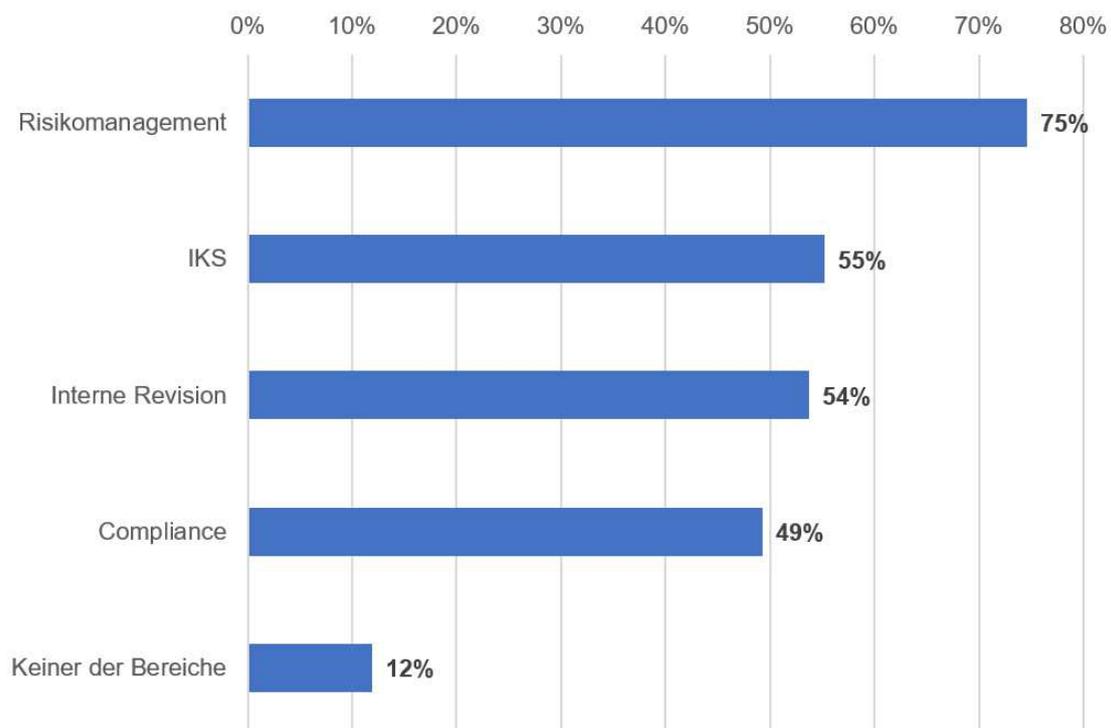

*Abbildung 10: GRC-Bereiche mit IT-Anwendungen (n=67)*

In Abbildung 10 ist zu sehen, dass laut Studienteilnehmerinnen und -teilnehmern die meisten IT-Anwendungen im Bereich Risikomanagement existierten (75 %), gefolgt von IKS (55 %), Interne Revision (54 %) und Compliance (49 %). Es gaben 12 % der Befragten an, dass keiner der GRC-Bereiche über IT-Anwendungen verfügt.

Wenn in Frage 8 ein Bereich ausgewählt wurde, dann wurden die folgenden Fragen 9 bis 12 entsprechend des ausgewählten Bereiches gestellt. Eine detaillierte Darstellung findet sich in den folgenden Abschnitten.

## 9. Bitte geben Sie die Software an, die Sie im Risikomanagement einsetzen.

**Frage 9**: Bitte geben Sie die Software an, die Sie im Risikomanagement einsetzen.

Die Frage wurde angezeigt, wenn in Frage 8 der Bereich Risikomanagement ausgewählt wurde und somit angegeben wurde, dass dort IT-Anwendungen existieren. Es handelte sich um ein Freitextfeld.

Von den n=50 Personen, die in Frage 8 angegeben haben, dass IT-Anwendungen im Risikomanagement existieren, haben n=42 Personen Frage 9 beantwortet. Da es sich um ein Freitextfeld handelte und mehrere Eingaben möglich waren, liegen insgesamt 51 Softwarenennungen vor. Es wurden von den Befragten maximal 3 Nennungen von IT-Anwendungen angegeben.

| Anzahl der Software-nennungen | Personenanzahl |
|---|---|
| 1 | 36 |
| 2 | 3 |
| 3 | 3 |
| Gesamt | 42 |

*Tabelle 1: Anzahl der Personen, die bei Frage 9 eine, zwei oder drei IT-Anwendungen genannt haben.*

In Tabelle 1 ist ein Überblick über die Anzahl der Personen enthalten, die jeweils eine, zwei oder drei Anwendungen genannt haben.

Für eine Zuordnung wurden die genannten IT-Anwendungen in unterschiedliche Kategorien einsortiert. Tabelle 2 beinhaltet einen Überblick über die Kategorien, denen die aufgeführten IT-Anwendungen zugeordnet wurden. Dabei wurde jede Nennung berücksichtigt, d.h. wenn eine IT-Anwendung von unterschiedlichen Personen angegeben wurde, wurde sie so oft, wie angegeben, mitgezählt und nicht zu einer Nennung zusammengefasst.

Dabei ist anzumerken, dass aus den Angaben nicht immer eindeutig hervorging, welche IT-Anwendung tatsächlich eingesetzt wird. So wurde in vielen Fällen ein Hersteller genannt und keine konkrete Softwarelösung. In diesen Fällen fand eine Zuordnung zu Kategorie 1 statt, wenn der Hersteller eine entsprechende Software im Angebot hat. Einige

IT-Anwendungen sind modular aufgebaut, so dass nicht klar ist, ob eine komplette GRC-Anwendung vorhanden ist oder nur ein Bereich mit der IT-Anwendung abgedeckt wird. Aus diesem Grund wird auch keine weitere Unterteilung der Kategorie 1 durchgeführt.

Kategorie 2 beinhaltet IT-Anwendungen, die ganz konkret als Eigenentwicklung oder individuelle Branchenlösung benannt wurden. Wohingegen Kategorie 3 Tabellenkalkulationen, weitere Office Produkte, Tools für statistische Analysen und/oder Simulationen, als auch Add-Ins für Tabellenkalkulationen beinhaltet.

Ging aus der Beschreibung keine eindeutige Zuordnung hervor, so wurde die IT-Anwendung der Kategorie 4 zugeordnet. Dies geschah z.B. bei der Bezeichnung „GRC". Das bedeutet, dass bei den Kategorie 4 zugeordneten IT-Anwendungen jedoch stets eine konkrete Angabe vorlag.

Schließlich erfolgten auch Angaben wie „nicht bekannt", „keine Angabe" oder Ähnliches. Diese Angaben sind in Kategorie 5 zusammengefasst. Das kommt zwar in Frage 9 tatsächlich nicht zum Tragen, da diese Einteilung aber auch für die folgenden Fragen 10 bis 12 verwendet wird, wird Kategorie 5 bereits hier eingeführt.

| Kategorie | Beschreibung |
|---|---|
| 1 | IT-Anwendungen, die ganz oder teilweise den entsprechenden Bereichen Risikomanagement, IKS, Interne Revision oder Compliance zugeordnet werden können und komplette GRC-Softwareanwendungen |
| 2 | IT-Anwendungen, die als Eigenentwicklung oder individuelle Branchenlösung bezeichnet werden |
| 3 | Tabellenkalkulationen, weitere Office-Produkte, Tools für statistische Analysen und/oder Simulationen, auch als Add-Ins für Tabellenkalkulation |
| 4 | IT-Anwendungen, die keiner der Kategorien 1 bis 3 zugeordnet werden können, weil das angegebene Softwareprodukt nicht identifizierbar ist, eine konkrete Angabe jedoch erfolgt ist |
| 5 | Angaben wie „nicht bekannt" oder „keine Angabe" |

*Tabelle 2: Kategorien zur Einteilung der aufgeführten IT-Anwendungen*

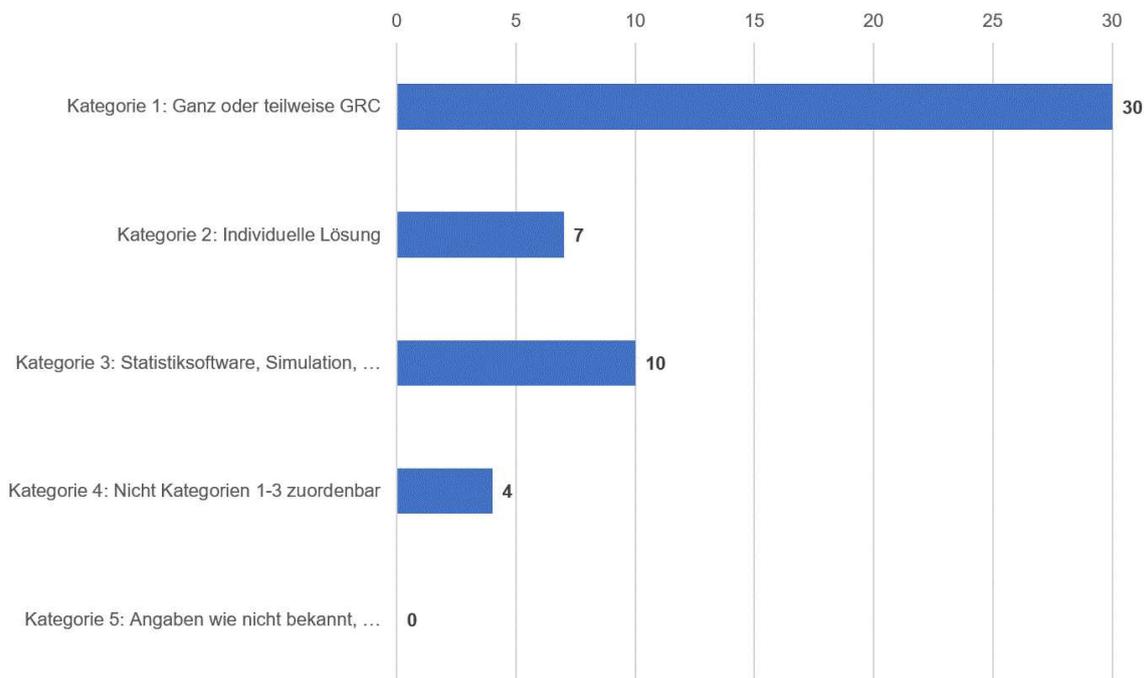

*Abbildung 11: IT-Anwendungen im Bereich Risikomanagement*

In Abbildung 11 ist die Übersicht über die Kategorien und die Nennungen pro Kategorie aufgeführt. Nennungen der Kategorie 2 folgen alle als Einzelnennung ohne weitere Angabe von zusätzlichen IT-Anwendungen. Im Fall von Kategorie 3 gab es 4 Einzelnennungen und 6 Angaben von IT-Anwendungen der Kategorie, bei denen zusätzlich auch ein oder zwei Produkte gleicher oder anderer Kategorie genannt wurden.

### *10. Bitte geben Sie die Software an, die Sie in Compliance einsetzen.*

**Frage 10**: Bitte geben Sie die Software an, die Sie in Compliance einsetzen.

Die Frage wurde angezeigt, wenn in Frage 8 der Bereich Compliance ausgewählt wurde und somit angegeben wurde, dass dort IT-Anwendungen existieren. Es handelte sich um ein Freitextfeld.

Von den n=33 Personen, die in Frage 8 angegeben haben, dass IT-Anwendungen in Compliance existieren, haben n=25 Personen Frage 10 beantwortet. Da es sich um ein Freitextfeld handelte und mehrere Eingaben möglich waren, liegen insgesamt 28 Softwarenennungen vor. Es wurden von den Befragten maximal zwei Nennungen von IT-Anwendungen angegeben.

| Anzahl der Software-nennungen | Personenanzahl |
|---|---|
| 1 | 22 |
| 2 | 3 |
| Gesamt | 25 |

*Tabelle 3: Anzahl der Personen, die bei Frage 10 eine oder zwei IT-Anwendungen genannt haben.*

In Tabelle 3 ist ein Überblick über die Anzahl der Personen enthalten, die jeweils eine oder zwei IT-Anwendungen genannt haben.

Die Einteilung in Kategorien erfolgt analog zu Tabelle 2 in Frage 9. Auch ist aus den dort genannten Gründen nicht immer eindeutig gegeben, welche IT-Anwendungen genau eingesetzt werden.

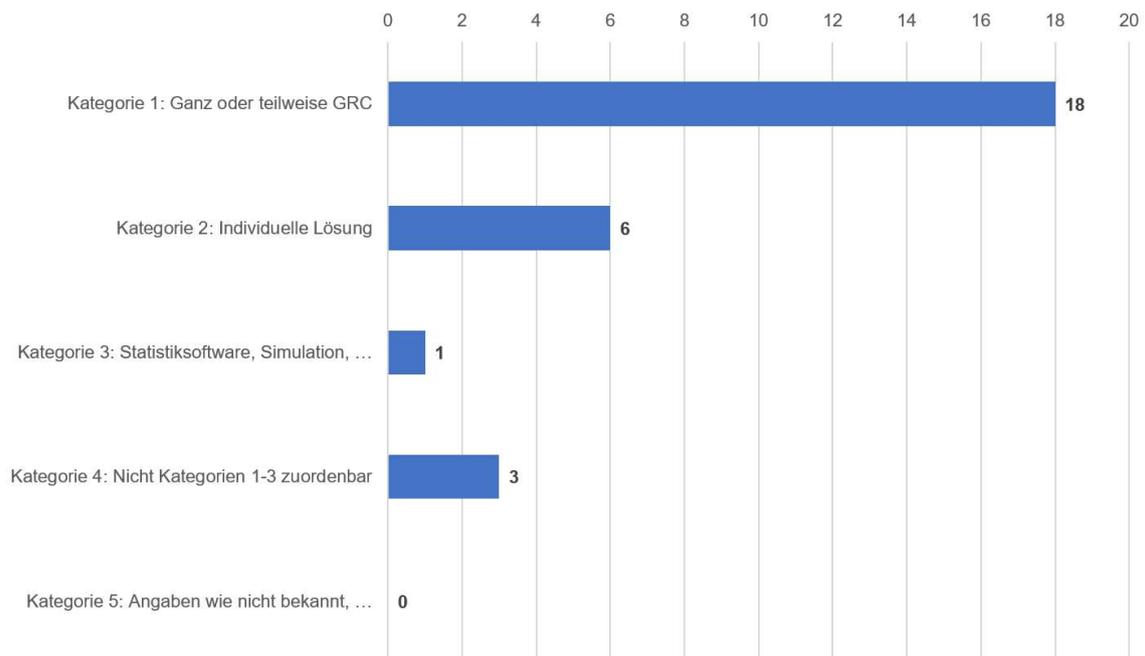

*Abbildung 12: IT-Anwendungen im Bereich Compliance*

Ein Überblick über die Einteilung der Nennungen in die Kategorien nach Tabelle 2 kann Abbildung 12 entnommen werden. Die einzige Meldung in Kategorie 3 erfolgte nicht als Einzelnennung, sondern zusammen mit einer Angabe aus Kategorie 1. Die Angaben zu Kategorie 2 erfolgten auch hier alle als Einzelnennung.

## 11. Bitte geben Sie die Software an, die Sie im IKS einsetzen.

**Frage 11**: Bitte geben Sie die Software an, die Sie im IKS einsetzen.

Die Frage wurde angezeigt, wenn in Frage 8 der Bereich IKS ausgewählt wurde und somit angegeben wurde, dass dort IT-Anwendungen existieren. Es handelte sich um ein Freitextfeld.

Von den n=37 Personen, die in Frage 8 angegeben haben, dass IT-Anwendungen im IKS existieren, haben n=31 Personen Frage 11 beantwortet. Da es sich um ein Freitextfeld handelte und mehrere Eingaben möglich waren, liegen insgesamt 36 Softwarenennungen vor. Es wurden von den Befragten maximal drei Nennungen von IT-Anwendungen angegeben.

| Anzahl der Software-nennungen | Personenanzahl |
|---|---|
| 1 | 27 |
| 2 | 3 |
| 3 | 1 |
| Gesamt | 31 |

*Tabelle 4: Anzahl der Personen, die bei Frage 11 eine, zwei oder drei IT-Anwendungen genannt haben*

In Tabelle 4 ist ein Überblick über die Anzahl der Personen enthalten, die jeweils eine, zwei oder drei IT-Anwendungen genannt haben.

Die Einteilung in Kategorien erfolgt analog zu Tabelle 2 in Frage 9. Auch ist aus den dort genannten Gründen nicht immer eindeutig gegeben, welche IT-Anwendungen genau eingesetzt werden.

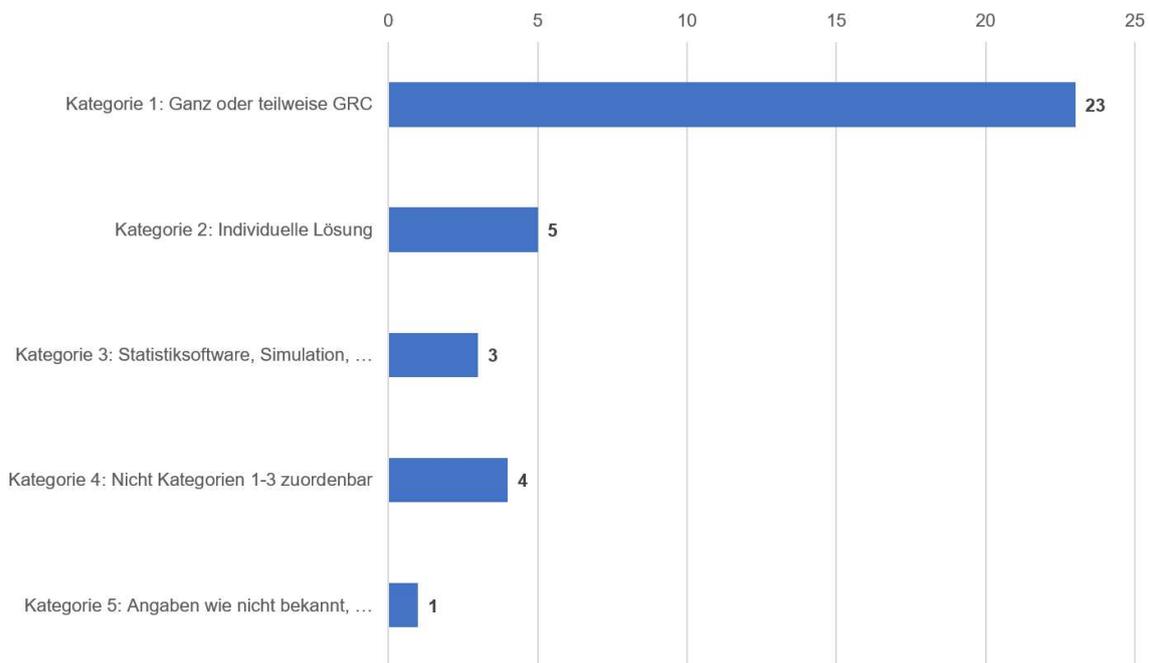

*Abbildung 13: IT-Anwendungen im Bereich IKS*

Zwei der drei Nennungen aus Kategorie 3 erfolgten zusammen mit anderen Angaben zu IT-Anwendungen. In einem Fall handelte es sich um eine Einzelnennung.

## 12. Bitte geben Sie die Software an, die Sie in der Internen Revision einsetzen.

**Frage 12**: Bitte geben Sie die Software an, die Sie in der Internen Revision einsetzen.

Die Frage wurde angezeigt, wenn in Frage 8 der Bereich Interne Revision ausgewählt wurde und somit angegeben wurde, dass dort IT-Anwendungen existieren. Es handelt sich um ein Freitextfeld.

Von den n=36 Personen, die in Frage 8 angegeben haben, dass IT-Anwendungen in der Internen Revision existieren, haben n=30 Personen Frage 12 beantwortet. Da es sich um ein Freitextfeld handelte und mehrere Eingaben möglich waren, liegen insgesamt 41 Softwarenennungen vor. Es wurden von den Befragten maximal sechs Nennungen von IT-Anwendungen angegeben.

| Anzahl der Software-nennungen | Personenanzahl |
|---|---|
| 1 | 25 |
| 2 | 3 |
| 3 | 0 |
| 4 | 1 |
| 5 | 0 |
| 6 | 1 |
| Gesamt | 41 |

*Tabelle 5: Anzahl der Personen, die bei Frage 12 zwischen ein und sechs IT-Anwendungen genannt haben*

In Tabelle 5 ist ein Überblick über die Anzahl der Personen enthalten, die jeweils zwischen ein bis zu sechs IT-Anwendungen genannt haben.

Die Einteilung in Kategorien erfolgt analog zu Tabelle 2 in Frage 9. Auch ist aus den dort genannten Gründen nicht immer eindeutig gegeben, welche IT-Anwendungen genau eingesetzt werden.

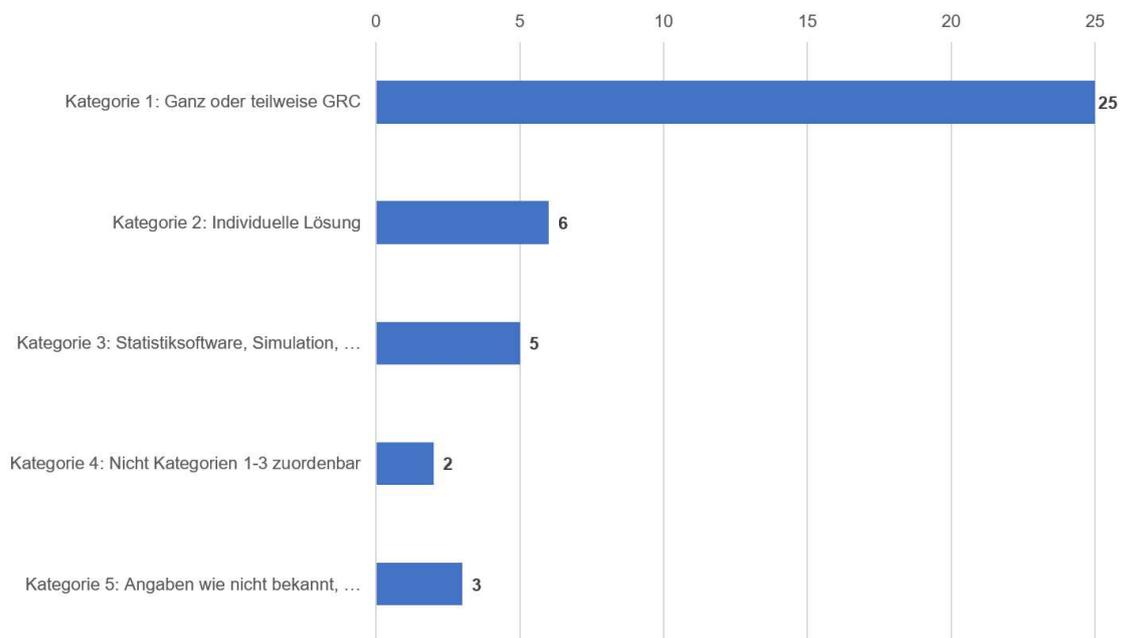

*Abbildung 14: IT-Anwendungen im Bereich Interne Revision*

Alle Angaben der Kategorie 3 in Abbildung 14 sind auf die beiden Rückmeldungen mit vier bzw. sechs Angaben zu IT-Anwendungen zurückzuführen. Bei der Kategorie 2 liegen wiederum ausschließlich Einzelnennungen vor.

## *9. bis 12. Gemeinsamer Überblick*

Da die Fragen 9 bis 12 in einem Zusammenhang stehen, wird in diesem Abschnitt ein kurzer Überblick über die gemeinsame Beantwortung der Fragen gegeben.

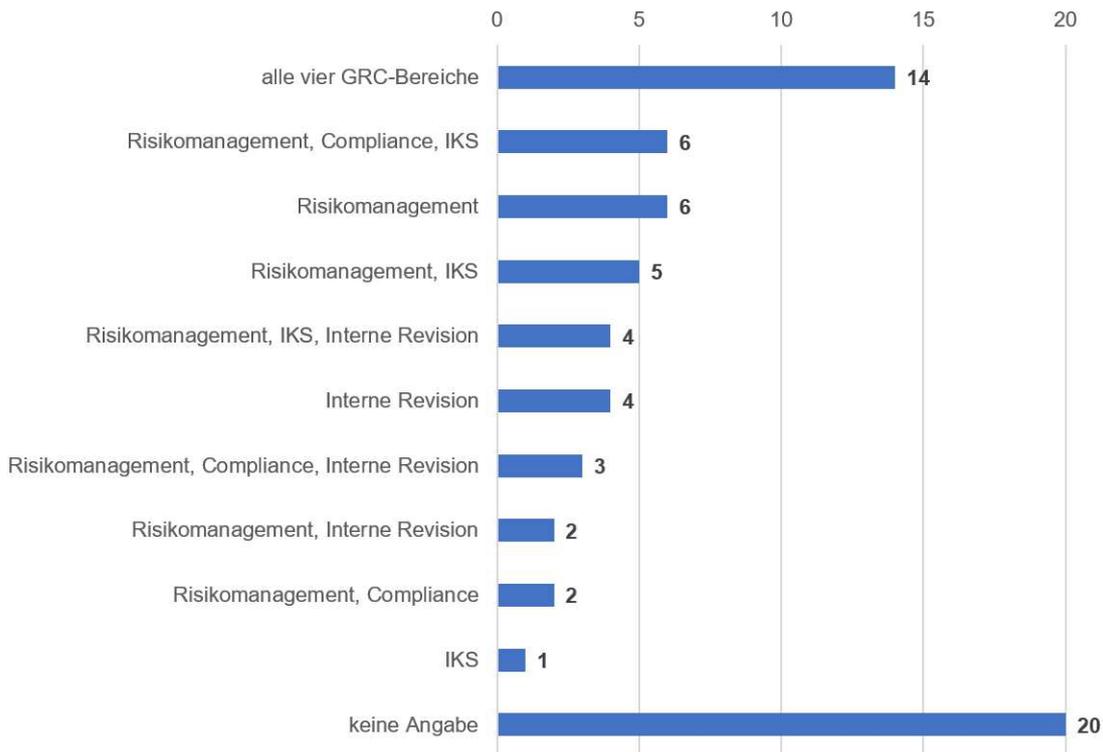

*Abbildung 15: Kombination der Angaben zu IT-Anwendungen (n=67)*

Wie in Abbildung 15 zu sehen, haben insgesamt 47 Personen Angaben zu IT-Anwendungen getätigt. Aus Abbildung 9 (Frage 8) geht hervor, dass diese Fragen insgesamt 57 Personen vorlagen. Zu den 10 Personen, die die Fragen auf Grund der Beantwortung von Frage 8 gar nicht erst vorgelegt bekommen haben, kommen somit noch 10 Personen, die für keinen der GRC-Bereiche Eingaben getätigt haben.

In der Abbildung sind nur die Bereiche und Kombinationen von Bereichen aufgeführt, für die pro Bereich irgendeine IT-Anwendung angegeben wurde. Von diesen 47 Fragebögen mit Angaben zu IT-Anwendungen wurden in den meisten Fällen (30 %) für alle vier GRC-Bereiche IT-Anwendungen aufgeführt. Angaben wie „unbekannt", „keine Angabe", oder ähnliche Angaben wurden dabei nicht als IT-Anwendung berücksichtigt.

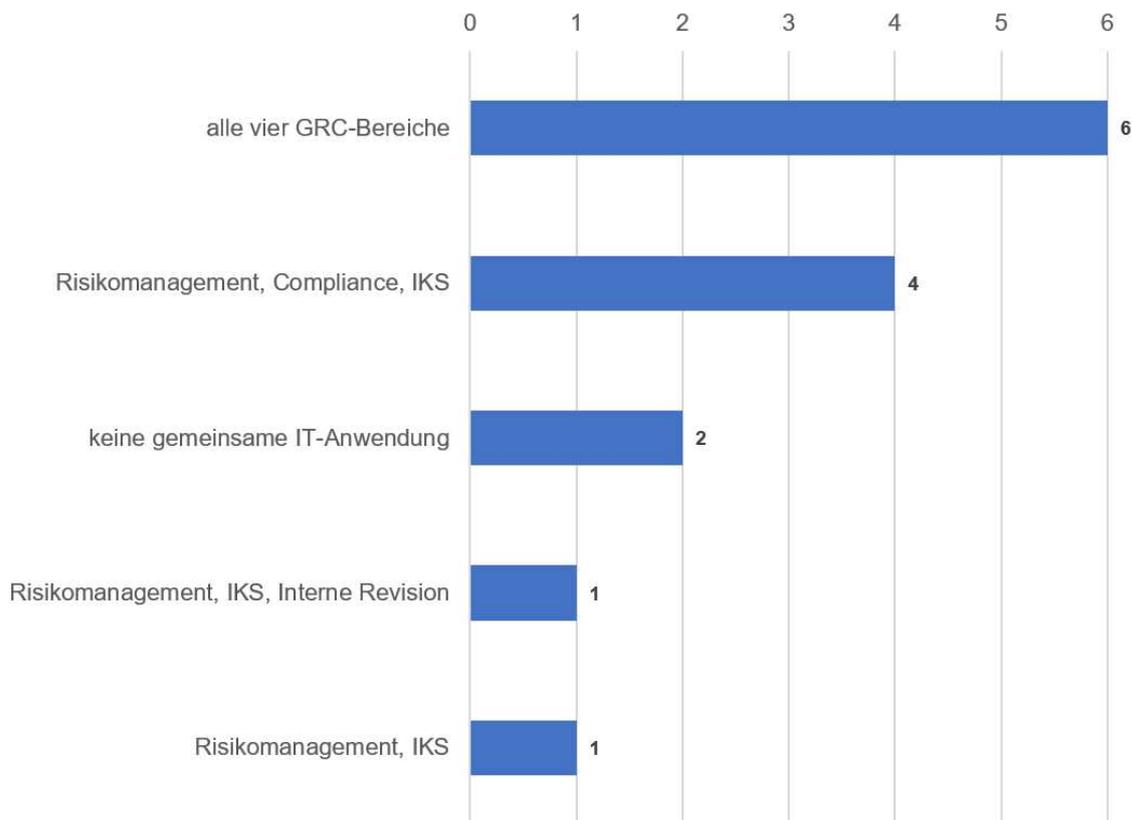

*Abbildung 16: Kombinationen gleicher IT-Anwendungen bei allen vier GRC-Bereichen (n=14)*

Abbildung 16 betrachtet die n=14 Angaben, bei denen für alle vier GRC-Bereiche IT-Anwendungen genannt wurden. Aufgeführt sind die Ergebnisse, bei denen für jeden Bereich mindestens eine der genannten IT-Anwendungen übereinstimmt. Immerhin n=6 Personen geben für alle vier Bereiche identische IT-Anwendungen an. Hierbei handelte es sich jedoch in 4 Fällen um Anwendungen der Kategorie 2 (IT-Anwendungen, die als Eigenentwicklung oder individuelle Branchenlösung bezeichnet werden).

## 13. Wie beurteilen Sie den Automatisierungsgrad der IT-Anwendungen in GRC bezogen auf den Datenaustausch?

**Frage 13**: Wie beurteilen Sie den Automatisierungsgrad der IT-Anwendungen in GRC bezogen auf den Datenaustausch?

Diese Frage wurde nur angezeigt, wenn in Frage 8 mindestens einer der Bereiche Risikomanagement, Compliance, IKS und Interne Revision ausgewählt wurde und somit angegeben wurde, dass in mindestens einem dieser Bereiche IT-Anwendungen existieren.

Angezeigt wurde eine Skala mit den Extremwerten vollständig manuell (1) und vollständig automatisiert (11) zur Auswahl des Grades der Automatisierung des Datenaustauschs. Mit einem Schieberegler konnte ein Wert zwischen diesen beiden Extremwerten ausgewählt werden. In der Auswertung wurden die Zahlen in 10 %-Schritten dargestellt. D.h. eine Angabe von 1 bedeutet 0 % Automatisierung, 2 entspricht 10 % Automatisierung und bei Angabe von 11 bedeutet dies eine Automatisierung von 100 %.

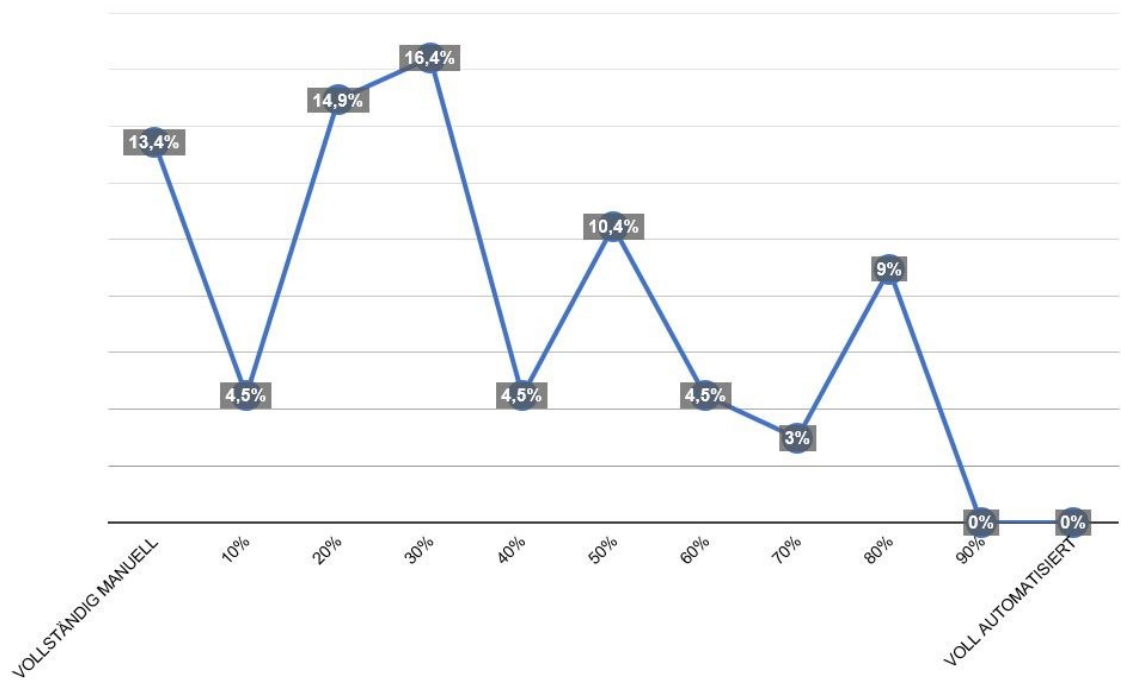

*Abbildung 17: Angabe zum Grad der Automatisierung des Datenaustausches (n=67)*

Abbildung 17 gibt den Grad der Automatisierung des Datenaustausches für alle n=67 Fragebögen an. Dabei wurde von n=54 Personen eine Angabe getätigt, die aufgeführten Angaben summieren sich somit auf 80,6 %. Die meisten Angaben befinden sich im unteren Bereich bei einem Automatisierungsgrad kleiner 40 %. Die Angaben 90 % und 100 % wurden nicht ausgewählt. Immerhin 13,4 % gaben an, dass der Datenaustausch vollständig manuell vonstattengeht. Hier wäre es interessant zu erfahren, wie genau vollständig manuell von den Befragten interpretiert wird.

## 2.2  Teil 2: KI-Potentiale in GRC

Werden in Teil 1 der Studie die allgemeinen Rahmenbedingungen geklärt, so soll in Teil 2 der Studie genauer auf die Potentiale von Künstlicher Intelligenz in Governance, Risk und Compliance eingegangen werden. Zu diesem Zweck verbinden die Fragen direkt diese beiden Bereiche.

### 14. Welche Anwendungsgebiete sehen Sie für den Einsatz von KI-Verfahren in GRC?

**Frage 14**: Welche Anwendungsgebiete sehen Sie für den Einsatz von KI-Verfahren in GRC?

Die Frage wurde mit folgendem Hinweis versehen:

„Im Folgenden sollen Potentiale identifiziert werden, die in den einzelnen GRC-Bereichen konkret in Ihrem Unternehmen vorhanden sind."

Dabei konnte für die Potentiale

- Systematische Einordnung von Daten in Klassen
- Entdeckung von ähnlichen Strukturen in Daten durch Clustering
- Prognose von Entwicklungen durch Regression
- Erlernen von lohnenswerten Handlungsstrategien

jeweils einer oder mehrere der Bereiche Risikomanagement, Compliance und IKS ausgewählt werden.[4]

---

[4] Die Potentiale orientieren sich an den Anwendungsgebieten der unterschiedlichen Lernstile, so die Klassifikation und Regression im überwachten Lernen, Clustering im unüberwachten Lernen und dem Erlernen von Handlungsstrategien im Reinforcement Learning. Vgl. hierzu beispielsweise auch Jo, 2021, Abschnitte 1.2 und 1.3.

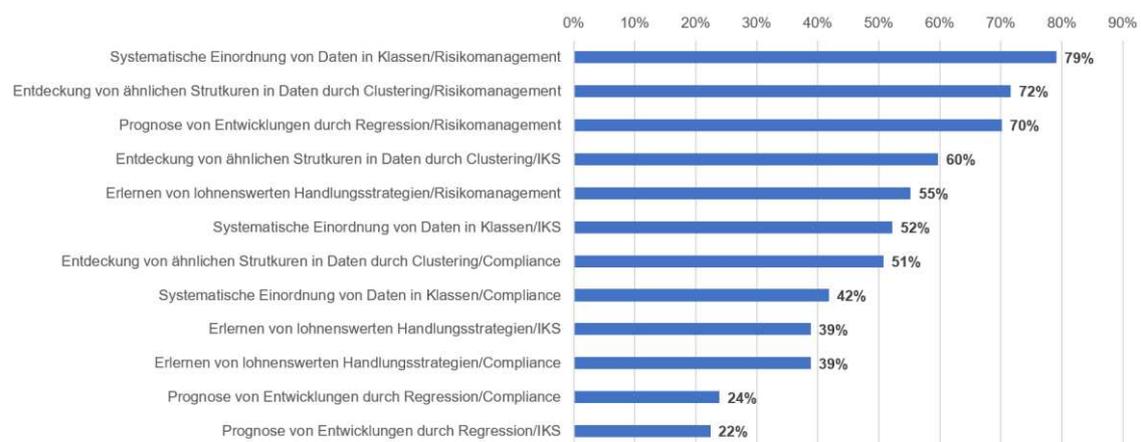

*Abbildung 18: Anwendungsgebiete für den Einsatz von KI-Verfahren (n=67)*

In Abbildung 18 sind für alle n=67 Fragebögen die Potentiale und die drei GRC-Bereiche Risikomanagement, Compliance und IKS, in denen diese Anwendungsgebiete von den Befragten gesehen wurden, aufgeführt. Bezogen auf den Bereich Risikomanagement finden sich dabei alle vier Potentiale in den fünf häufigsten Nennungen.

### 15. Falls Sie Potentiale bzw. Beispiele für das Risikomanagement sehen, skizzieren Sie diese bitte.

**Frage 15**: Falls Sie Potentiale bzw. Beispiele für das Risikomanagement sehen, skizzieren Sie diese bitte.

Es handelte sich um ein Freitextfeld. Bei Bedarf wurden zusätzliche Freitextfelder eingeblendet.

Die Frage wurde von 26 Personen beantwortet. Dabei wurde in 22 Fällen eine Angabe getätigt, in 3 Fällen zwei Angaben und in einem Fall drei Angaben, so dass insgesamt 31 Angaben zu KI-Potentialen bzw. Beispielen für das Risikomanagement vorliegen.

Auf Grundlage der vorliegenden Angaben wurde eine erste Einteilung mit fünf Kategorien ermittelt, denen die Angaben zugeordnet werden können. Diese Kategorien sind in Tabelle 6 aufgeführt. Diese Einteilung wurde auch bei den Fragen 16 und 17 angewendet.

| Kategorie | Beschreibung |
|---|---|

| 1 | KI-gestützte Datenanalyse, beispielsweise zur Mustererkennung, Klassifikation oder Clustering u.a. mit den Zielsetzungen Risikoidentifikation und -bewertung, Betrugserkennung, Herausfinden von Inkonsistenzen, Qualitätssicherung, Kontrolle, unabhängige Bewertung |
|---|---|
| 2 | Frühwarnsystem, Prognosen, Predictive Analytics |
| 3 | Integration von Systemen, Anwendungen und Bereichen |
| 4 | Modellierung, Simulation, Szenario-Analyse, Algorithmen |
| 5 | Angaben, die nicht direkt mit KI-Potentialen und Beispielen in Verbindung gebracht werden können |

*Tabelle 6: Kategorien zur Einteilung der aufgeführten Potentiale*

Die Einteilung der Antworten in die Kategorien erfolgte nach bestem Wissen und Gewissen, jedoch unterliegen einige Zuordnungen einem gewissen Interpretationsspielraum, da im Rahmen der Umfrage natürlich keine Möglichkeit zur Nachfrage bestand.

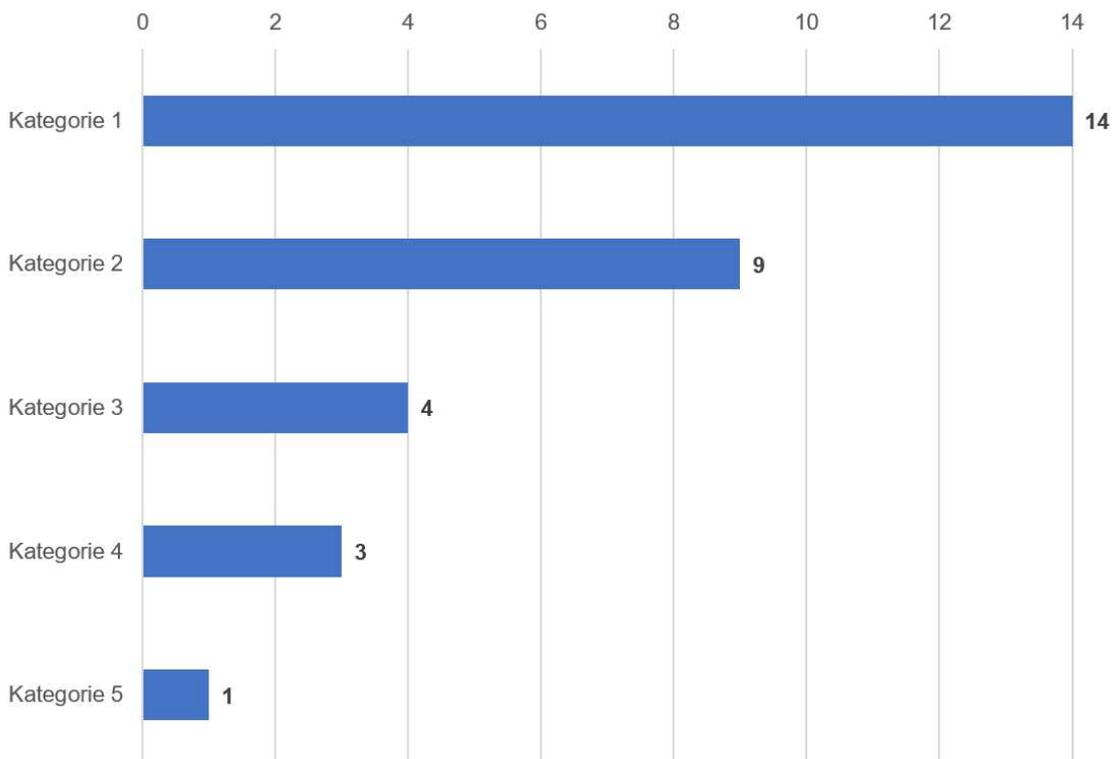

*Abbildung 19: Einteilung der Potentiale bzw. Beispiele für das Risikomanagement gemäß den Kategorien aus Tabelle 6*

In Abbildung 19 ist erkennbar, dass ein großer Teil der Angaben (45 %) sich auf Mustererkennung, Klassifikation oder auch Clustering bezog. Da die Frage im Anschluss an Frage 14 mit einer Auflistung möglicher Potentiale gestellt wurde, kann dies als Erweiterung der dortigen Angaben gesehen werden. Auch in Frage 16 (Compliance) und 17 (IKS) nimmt diese Kategorie einen großen Raum ein.

Auch Kategorie 2 mit den Angaben zu Prognosen und Frühwarnsystemen ist aus Frage 14 ableitbar. Diese Kategorie wurde in den folgenden Fragen zu Compliance und IKS nicht genannt, stellt also ein Alleinstellungsmerkmal für die Einsatzmöglichkeiten im Risikomanagement bezogen auf diese Studie dar. Mit Abstand folgten die Kategorien 3 und 4.

### 16. Falls Sie Potentiale bzw. Beispiele für Compliance sehen, skizzieren Sie dies bitte.

**Frage 16**: Falls Sie Potentiale bzw. Beispiele für Compliance sehen, skizzieren Sie diese bitte.

Es handelte sich um ein Freitextfeld. Bei Bedarf wurden zusätzliche Freitextfelder eingeblendet.

Die Frage wurde von 18 Personen beantwortet, die alle ausschließlich eine einzelne Angabe getätigt haben, so dass insgesamt auch 18 Angaben zu Potentialen bzw. Beispielen für Compliance vorliegen.

Die Antworten aus den Freitextfeldern wurden wiederum gemäß Tabelle 6 kategorisiert und in Abbildung 20 aufgeführt.

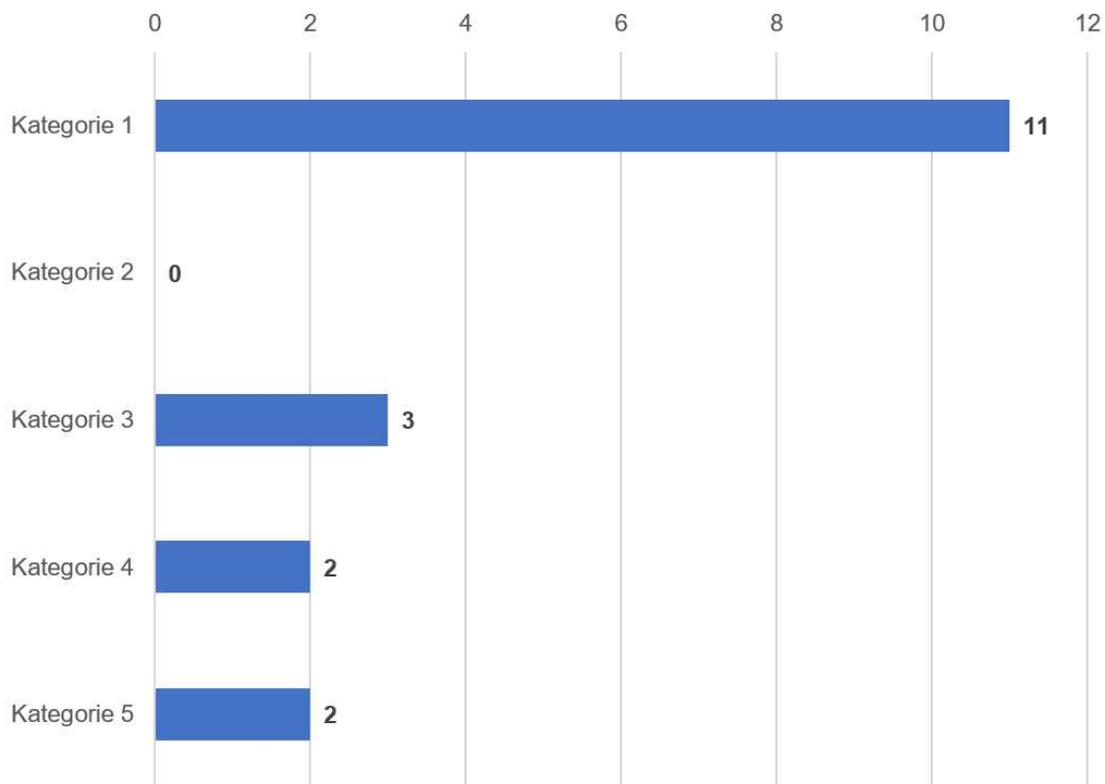

*Abbildung 20: Einteilung der Potentiale bzw. Beispiele für Compliance gemäß den Kategorien aus Tabelle 6*

Bei Kategorie 1 waren wiederum die meisten Zuweisungen von Antworten zu verzeichnen (61 %). Kategorie 2 war in diesem Fall nicht vertreten.

## 17. Falls Sie Potentiale bzw. Beispiele für das IKS sehen, skizzieren Sie diese bitte.

**Frage 17**: Falls Sie Potentiale bzw. Beispiele für IKS sehen, skizzieren Sie diese bitte.

Es handelte sich um ein Freitextfeld. Bei Bedarf wurden zusätzliche Freitextfelder eingeblendet.

Die Frage wurde von 20 Personen beantwortet, die alle ausschließlich eine einzelne Angabe getätigt haben, so dass insgesamt 20 Angaben zu Potentialen bzw. Beispielen für das IKS vorliegen.

Die Antworten aus den Freitextfeldern wurden wiederum gemäß Tabelle 6 kategorisiert und in Abbildung 21 aufgeführt.

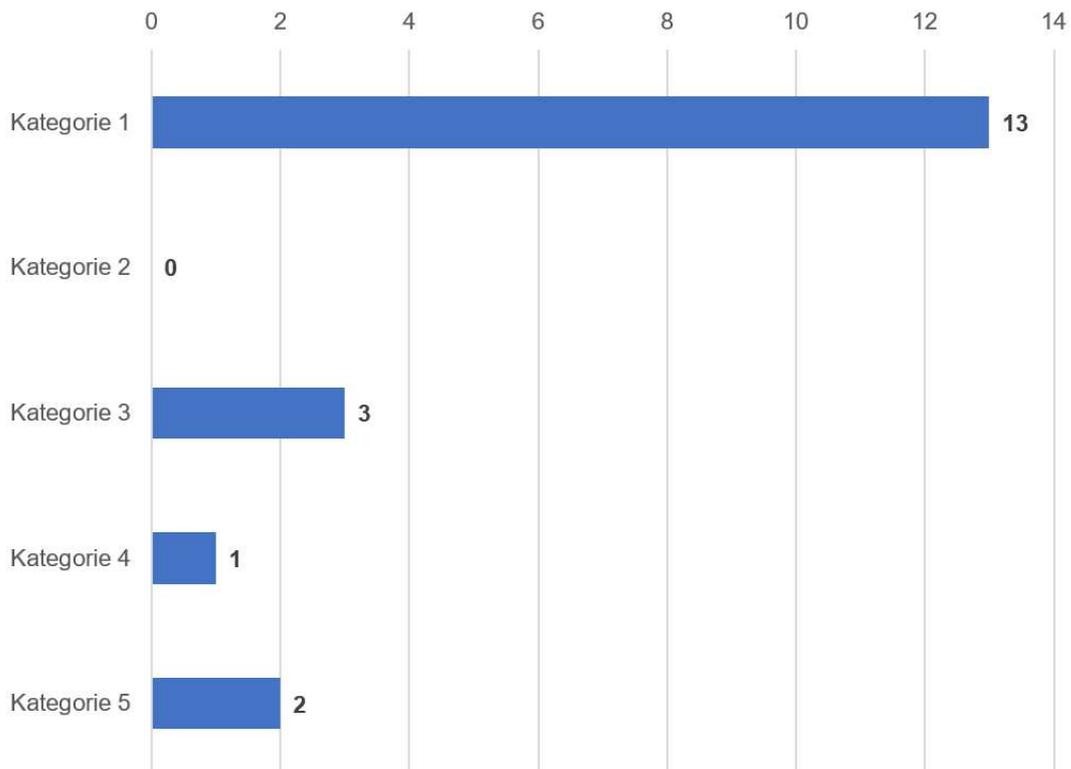

*Abbildung 21: Einteilung der Potentiale bzw. Beispiele für das IKS gemäß den Kategorien aus Tabelle 6*

Im Ergebnis lässt sich das gleiche festhalten, wie bereits bei Frage 16. Kategorie 1 wurde als Potential erkannt (65 %), Kategorie 2 ist nicht vertreten.

### *15. bis 17. Gemeinsamer Überblick*

Die Beantwortung der Fragen 15 bis 17 steht in einem engen Zusammenhang, da diese Fragen nicht unabhängig voneinander stehen, sondern jeweils hintereinander von den Befragten beantwortet wurden.

Dabei beantworteten 6 Personen die Fragen 15, 16 und 17 zu den unterschiedlichen GRC-Bereichen identisch. Hiervon fielen drei Antworten in Kategorie 3, zwei Antworten in Kategorie 1 und eine Antwort in Kategorie 4.

10 Personen verwendeten unterschiedliche Kategorien bei der Beantwortung der Fragen 15, 16 und 17.

### 18. Wird KI bereits in GRC eingesetzt?

**Frage 18**: Wird KI bereits in GRC eingesetzt?

Antwortmöglichkeiten waren: ja, nein, nicht bekannt. (n=67)

Die Frage unterscheidet sich zu Frage 1 insoweit, dass sie sich konkret auf den Bereich GRC und nicht auf das gesamte Unternehmen bezieht.

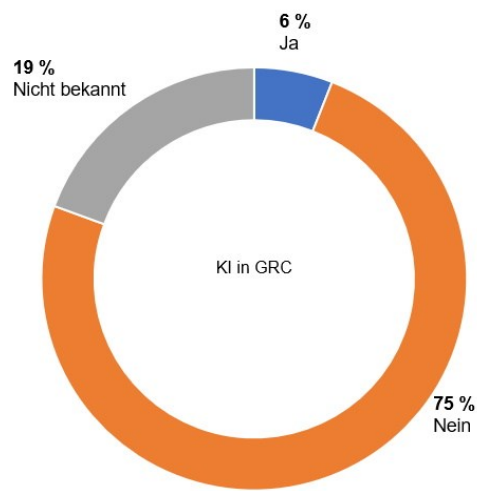

*Abbildung 22: Einsatz von Künstlicher Intelligenz in GRC (n=67)*

Wie in Abbildung 22 dargestellt, gaben von den n=67 Befragten die Mehrheit (75 %) an, dass Künstliche Intelligenz in GRC nicht eingesetzt wird. Es gaben 6 % an, dass Künstliche Intelligenz eingesetzt wird, 19 % gaben an, dass ihnen der Einsatz von Künstlicher Intelligenz in GRC nicht bekannt sei.

### 19. Für welche Bereiche oder Systeme mit GRC-Bezug sehen Sie Potentiale für den Einsatz von KI?

**Frage 19**: Für welche Bereiche oder Systeme mit GRC-Bezug sehen Sie Potentiale für den Einsatz von KI?

Es standen die Antwortmöglichkeiten

- Interne Revision
- IT-Management
- Information Security Management System
- Weitere Bereiche oder Systeme (mit Texteingabe)

zur Verfügung. Mehrfachauswahl war möglich.

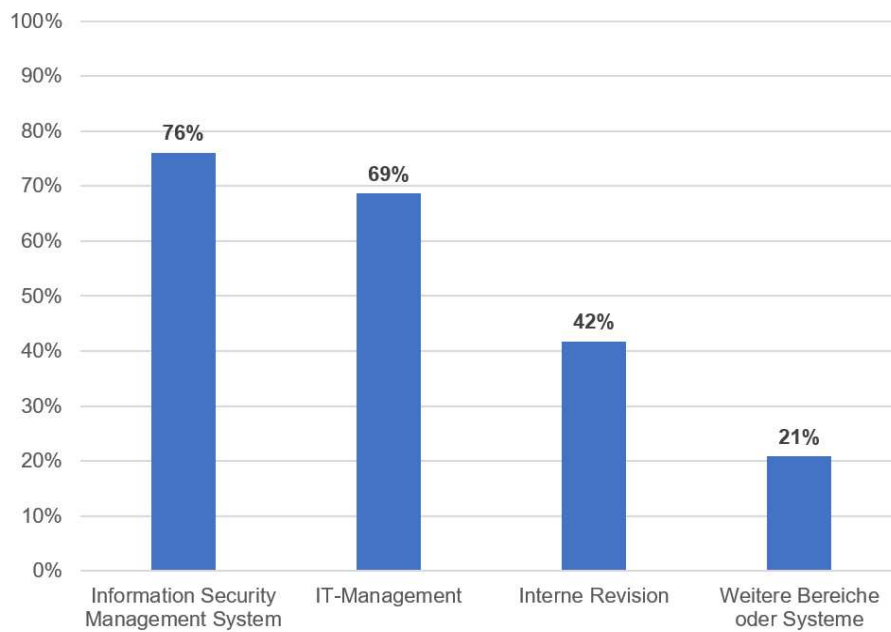

*Abbildung 23: Weitere Potentiale für den Einsatz von KI (n=67)*

Der Bereich Information Security Management System wurde in den meisten Fällen ausgewählt (76 %), gefolgt von IT-Management (69 %) und Interne Revision (42 %). Weitere Bereiche oder Systeme wurden von 21 % ausgewählt.

Von den weiteren Bereichen oder Systemen wurden folgende konkret benannt: Controlling (2), Qualitätsmanagement (2), Umwelt-, IT- und Energiemanagement-Systeme, SOX-Compliance, Fraudmanagement, Datenschutz, Geldwäsche, BCM, Lieferketten, Produktionssteuerung, Maschinenvernetzung, Bestandsmanagement, Überwachte Lernprozesse sowie „alle Unternehmensfunktionen".

## 20. Welche Herausforderungen sehen Sie in Ihrem Unternehmen für den erfolgreichen Einsatz von KI speziell in GRC?

**Frage 20**: Welche Herausforderungen sehen Sie in Ihrem Unternehmen für den erfolgreichen Einsatz von KI speziell in GRC?

Es standen die Auswahlmöglichkeiten

- Personalbeschaffung und -weiterbildung bezogen auf KI-Kompetenzen
- Datenqualität und -quantität
- Homogene IT-Systemlandschaft
- Finanzierbarkeit
- Tone of the top
- Transparenz gesetzlicher Vorgaben
- Robustheit gegenüber potentiellen regulatorischen Anforderungen
- Identifikation von Anwendungspotentialen
- Integration von KI-Technologien in die GRC-Funktionen und -Prozesse
- Erweiterung der Kontrollfunktion in GRC auf eingesetzte KI-Technologien
- Weitere Herausforderungen (mit Texteingabe)

zur Verfügung. Mehrfachauswahl war möglich.

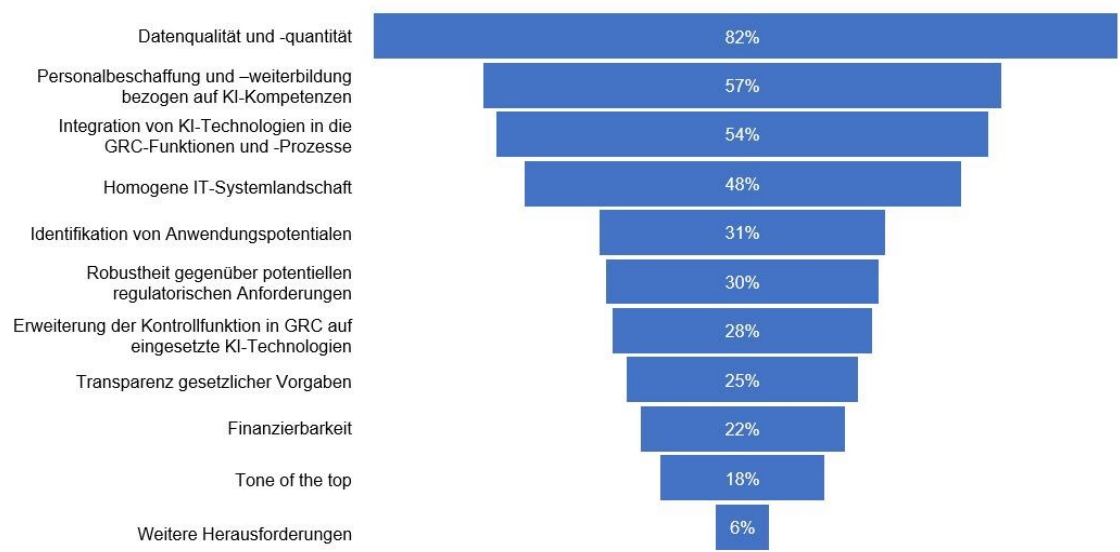

*Abbildung 24: Herausforderungen für den erfolgreichen Einsatz von KI in GRC (n=67)*

Von den n=67 Befragten haben alle bis auf eine Person mindestens eine Herausforderung für den erfolgreichen Einsatz von KI in GRC angegeben. Ein Überblick über die ausgewählten Antworten ist in Abbildung 24 aufgeführt. Der Punkt Datenqualität und -quantität, der eine essentielle Grundlage für den erfolgreichen Einsatz von KI darstellt, wurde mit 82 % am häufigsten genannt. Interessanterweise wurde die Finanzierbarkeit mit 22 % im unteren Bereich aufgeführt.

Als weitere Herausforderungen wurden konkret genannt: Es scheitert meistens an Hierarchien und dem Silodenken, Mangelnde Akzeptanz bei den Risk-Ownern, wenig praktischer Nutzen, dringlichere Themen, clear Return of Investment for management.

## 2.3  Teil 3: Allgemeine Angaben zum Unternehmen

Um die Antworten besser einordnen zu können, war es von Interesse allgemeine Informationen darüber zu erhalten, wer den Fragebogen ausgefüllt hat, und auf welche Rahmenbedingungen sich die Antworten beziehen. Aus diesem Grund wurden in Teil 3 der Studie allgemeine Angaben erhoben.

### *21. – 23. Einordnung Kleinstunternehmen, kleine und mittlere Unternehmen.*

Die Fragen 21, 22 und 23 dienten der Einordnung des Unternehmens. Herangezogen wurde die Definition der EU-Kommission zur Festlegung, ob es sich bei dem Unternehmen um ein Kleinstunternehmen oder ein kleines oder mittleres Unternehmen (KMU) handelt.[5] Hierzu wurden die Fragen wie folgt formuliert:

**Frage 21: Wie viele Beschäftigte hatte Ihr Unternehmen im letzten Geschäftsjahr vor der COVID-19 Pandemie?**

Mögliche Angaben waren:

- Weniger als 10 Beschäftigte

---

[5] Vgl. Eurostat.

- 10 bis 49 Beschäftigte
- 50 bis 249 Beschäftigte
- 250 oder mehr Beschäftigte
- Keine Angabe

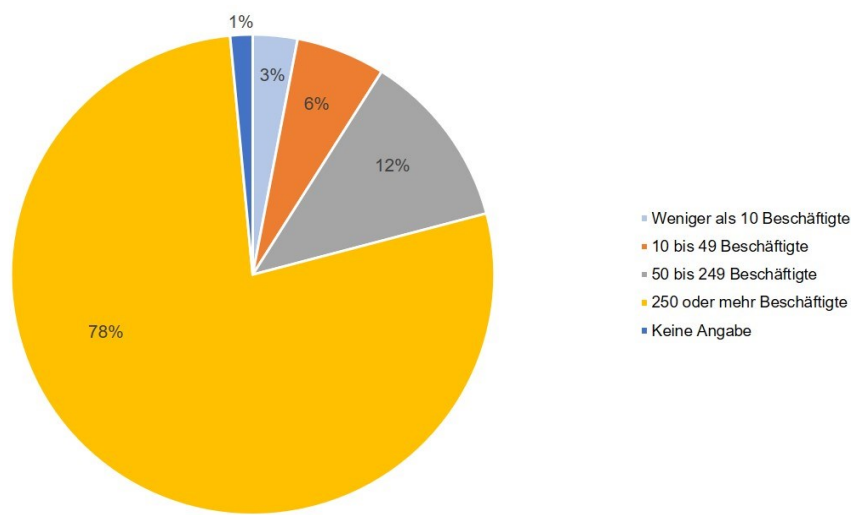

*Abbildung 25: Anzahl der Beschäftigten im letzten Geschäftsjahr vor der COVID-19 Pandemie (n=67)*

Abbildung 25 ist zu entnehmen, dass ein Großteil (78 %) der Befragten (n=67) angab, dass das Unternehmen 250 oder mehr Beschäftigte besitzt. Diese Unternehmen sind damit keine Kleinstunternehmen, kleine oder mittlere Unternehmen.

**Frage 22: Wie hoch war der Jahresumsatz des letzten Geschäftsjahres vor der COVID-19 Pandemie?**

Mögliche Antworten waren:

- Bis 2 Mio. €
- Mehr als 2 Mio. € bis 10 Mio. €
- Mehr als 10 Mio. € bis 50 Mio. €
- Mehr als 50 Mio. €
- Keine Angabe

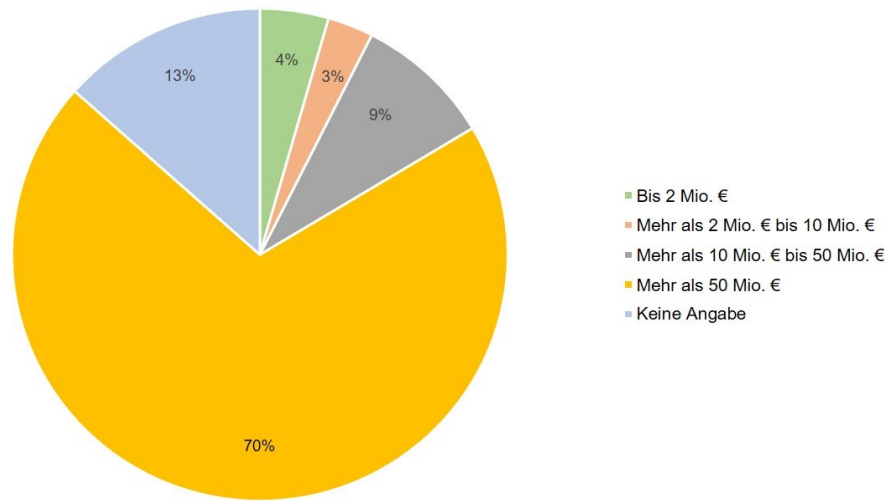

*Abbildung 26: Höhe des Jahresumsatzes des letzten Geschäftsjahres vor der COVID-19 Pandemie (n=67)*

Abbildung 26 ist zu entnehmen, dass ein Großteil der Befragten (70 %) in einem Unternehmen arbeitete, dessen Jahresumsatz des letzten Geschäftsjahres vor der COVID-19 Pandemie bei mehr als 50 Mio. € liegt. Dieser Wert beschreibt einen Schwellwert für die Einordnung zu Kleinstunternehmen, kleine und mittlere Unternehmen.

**Frage 23: Wie hoch war die Bilanzsumme des letzten Geschäftsjahres vor der COVID-19 Pandemie?**

Mögliche Antworten waren:

- Bis 2 Mio. €
- Mehr als 2 Mio. € bis 10 Mio. €
- Mehr als 10 Mio. € bis 43 Mio. €
- Mehr als 43 Mio. €
- Keine Angabe

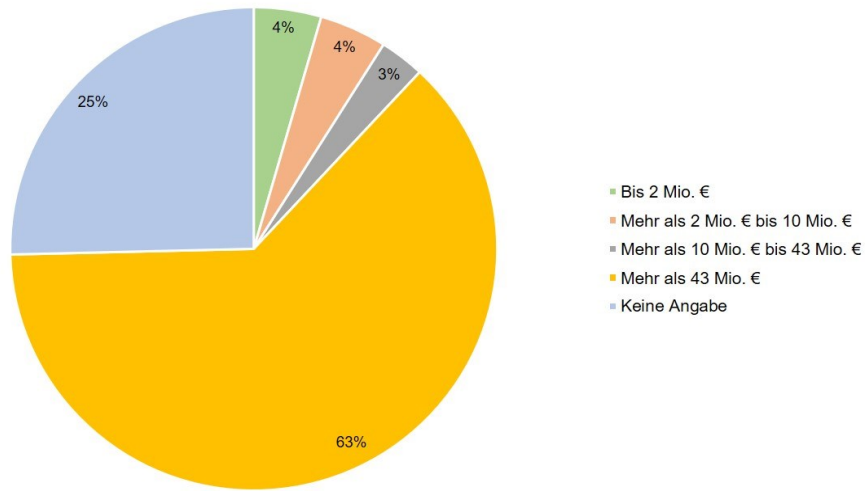

*Abbildung 27: Höhe der Bilanzsumme des letzten Geschäftsjahres vor der COVID-19 Pandemie (n=67)*

Der Schwellwert für die Einordnung zu Kleinstunternehmen, kleine und mittlere Unternehmen beträgt 43 Mio. €. Auch hier lag die überwiegende Mehrheit (63 %) über diesem Wert.

**Einteilung KMU**

Wurden die Fragen 21 bis 23 insgesamt im Kontext ausgewertet, so zeigte sich unter Anwendung der Definition der EU-Kommission für KMU[6] die Verteilung gemäß Abbildung 28. Die überwiegende Mehrheit der Rückmeldungen konnten nicht einem KMU zugeordnet werden.

---

[6] Vgl. Eurostat.

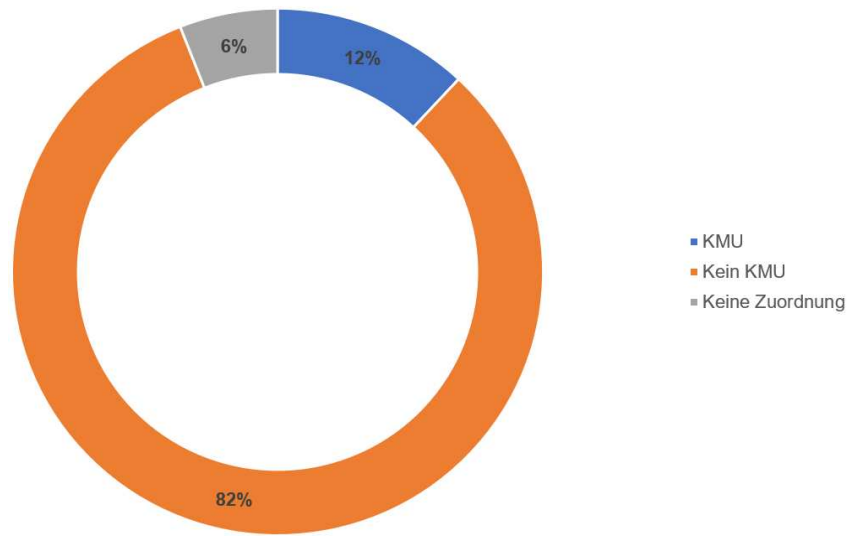

*Abbildung 28: Einteilung KMU (n=67)*

## 24. Welche Rechtsform hat Ihr Unternehmen?

**Frage 24**: Welche Rechtsform hat Ihr Unternehmen?

Mögliche Antworten waren:

- AG (börsennotiert)
- AG (nicht börsennotiert)
- Einzelunternehmer
- GmbH
- KG
- KGaA
- OHG
- Andere Rechtsform (mit Texteingabe)

Diese Frage besitzt Relevanz bezüglich möglicher rechtlicher Vorgaben in Abhängigkeit der Rechtsform eines Unternehmens.

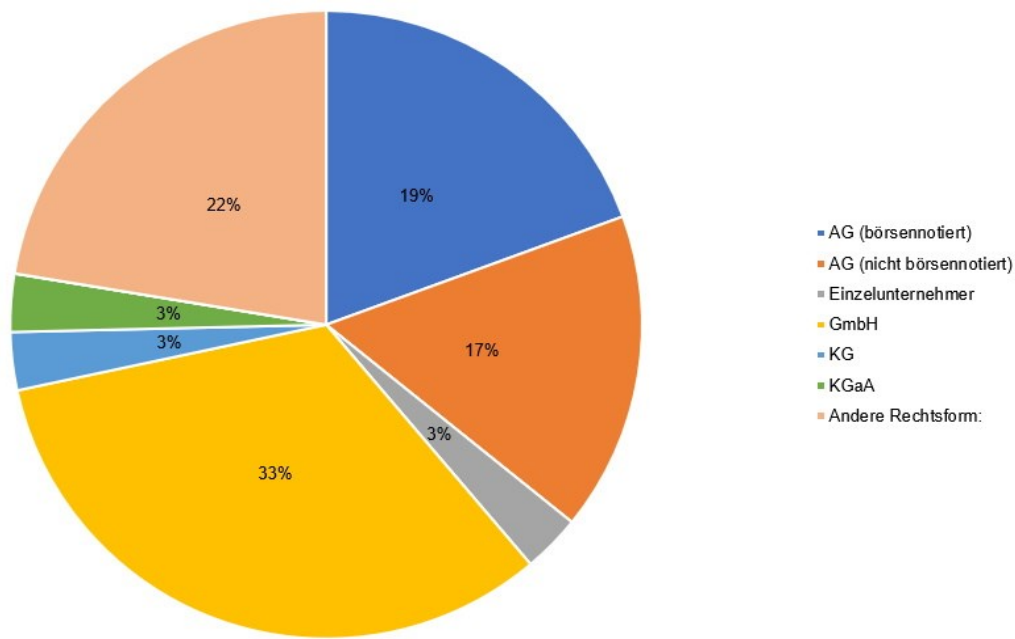

*Abbildung 29: Rechtformen (n=67)*

Unter „Andere Rechtsform" wurden die folgenden Eingaben getätigt: Gemeinschaftseinrichtung, VVaG (3), K.ö.R., GbR, eG, AöR, Stiftung, SE & Co. KG, Lehrinstitut, Öffentliche Rechtsform.

Wie aus Abbildung 29 hervorgeht, handelte es sich bei einem Großteil der Unternehmen um börsennotierte oder nicht börsennotierte Aktiengesellschaften (insgesamt 36 %) und um Gesellschaften mit beschränkter Haftung (33 %).

### 25. Welcher Branche gehört Ihr Unternehmen an?

**Frage 25**: Welcher Branche gehört Ihr Unternehmen an?

Ausgewählt werden konnten Wirtschaftszweige gemäß NACE Rev. 2: Statistische Systematik der Wirtschaftszweige in der Europäischen Gemeinschaft[7].

---

[7] Vgl. Eurostat, 2017. Die Darstellung ergibt sich dort aus der Tabelle auf S. 44 (grobes SNA/ISIC-Aggregat A*10/11) in Kombination mit der Tabelle auf S. 61 (grobe Struktur der NACE Rev. 2).

- Land- und Forstwirtschaft, Fischerei
- Verarbeitendes Gewerbe/Herstellung von Waren, Bergbau und Gewinnung von Steinen und Erden, sonstige Industrie
    - Bergbau und Gewinnung von Steinen und Erden
    - Verarbeitendes Gewerbe/Herstellung von Waren
    - Energieversorgung
    - Wasserversorgung, Abwasser- und Abfallentsorgung und Beseitigung von Umweltverschmutzungen
- Baugewerbe/Bau
- Handel, Verkehr und Lagerei
    - Handel, Instandhaltung und Reparatur von Kraftfahrzeugen
    - Verkehr und Lagerei
    - Gastgewerbe/Beherbergung und Gastronomie
- Information und Kommunikation
- Erbringung von Finanz- und Versicherungsdienstleistungen
- Grundstücks- und Wohnungswesen
    - Grundstückswesen
    - Wohnungswesen
- Erbringung von freiberuflichen, wissenschaftlichen und technischen Dienstleistungen sowie von sonstigen wirtschaftlichen Dienstleistungen
    - Erbringung von freiberuflichen, wissenschaftlichen und technischen Dienstleistungen
    - Erbringung von sonstigen wirtschaftlichen Dienstleistungen
- Öffentliche Verwaltung, Verteidigung, Sozialversicherung, Erziehung und Unterricht, Gesundheits- und Sozialwesen
    - Öffentliche Verwaltung, Verteidigung, Sozialversicherung
    - Erziehung und Unterricht
    - Gesundheits- und Sozialwesen
- Sonstige Dienstleistungen
    - Kunst, Unterhaltung und Erholung
    - Erbringung von sonstigen Dienstleistungen

- Private Haushalte mit Hauspersonal, Herstellung von Waren und Erbringung von Dienstleistungen durch private Haushalte für den Eigenbedarf ohne ausgeprägten Schwerpunkt
- Exterritoriale Organisationen und Körperschaften

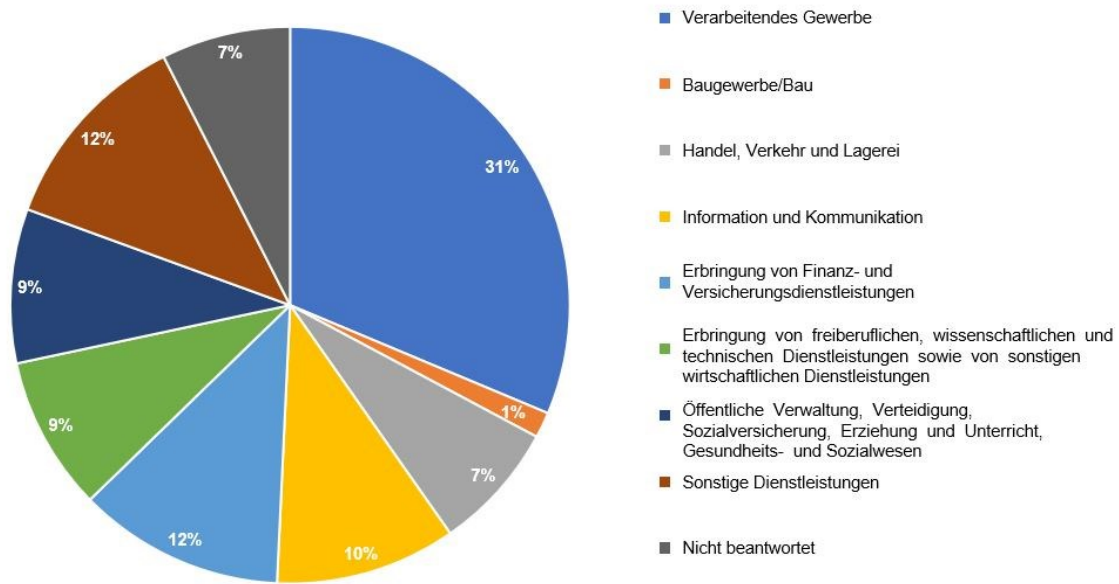

*Abbildung 30: Branche nach NACE Rev.2 (n=67)*

In Abbildung 30 werden die Angaben nach den Oberbegriffen aggregiert aufgeführt. Die Zahlen beziehen sich auf die Gesamtanzahl n=67. 5 Personen (7 %) hatten die Frage nicht beantwortet. Ein großer Anteil der befragten Personen (31 %) kam aus dem Bereich „Verarbeitendes Gewerbe". Gefolgt von „Erbringung von Finanz- und Versicherungsdienstleistungen" sowie „Sonstige Dienstleistungen", beide (12 %).

### *26. Der Unternehmensstandort, an dem sich Ihr Arbeitsplatz befindet*

**Frage 26**: Der Unternehmensstandort, an dem sich Ihr Arbeitsplatz befindet, liegt in:

- Deutschland,
- Österreich,
- Schweiz?

Wurde in Frage 26 Deutschland angegeben, so wurde in Frage 27 nach den ersten beiden Ziffern der Postleitzahl gefragt.

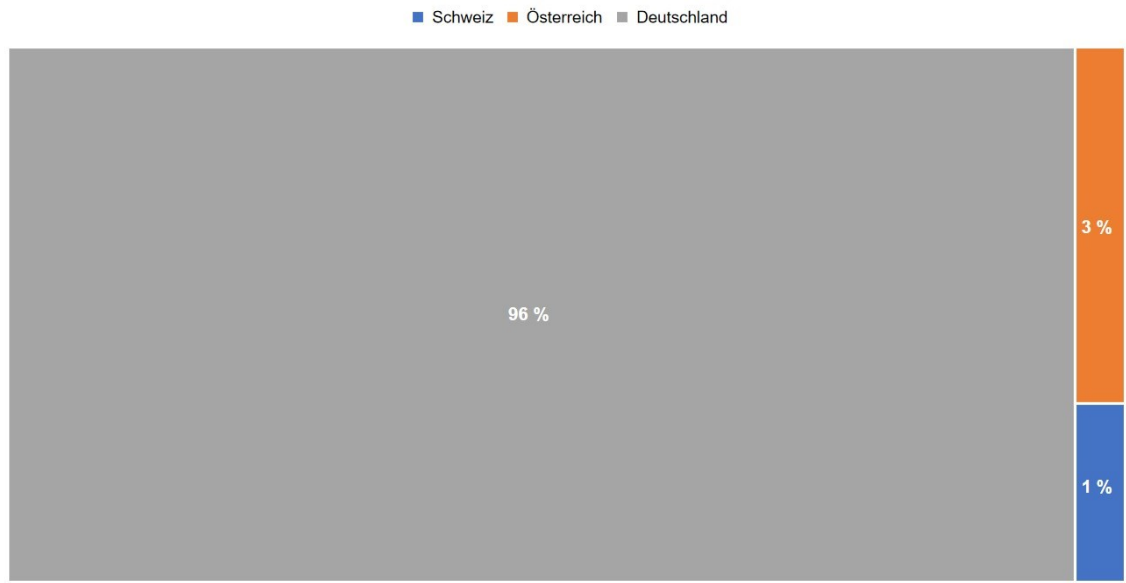

*Abbildung 31: Unternehmensstandort (n=67)*

Frage 26 wurde von allen Befragten (n=67) beantwortet. Bei der überwiegenden Mehrheit der Befragten (96 %) lag, wie Abbildung 31 verdeutlicht, der Unternehmensstandort in Deutschland.

## 27. Bitte geben Sie die ersten beiden Ziffern der Postleitzahl Ihres Unternehmensstandortes an.

**Frage 27**: Bitte geben Sie die ersten beiden Ziffern der Postleitzahl Ihres Unternehmensstandortes an.

Die Frage wurde nur angezeigt, wenn in Frage 26 Deutschland ausgewählt wurde.

Diese Frage sollte dazu dienen, die Regionen herauszuarbeiten, aus denen die Befragten in Deutschland kamen, ohne die Anonymität der Befragung aufzuheben. Von den n=64 Befragten, welche Frage 26 mit Deutschland beantwortet haben, tätigten 62 eine Angabe zur Postleitzahl. Geht man von einer groben Einteilung in Nord, Süd und Mitte aus, gemäß Tabelle 7, so kann die dort ebenfalls aufgeführte Verteilung ermittelt werden.

| Angabe Region | Postleitzahlen beginnend mit | Anzahl |
|---|---|---|
| Nord | 1, 2 | 12 |
| Mitte | 0, 3, 4, 5 | 33 |

| Süd | 6, 7, 8, 9 | 17 |

*Tabelle 7: Regionen in Deutschland laut Postleitzahl (n=67), 2 Personen machten keine Angabe*

## 28. Welche Position haben Sie, die befragte Person, im Unternehmen?

**Frage 28**: Welche Position haben Sie, die befragte Person, im Unternehmen?

Mögliche Antwortoptionen waren:

- Vorstand/Geschäftsführung
- Abteilung Risikomanagement
- Abteilung Compliance
- Abteilung IKS
- Abteilung Interne Revision
- Abteilung GRC
- Abteilung Risikocontrolling
- Abteilung Controlling
- Keine der oben aufgeführten, aber Position mit GRC-Bezug
- Keine der oben aufgeführten, aber Position ohne GRC-Bezug

Mehrfachauswahl war möglich. Wurden die Antwortmöglichkeiten Abteilung Risikomanagement bis Abteilung Controlling ausgewählt, wurde in einem Untermenü nachgefragt, ob es sich um eine Leitungsposition oder die Position einer Mitarbeiterin oder eines Mitarbeiters handelt.

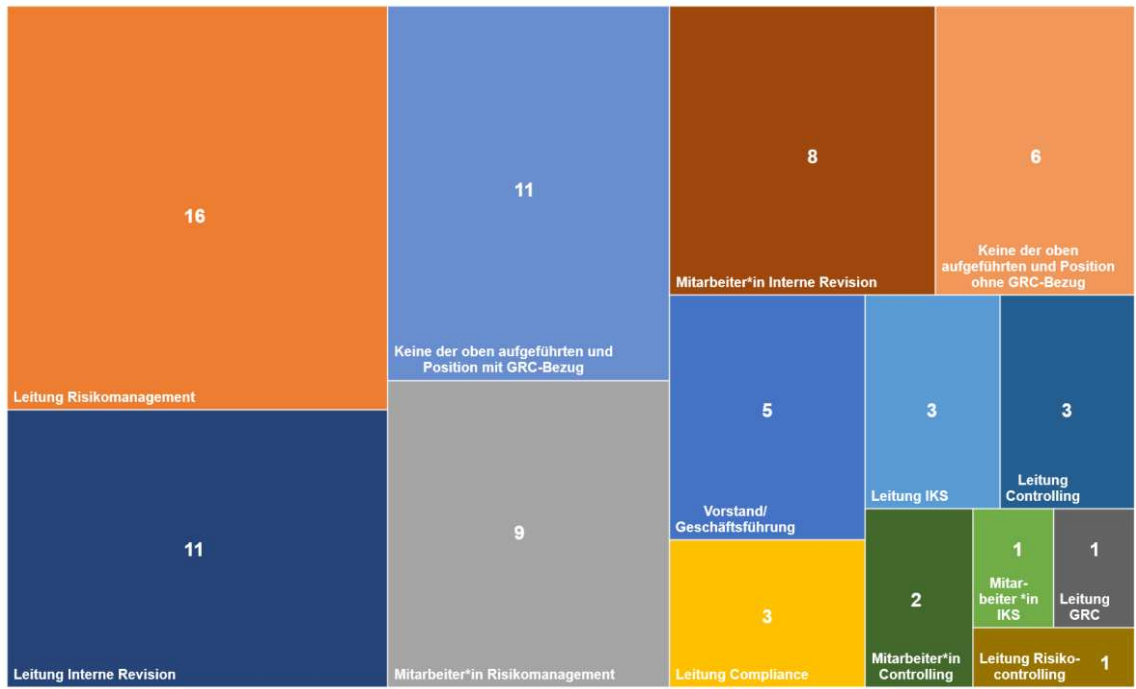

*Abbildung 32: Position der befragten Person im Unternehmen (n=67, Mehrfachauswahl möglich)*

Bei dieser Frage handelt es sich um eine Pflichtabfrage. Sie wurde von allen Befragten beantwortet. Auf Grund der Möglichkeit, mehrere Angaben zu tätigen, sind in Abbildung 32 insgesamt 80 und damit mehr als 67 Angaben aufgeführt. Wie aus der Abbildung ersichtlich ist, war der Bereich Risikomanagement stark vertreten.

## 29. Besitzen Sie eine Expertise in IT-Revision, IT-Sicherheitsmanagement oder IT-Governance?

**Frage 29**: Besitzen Sie eine Expertise in IT-Revision, IT-Sicherheitsmanagement oder IT-Governance?

Antwortmöglichkeiten waren ja oder nein. Insgesamt 46 % der Befragten (n=67) bejahten die Frage, wohingegen 54 % der Befragten keine Expertise in IT-Revision, IT-Sicherheitsmanagement oder IT-Governance besaßen.

***30. Den Fragebogenteil zu unserer Studie haben Sie nun beendet. Möchten Sie abschließend noch etwas zum Thema oder zur Studie anmerken?***

**Frage 30**: Den Fragebogenteil zu unserer Studie haben Sie nun beendet. Möchten Sie abschließend noch etwas zum Thema oder zur Studie anmerken?

Es handelte sich um ein Freitextfeld.

Direkt zum Inhalt der Studie gab es drei Anmerkungen. In zwei Fällen erfolgte ein Hinweis zur Präzisierung der Branche aus Frage 24. In einem Fall gab es den Hinweis, dass die Studie in der eigentlichen Thematik „zu akademisch gedacht, zu weit weg von der betrieblichen Praxis" sei.

## 3     Auswertung

Als Ergebnisse aus der Studie können mehrere Sachverhalte festgehalten werden. Betrachtet man die Ergebnisse aus Frage 1 und Frage 18, so lässt sich feststellen, dass KI in Unternehmen zwar angewendet wird (52 %), jedoch nur zum kleinen Teil im Bereich GRC (6 %). Gleichzeitig zeigen die Angaben zu Frage 14, dass die Befragten Potentiale der Anwendung von KI in GRC in unterschiedlichen Bereichen sehen. Hier besteht also Potential, KI-Anwendungen für den Einsatz in GRC zu analysieren und zu optimieren und entsprechende Projekte zu implementieren und zu begleiten.

Den Ergebnissen zu Frage 4 ist zu entnehmen, dass GRC-Bereiche durchaus miteinander verzahnt sind. So geben hier 78 % an, dass eine Verzahnung vorliegt. An dieser Stelle wären weitere Untersuchungen relevant, wie diese Verzahnung im Hinblick auf die betroffenen Bereiche sowie Art und Struktur der Verzahnung (Fragen 5 bis 7) eine Grundlage für einen Datenaustausch schaffen kann, so dass Methoden der Künstlichen Intelligenz sinnvoll angewendet werden können.

Aus den Rückmeldungen zu Frage 8 geht hervor, dass grundsätzlich IT-Anwendungen in den GRC-Bereichen verwendet werden. Die Resultate aus den Fragen 9 bis 12 geben einen ersten Überblick über gemeinsam genutzte IT-Anwendungen in den GRC-Bereichen. Um klare Handlungsstrukturen abzuleiten sind jedoch mehr Informationen über die Art der Datenhaltung und den konkreten Datenaustausch zwischen den Systemen notwendig. In diesem Zusammenhang wäre auch interessant, welche Elemente die persönliche Einschätzung des Automatisierungsgrads im Unternehmen bestimmen, der von den Befragten in den Antworten zu Frage 13 bezogen auf die IT-Anwendungen in GRC allgemein nicht als sehr hoch angesehen wird.

Immer unter Berücksichtigung der Tatsache, dass es sich nicht um eine repräsentative Studie handelt, geben die Ergebnisse zu Frage 14 eine Richtung vor, in die sich der Einsatz von KI in GRC als erstes bewegen könnte. Die aufgeführten KI-Verfahren werden sehr stark im Risikomanagement als Anwendungsgebiet gesehen. Dies führen die Rückmeldungen aus den Fragen 15 bis 17 weiter. Aufbauend auf den dort aufgeführten Informationen zu Potentialen und Beispielen könnten weitergehende Untersuchungen zu Einsatzzielen und deren Priorisierung führen.

Die vier häufigsten Antworten in Frage 20 zu Herausforderungen zum erfolgreichen Einsatz von KI in GRC, Datenqualität und -quantität (82 %), Personalbeschaffung und -weiterbildung bezogen auf KI-Kompetenzen (57 %), Integration von KI-Technologien in die GRC-Funktionen und -Prozesse (54 %), Homogene IT-Systemlandschaft (48 %) stehen allgemein für Herausforderungen im Bereich Digitalisierung. Gerade KI-Prozesse benötigen qualitativ hochwertige Daten in großen Mengen. Und der aktuelle Mangel an Fachpersonal macht sich natürlich auch bei der notwendigen Neu- oder Umstrukturierung von Prozessen zum Einsatz von KI bemerkbar.

Letztendlich beruhen die Ergebnisse der Studie zu 82 % auf Aussagen von Personen in Unternehmen, die keine KMU darstellen. Hinzu kommen die in Frage 24 ermittelten Rechtsformen, die in engem Zusammenhang mit dem Rahmen der regulatorischen Anforderungen sowie Berichtspflichten stehen. Ein genauerer Blick in die Rahmenbedingungen und Bedürfnisse von KMU wäre ein weiterer Schritt, um den Einsatz von KI in GRC zu bewerten.

# 4 Zusammenfassung und Ausblick

In der vorliegenden Arbeit wurde ein erster Überblick über die Ergebnisse einer Studie zu Anwendungspotentialen von KI in GRC vorgelegt, die im Rahmen einer anonymen Online-Befragung zwischen September 2021 bis Dezember 2021 durchgeführt wurde.

Erläutert wurden die Fragen, die im Rahmen der Studie gestellt wurden. Die Ergebnisse zu den einzelnen Fragen wurden aufgeführt und kommentiert. Dabei dient diese Studie als Grundlage für weitere, komplexere Analysen der vorliegenden Daten sowie zur Ermittlung weiterer Forschungsansätze und Diskussionen zu diesem Thema.

Auch wenn die vorliegende Studie nicht repräsentativ ist, so zeigt sie doch ein paar interessante Ansätze für weitere Forschungsfragen. Wie genau wirkt sich die Verzahnung von GRC-Bereichen auf den Datenaustausch aus? Wie entwickeln sich die Datenbasis und der Datenaustausch in den GRC-Bereichen? Wie sehen die Herausforderungen beim Einsatz von KI in GRC im Detail aus und, wie kann ihnen begegnet werden? Welche KI-Technologien können in GRC sinnvoll eingesetzt werden? Wann ist ein Unternehmen bereit für den Einsatz von KI? Welche konkreten Rahmenbedingungen müssen beachtet werden, wenn KI-Technologien in GRC eingesetzt werden und davon abgegrenzt, wenn im Unternehmen angewandte KI-Technologien in GRC kontrolliert werden? Wie kann die Fragestellung der Anwendungspotentiale von KI in GRC sinnvoll auf KMU übertragen werden?

# Abbildungsverzeichnis

## Tabellenverzeichnis

# Quellenverzeichnis

Armbrüster, 2022: Armbrüster, C. „Digitalisierung und Nachhaltigkeit – rechtliche Herausforderungen für den Versicherungssektor, insbesondere beim Einsatz von Künstlicher Intelligenz", *ZVersWiss* **111**, 19–31 (2022), https://doi.org/10.1007/s12297-022-00518-3.

Eurostat: o.V. „Kleine und mittlere Unternehmen (KMU)", online verfügbar unter https://ec.europa.eu/eurostat/de/web/structural-business-statistics/small-and-medium-sized-enterprises, zuletzt abgerufen am 05.09.2022.

Eurostat, 2017: Europäische Kommission „NACE Rev. 2 – statistische Systematik der Wirtschaftszweige in der Europäischen Gemeinschaft", Publications Office, 2017.

Jo, 2021: Jo, T. „Machine Learning Foundations – Supervised, Unsupervised, and Advanced Learning", Springer Nature, 2021.

Leiner, 2019: Leiner, D. J. „Sosci Survey (Version 3.3.06)" [Computer software], verfügbar unter https://www.soscisurvey.de.

Ransbotham et al., 2017: Ransbotham, S., Kiron, D., Gerbert, P., Reeves, M. „Reshaping Business with Artificial Intelligence: Closing the Gap Between Ambition and Action", MIT Sloan Management Review and The Boston consulting Group, September 2017.

Tarafdar et al., 2019: Tarafdar, M, Beath, C., Ross, J. „Using AI to Enhance Business Operations", MIT Sloan Management Review 60(4), 37-44 (2019).

Eva Ponick, Gabriele Wieczorek

# Artificial Intelligence in Governance, Risk and Compliance

Results of a study on potentials for the application of artificial intelligence (AI) in governance, risk and compliance (GRC)

2021

**Contact Details:**

Hochschule Hamm-Lippstadt
University of Applied Sciences
Marker Allee 76-78
59063 Hamm
Germany

Prof. Dr. Eva Ponick        Prof. Dr. Gabriele Wieczorek
eva.ponick@hshl.de          gabriele.wieczorek@hshl.de

# Abstract

The digital transformation leads to fundamental change in organizational structures. To be able to apply new technologies not only selectively, processes in companies must be revised and functional units must be viewed holistically, especially with regard to interfaces.

Target-oriented management decisions are made, among other things, on the basis of risk management and compliance in combination with the internal control system as governance functions. The effectiveness and efficiency of these functions is decisive to follow guidelines and regulatory requirements as well as for the evaluation of alternative options for acting with regard to activities of companies. GRC (Governance, Risk and Compliance) means an integrated governance-approach, in which the mentioned governance functions are interlinked and not separated from each other.

Methods of artificial intelligence represents an important technology of digital transformation. This technology, which offers a broad range of methods such as machine learning, artificial neural networks, natural language processing or deep learning, offers a lot of possible applications in many business areas from purchasing to production or customer service. Artificial intelligence is also being used in GRC, for example for processing and analysis of unstructured data sets.

This study contains the results of a survey conducted in 2021 to identify and analyze the potential applications of artificial intelligence in GRC.

# Table of Contents

# 1 Introduction

Artificial intelligence (AI) represents an important factor in automating processes within the framework of digital transformation. Today, many proven methods are available that can be deployed in different areas and processes in companies. But even though the use of artificial intelligence is often seen as an added value, implementation and integration of these methods in the process structures of the companies lag significantly behind.[1] This is because the deployment does not happen by itself but is based on certain framework conditions being met. Tarafdar et al., 2019, for example, examine the use of artificial intelligence to optimize business processes and identify five capabilities that companies should possess in order to successfully integrate artificial intelligence into their own processes. This includes "data science competence" which means the safe handling of data and the importance of data for the methods of artificial intelligence and "business domain proficiency" which includes detailed knowledge of the business units and their business processes as well as an idea about how the use of artificial intelligence can lead to improvements here. Furthermore "enterprise architecture expertise", the knowledge about how the elements of a company's architecture interact and business process modeling, the "operational IT backbone" and therefore existing IT structures and technologies as well as "digital inquisitiveness" to deal with these new technologies.

In the present study, the focus is now quite specifically on risk management, compliance and the internal control system as well as their integration. It is precisely managing opportunities and risks across business functions and segments against the backdrop of great uncertainty about future developments what makes an integrated GRC concept (GRC means governance, risk and compliance) so interesting looking at artificial intelligence applications. This is because, with this development, there is an expectation that the level of automatization and data exchange will increase. Quality and quantity of structured and not structured data sets play a major role for the successful use of artificial intelligence. However, this is not just a matter of incorporating AI in GRC, but rather the company-wide risk or rather GRC strategy must ensure that the use of artificial intelligence in other company processes like production or customer support proceeds in accordance with the rules and is analyzed from a risk perspective.[2]

---

[1] See Tarafdar et al., 2019 and Ransbotham et al., 2017.

From this point of view, the present study was conducted in 2021 in Germany in order to elaborate more precisely the interplay between GRC and AI. The survey on which this study is based took place online and anonymously. A questionnaire was created with Sosci Survey[3] an online tool for questionnaires and made available on [www.soscisurvey.de](www.soscisurvey.de). The survey is aimed specifically at people with a connection to GRC. The link to the survey was distributed through different sources (Allianz für Sicherheit in der Wirtschaft e.V. (ASW), the German Institute for Compliance (DICO), the German Institute for Internal Revision e.V. (DIIR), the Internationaler Controller Verein e.V. (ICV), the ISACA Germany Chapter e.V. (ISACA) and the Risk Management & Rating Association e.V. (RMA) as well as several regional and transregional AI platforms and institutions for economic development). Data sets were collected in the period from September 2021 until December 2021. In total 140 data sets were acquired, of which 67 were completed and were used as the basis for the evaluation. Due to the small number of interviews and because people were addressed with the aid of relevant associations, the result cannot be considered representative. Nevertheless, interesting statements are arising which can serve as a basis for further research in the context of the application of AI in GRC.

Specifically, the study should provide answers to the following questions

- What is the current status of using AI in GRC?
- Are the necessary conditions in place for using AI in GRC?
- What are the concrete potentials and application possibilities for AI in GRC that already exist in companies?

---

[2] See for example Armbrüster, 2022 for a discussion of legal challenges looking at AI in the insurance sector.
[3] See Leiner, 2019.

- What concrete potentials and application possibilities for AI do the respondents generally see with regard to GRC?

This paper provides an initial overview of the results of the study on potentials for the application of AI in GRC. The questions asked in the study as well as the survey results are listed below. The obtained data from the survey serve as basis for additional, more complex analysis and further research.

Chapter 2 lists the questions of the study and the results in the order used. The survey was divided into three parts:

- Status Quo on GRC and AI,
- Potentials of AI in GRC,
- General information about the company?

The main findings are presented in chapter 3. The results then are summarized in chapter 4. The study concludes with a summary in chapter 5.

## 2  Fundamental Results

### 2.1  Part 1: Status Quo on GRC and AI

Part 1 contains basic information on whether AI is already used in GRC. In addition, the questions from the questionnaire provide important information for the integration of the GRC core functions and thus to the framework conditions for the use of AI.

The survey was conducted in German. So, all questions and answers provided for selection listed in this chapter were translated from German.

*1. Do you use artificial intelligence in your company?*

**Question 1**: Do you use artificial intelligence in your company?

The following note had been added to the text:

"AI can be used e.g. for speech or text comprehension, image recognition or tone detection, knowledge-based systems, human-computer interaction or assistance systems, autonomous systems, sensor and robot technology, virtual reality and augmented reality."

Possible answers were: yes, no, don't know.

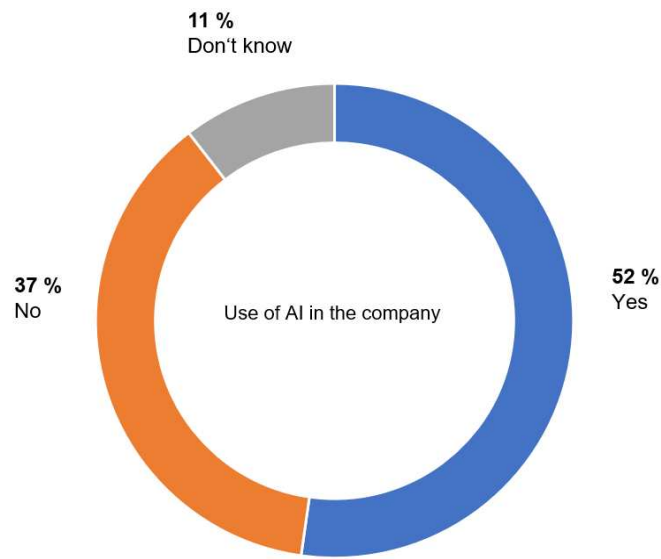

*Figure 1: Use of AI in the company (n=67)*

The information given refers to the completed and evaluated questionnaires (n=67). The majority of the respondents (52 %) state that AI is already in use in the company.

## *2. In which areas of application in the company the use of AI technologies is planned or already in use?*

**Question 2**: In which areas of application in the company the use of AI technologies is planned or already in use?

The following application areas should be evaluated:

- Virtual reality and augmented reality
- Autonomous systems, sensor and robot technology
- Human-machine interaction or assistance systems
- Knowledge-based systems
- Image recognition or tone detection
- Speech or text comprehension

If there are several projects in the company, then the highest state of development should be specified. Possible answers were: no project planned, project planned for introduction, project launched for application, project already implemented, don't know.

The results for this question are shown in the following figures depending on the answers of Question 1.

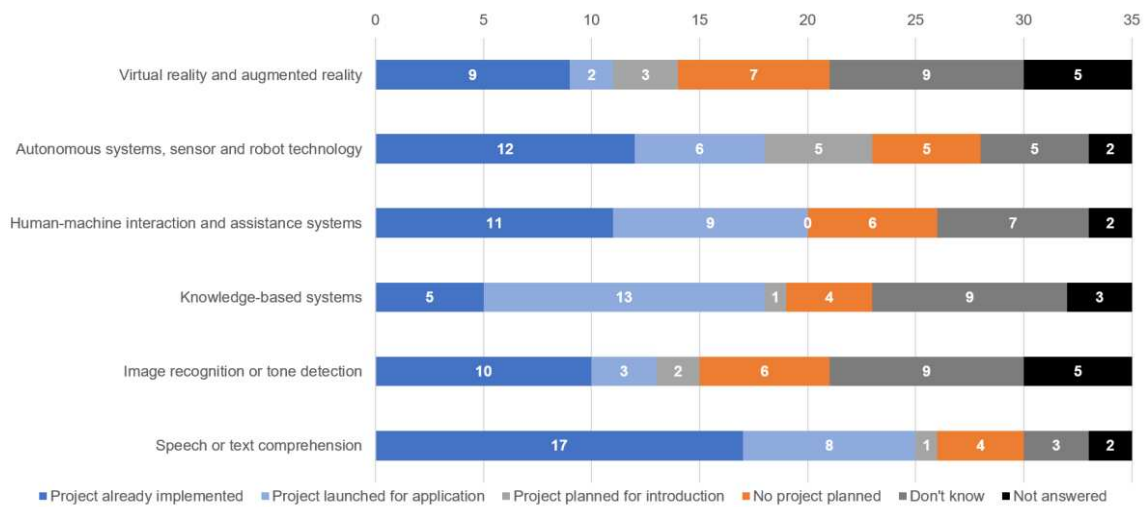

*Figure 2: Areas of application of AI projects with different implementation states in companies, if Question 1 (use of AI in the company) is answered with "Yes" (n=35)*

Looking at the companies which already use AI methods (n=35) in Figure 2 the point speech or text comprehension clearly stands out.

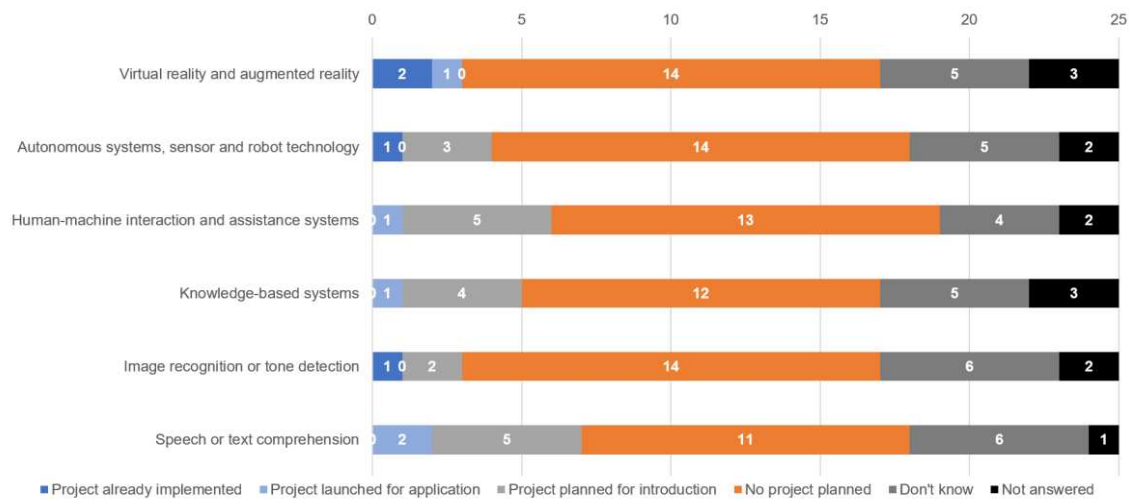

*Figure 3: Areas of application of AI projects with different implementation status in companies, if Question 1 (use of AI in the company) is answered with "No" (n=25)*

Figure 3 shows that in the cases in which no AI projects have been implemented in the company (n=25), no projects of this type are usually planned. The information on projects already implemented are striking because they are not consistent with the previous statement that there is no use of AI in the company. An explanation for this statement cannot be derived from the data set.

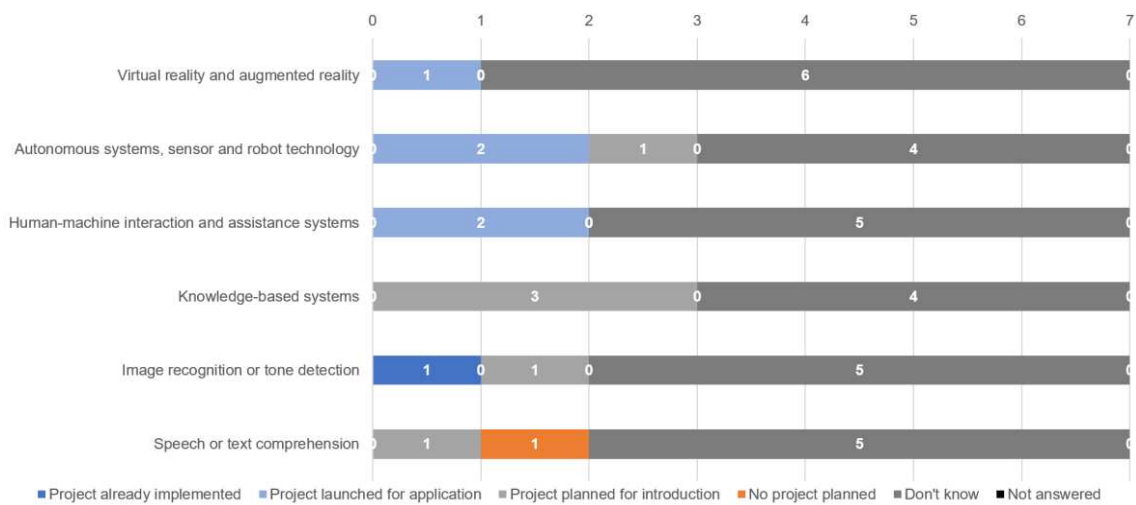

*Figure 4: Areas of application of AI projects with different implementation status in companies, if Question 1 (use of AI in the company) is answered with "Don't know" (n=7)*

Figure 4 presents the last of the three groups from Question 1 that answered the question of the use of AI in the company with "Don't know" (n=7). In accordance to this the majority stated that it is also not known if projects are planned for introduction or launched for application in the specified areas of application. One entry in image recognition or tone detection also stands out here, which specifies a project as already implemented. In connection with the corresponding results of the group of respondents stating that there is no use of AI in the company it would be of interest here how the question was actually interpreted by the participants of the survey.

## 3. Are the GRC business segments in your company already integrated in any way, i.e. is there at least an information exchange between two (or more) GRC business segments?

**Question 3**: Are the GRC business segments in your company already integrated in any way, i.e. is there at least an information exchange between two (or more) GRC business segments?

In order to create a unified image, the following description of the definition of integrated GRC was placed in front of the question:

"In the following GRC, or rather an integrated GRC approach, is understood to mean integrating the governance functions risk management, compliance and internal control system with the aim to raise synergistic benefits in dealing with risks. On the other hand, internal auditing as function with reference to GRC plays a superior role in integrated GRC management."

Moreover, it was pointed out that the answer would be essential for the further course of the survey.

Possible answers were: yes, no, not specified.

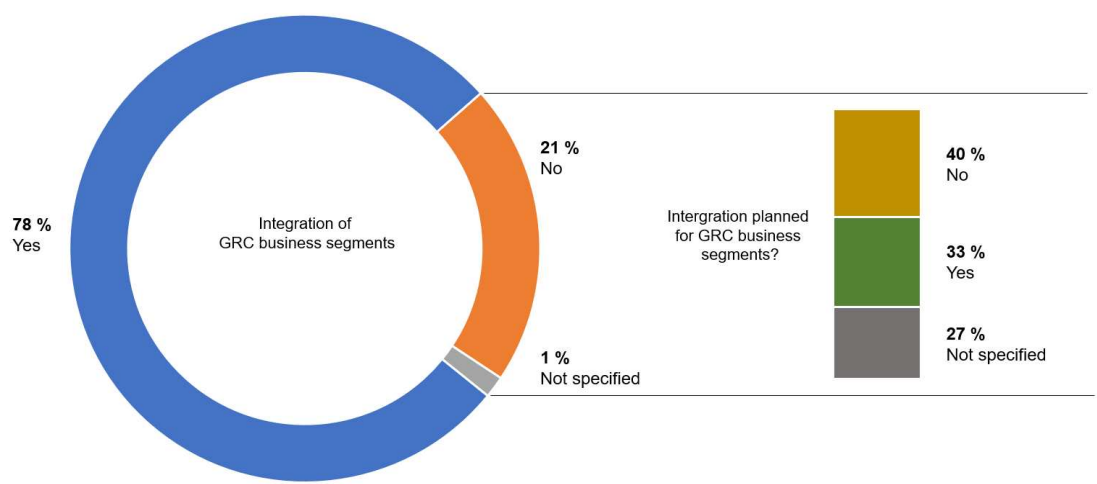

*Figure 5: Integration of the GRC business segments risk management, compliance and internal control system (n=67)*

Answering this question with „No" (n=14) or "Not specified" (n=1) led to the following question as to whether the integration of GRC business segments is planned. This information can be taken from the right section of Figure 5 and refers to n=15 persons. A clear trend with regard to planned integration cannot be inferred from the data.

## 4. Is the integration planned for GRC business segments?

**Question 4**: Is the integration planned for GRC business segments?

Question 4 was only displayed if Question 3 was answered with "No" or "Not specified". Possible answers were: yes, no, not specified.

The results of this question can be seen in Figure 5, Question 3.

## 5. Which business segments are integrated with each other?

**Question 5**: Which business segments are integrated with each other?

This question was *not* shown, if Question 3 was answered with "No" or "Not specified". Possible answers were:

- Risk management and compliance
- Risk management and internal control system
- Compliance and internal control system
- Risk management, compliance and internal control system.

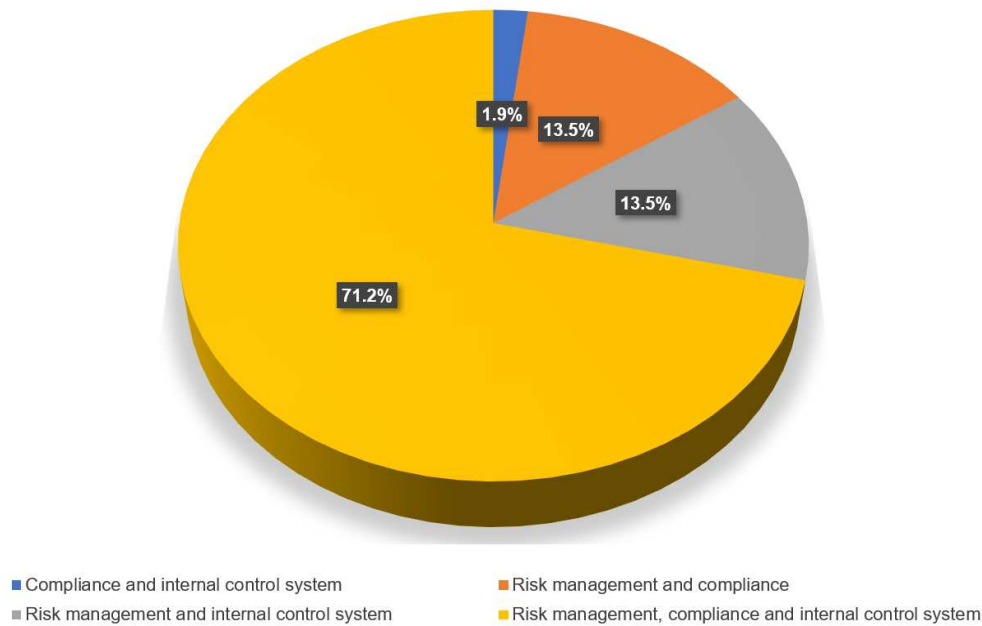

*Figure 6: Overview of the business segments risk management, compliance and/or internal control system integrated with each other (n=52)*

Only n=52 participants answered Question 3 with "Yes". Question 3 now refers to these participants and was not shown to the other n=15 participants. The partition of the answers regarding the composition of integrated business segments can be seen in Figure 6. A majority of 71 % stated that there is a threefold integration of all business segments risk management, compliance and internal control system.

## 6. How do you assess the operational integration of the GRC business segments in your company?

**Question 6**: How do you assess the operational integration of the GRC business segments in your company?

The question served to assess the existing operational integration. The following statements were given for evaluation:

- The GRC business segments have a uniform basic understanding of terms and definitions within the context of risk (e.g. risk, types of risk, …).
- Beyond a uniform basic understanding, the GRC business segments coordinate the content from risk analysis to risk control.

- The GRC business segments use coordinated, uniform methods in risk assessment.
- The cooperation of the GRC business segments is harmonized in terms of content and timing.
- Information gathering and analysis between the GRC business segments is redundancy-free.

This question was not shown, if Question 3 about the integration of the business segments was answered with "No" or "Not specified". The listed statements could be assessed on a 5-point scale (strongly disagree, rather disagree, partly, partly, rather agree, strongly agree)

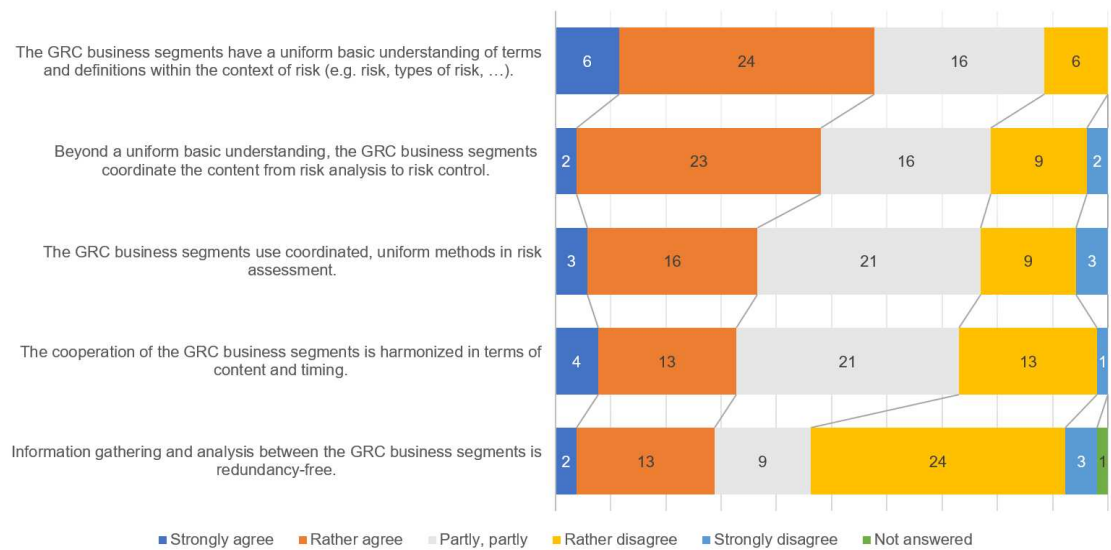

*Figure 7: Assessment of the operational integration of the GRC business segments (n=52)*

If Question 3 about whether the GRC business segments are integrated in any way was answered with "Yes", Question 6 should provide information about the structure of the integration. As shown in Figure 7 the majority of 58 % of those 52 participants who answered the question rather or strongly agree that at least a uniform basic understanding of terms and definitions in the context of risk exists.

29 % of the participants rather or strongly agree that information gathering and analysis between the GRC business segments is redundancy-free, which is a framework condition for successful use of AI in GRC. However, 52 % stated, that they rather disagree or strongly disagree to this.

## 7. Please select which superior elements or instruments for integration of the GRC business segments are already exist or planned in your company?

**Question 7**: Please select which superior elements or instruments for integration of the GRC business segments are already exist or planned in your company?

The following elements and instruments should be subject to evaluation:

- GRC committee
- Integrated GRC reporting
- Integrated GRC strategy
- GRC process
- GRC as integrated function
- GRC as integrated segment
- GRC handbook
- Integrated GRC software solution

This Question was not shown, if Question 3 whether the business segments are integrated was answered with "No" or "Not specified". Therefore n=52 questionnaiers were inluded in the evaluation.

For each element or instrument the following specifications were possible: not planned, in evaluation, planned or in implementation, already exists, don't know.

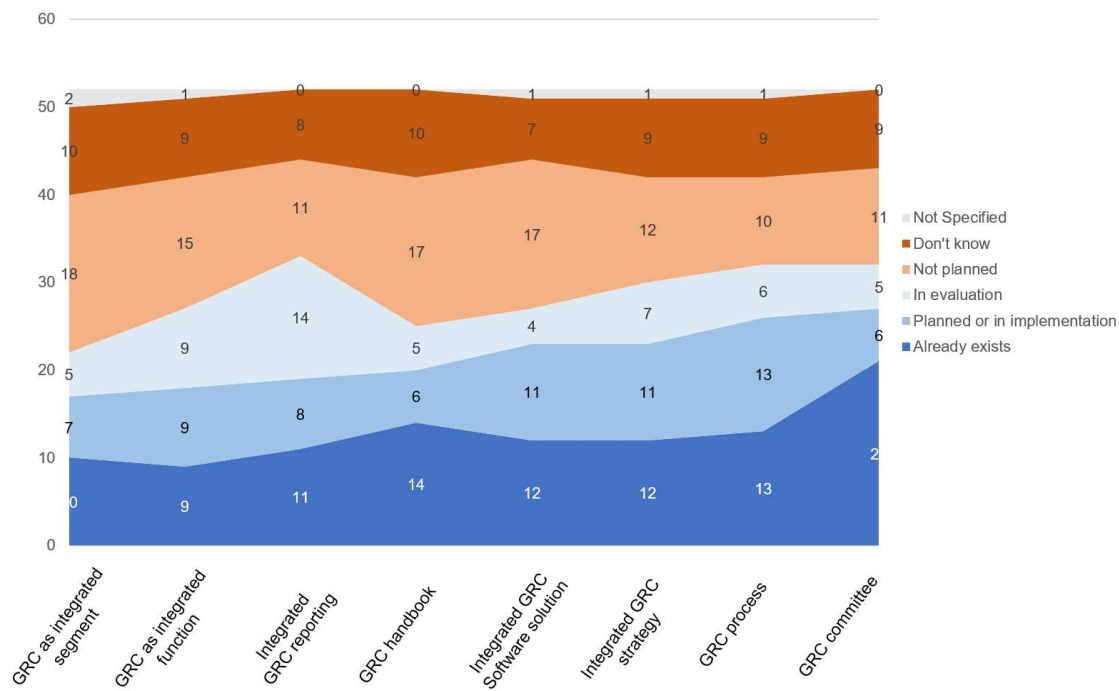

*Figure 8: Instruments and elements for integration of the GRC business segments (n=52)*

Considering already existing elements or instruments that are planned or in implementation then GRC as integrated segment stands out in particular. The percentage of those who said that this instrument already exists, is planned or is in implementation was the lowest at 33 %. Overall, 35 % even stated, that this instrument ist not planned at all which represents the highest value in this category.

Looking at the integrated GRC reporting, 35 % stated, that this is an instrument already existing, planned or in implementation. Only 21 % pointed out, that this instrument is not planned. However, at 27 %, the proportion or those who stated that this instrument is in evaluation was the highest. This raises the interesting question of the reason for this intensive evaluation.

Existence, planning and implementation of a GRC committee was indicated from 52 % of the participants under consideration, which represents the highest value. Here, too, only 21 % stated, that this instrument is not planned and only 10 % indicated, that the use of this instrument is in evaluation.

After all, 33 % indicated, that a GRC handbook and a GRC software solution in each case are not planned.

## 8. For which GRC business segments do IT applications exist?

**Question 8**: For which GRC business segments do IT applications exist?

This question was always asked no matter, what answers were given in other questions, and therefore refers to n=67 questionnaires.

Possible answers were:

- Risk management
- Compliance
- Internal control system
- Internal auditing
- None of these segments

Multiple choice was possible.

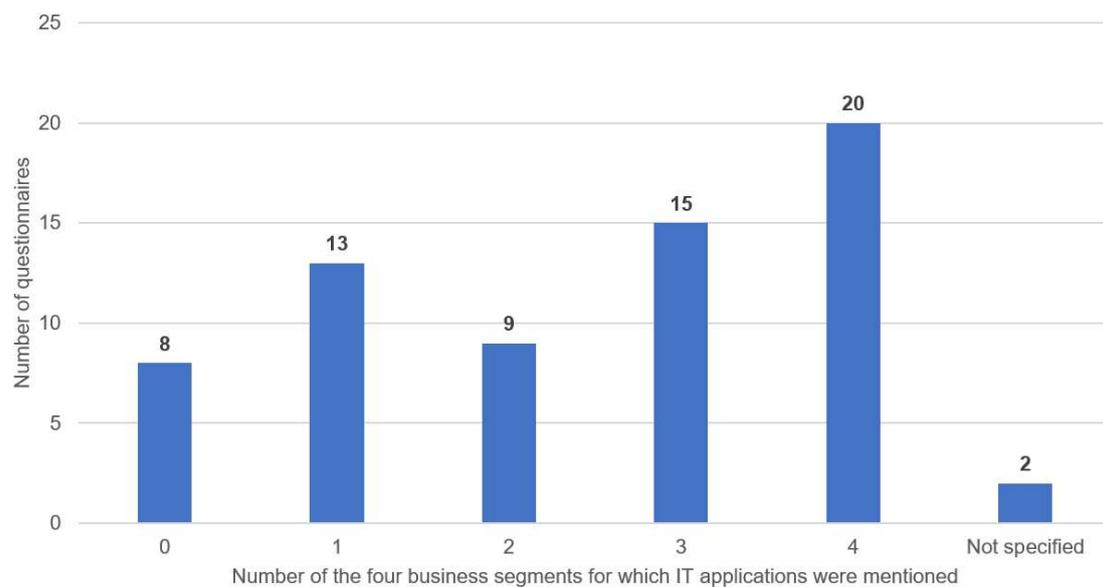

*Figure 9: Number of the GRC business segments with existing IT applications (n=67)*

A first overview, whether IT applications exist for one, two, three or all four GRC business segments, is shown in Figure 9. Most participants (30 %) made specifications for all four business segments. None of these segments (number 0 in Figure 9) was stated by 12 %, and in 3 % of the cases, no information was provided on this question.

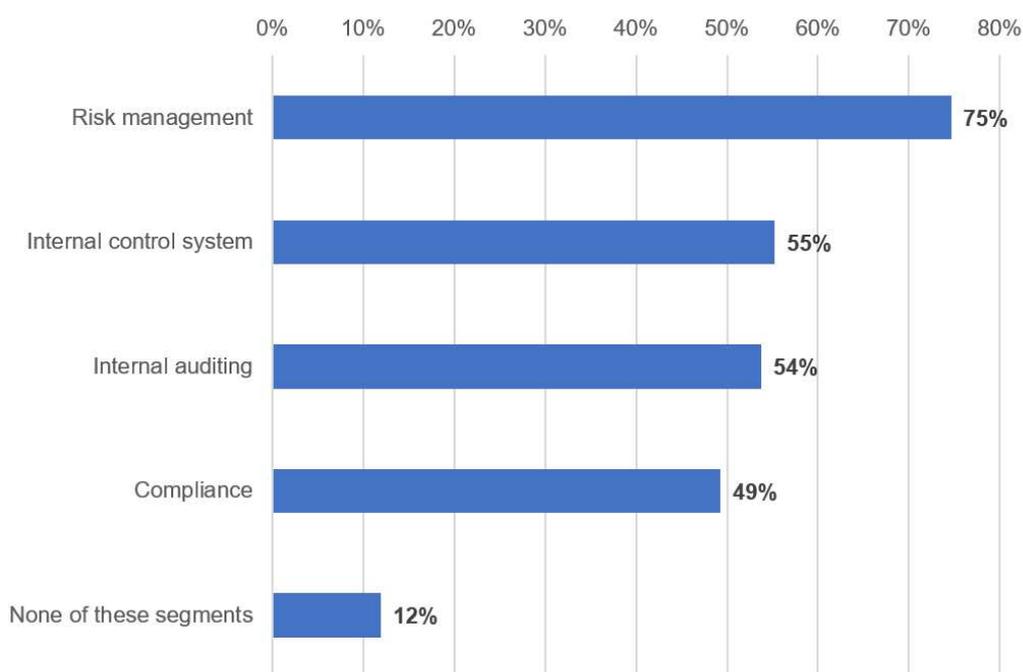

*Figure 10: GRC business segments with IT applications (n=67)*

Figure 10 shows that, according to the participants, most IT applications exist in risk management (75 %), followed by internal control system (55 %), internal auditing (54 %) and compliance (49 %). 12 % of the participants stated, that none of these GRC business segments has IT applications.

If in Question 8 one of the business segments was chosen, then Question 9 to 12 were asked accordingly for this segment. A detailed explanation can be found in the following sections.

### *9. Please specify the software that you use in risk management.*

**Question 9**: Please specify the software that you use in risk management.

This question was shown only if risk management was chosen in Question 8 and therefore the information was given, that IT applications exist for risk management. This was a free text input.

From n=50 participants who stated in Question 8 that IT applications exist for risk management, n=42 persons answered Question 9, too. Because it was possible to enter free text, and multiple inputs were possible, there are a total of 51 entries of software applications. The respondents gave a maximum of 3 mentions of IT applications.

| Number of software mentioned | Number questionnaires |
|---|---|
| 1 | 36 |
| 2 | 3 |
| 3 | 3 |
| Total | 42 |

*Table 1: Number of respondents, who mentioned one, two or three IT applications in Question 9*

Table 1 shows the number of participants who mentioned one, two or three IT applications.

For classification purpose, the IT applications were assigned to different categories. An overview of these categories of IT applications can be taken from Table 2. Every single specification was considered, that means if a statement of an IT application was made by different persons it has been counted as many times as indicated. It was not summarized into one statement.

It should be noted here, that it is not always clear from the information provided, which IT application is actually used. In many cases a software manufacturer was named but no specific application was mentioned. In these cases, an assignment to Category 1 took place when the manufacturer has a corresponding software on offer. Some IT applications are modular in design so that it is not clear whether it is used with all possible GRC modules or is limited to special application areas. For this reason, no additional subdivision of Category 1 is made.

IT applications labeled as own-developed or individual industry solution are summarized in Category 2. Whereas Category 3 includes other office products, tools for statistical analysis and/or simulation, also as an Add-In for spreadsheet programs.

If no clear specification was possible on the basis of the description, the IT application was assigned to Category 4. This happened e.g. for the term "GRC". However, this means that at least a concrete specification was available when assigning an IT application to Category 4.

Descriptions such as "don't know", "not specified" or something like that were summarized in Category 5. This was not the case for Question 9, but because this system of categories will be used also for Question 10 to 12, Category 5 is already introduced here.

| Category | Description |
| --- | --- |
| 1 | IT applications that can be allocated in whole or in part to the business segments risk management, internal control system, internal auditing, compliance and complete GRC IT solutions. |
| 2 | IT applications labeled das own-developed or individual industry solution. |
| 3 | Spreadsheet programs, other office-products, tools for statistical analysis and/or simulations, also as an Add-in for spreadsheets |
| 4 | IT applications, that do not belong to categories 1 to 3, because the mentioned software product cannot be identified, but a concrete indication has been made. |
| 5 | Information like "don't know" or "not specified" |

*Table 2: Categories for classification of the IT applications listed*

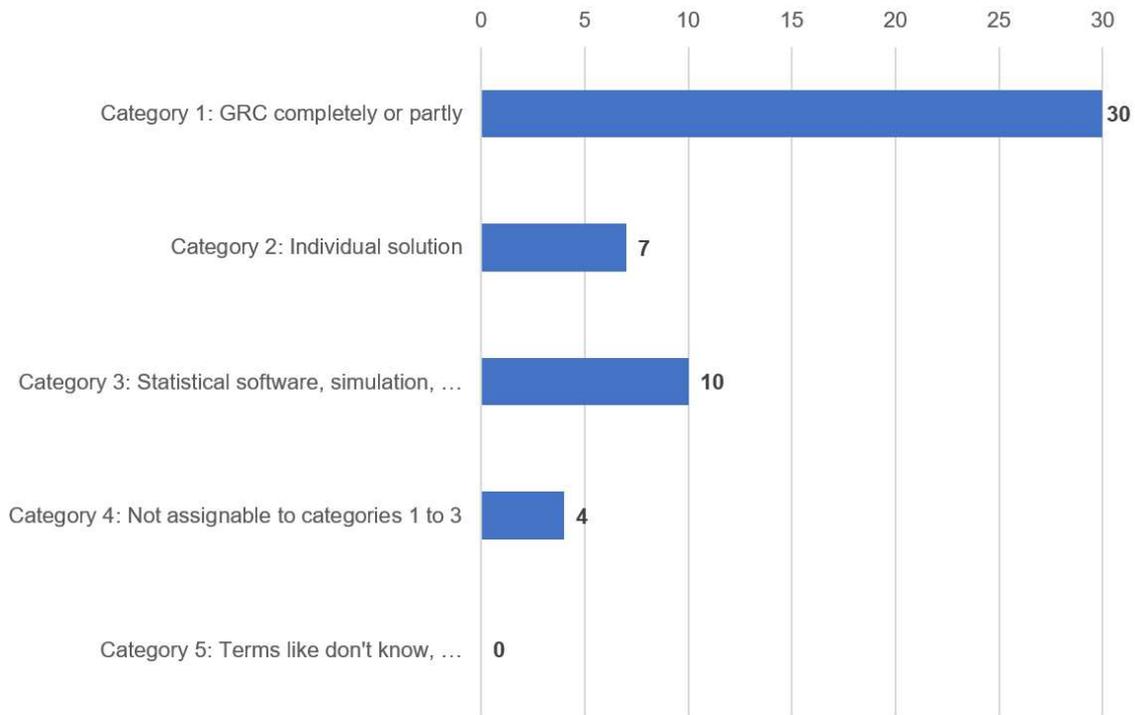

*Figure 11: IT applications in risk management*

The categorization of the mentioned IT applications can be seen in Figure 11. All information in Category 2 are single data and were made without further disclosures of additional IT applications. Category 3 includes 4 single appointments and 6 information of IT applications, where in addition one or two products of the same or other categories were mentioned.

## 10. Please specify the software that you use in compliance.

**Question 10**: Please specify the software that you use in compliance.

This question was shown only if compliance was chosen in Question 8 and therefore the information was given, that IT applications exist for compliance. This was a free text input.

From n=33 participants who stated in Question 8 that IT applications exist for compliance, n=25 persons answered Question 10, too. Because it was possible to enter free text, and multiple inputs were possible, there are a total of 28 entries of software applications. The respondents gave a maximum of 2 mentions of IT applications.

| Number of software mentioned | Number of questionnaires |
|---|---|
| 1 | 22 |
| 2 | 3 |
| Total | 25 |

*Table 3: Number of respondents, who mentioned one or two IT applications in Question 10*

Table 3 shows the number of participants who mentioned one or two IT applications.

The classification is analogous to Table 2 of Question 9. For the reasons stated there, too, it was not always clear which IT application exactly was involved.

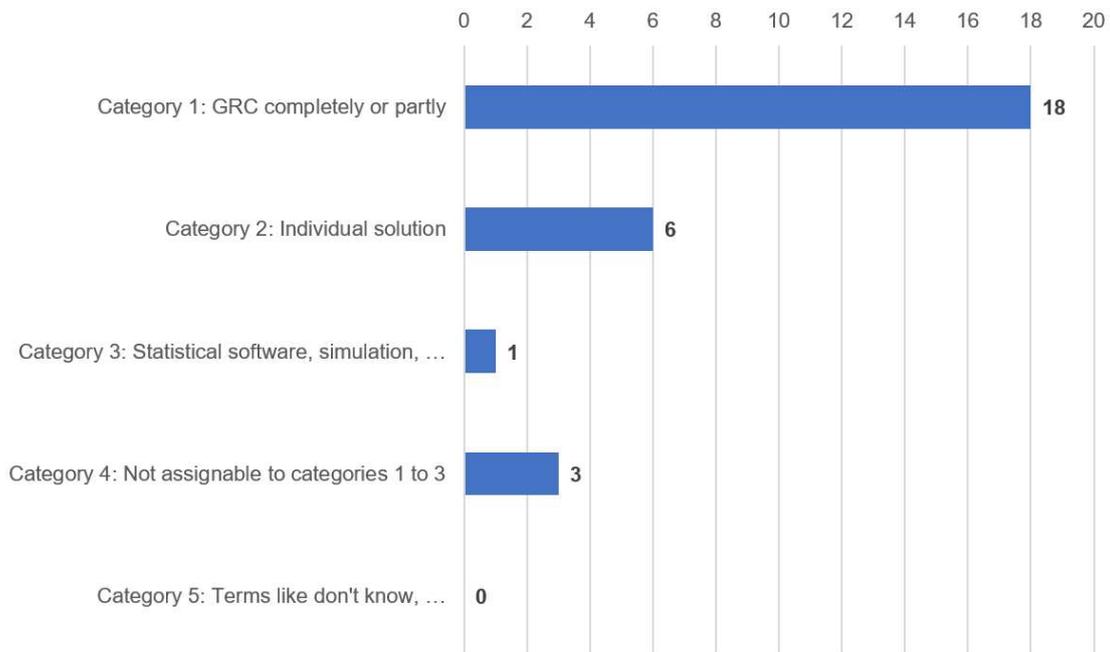

*Figure 12: IT applications in compliance*

The categorization of the mentioned IT applications analogous to Table 2 can be seen in Figure 12. Category 3 contains only one data. This was no single information but was given together with one application from Category 1. All information to Category 2 was given as single data without mentioning other applications.

## 11. Please specify the software that you use in the internal control system.

**Question 11**: Please specify the software that you use in the internal control system.

This question was shown only if internal control system was chosen in Question 8 and therefore the information was given, that IT applications exist for internal control system. This was a free text input.

From n=37 participants who stated in Question 8 that IT applications exist for internal control system, n=31 persons answered Question 11, too. Because it was possible to enter free text, and multiple inputs were possible, there are a total of 36 entries of software applications. The respondents gave a maximum of 3 mentions of IT applications.

| Number of software mentioned | Number of questionnaires |
|---|---|
| 1 | 27 |
| 2 | 3 |
| 3 | 1 |
| Total | 31 |

*Table 4: Number of respondents, who mentioned one, two or three IT applications in Question 11*

Table 4 shows the number of participants who mentioned one, two or three IT applications.

The classification is analogous to Table 2 of Question 9. For the reasons stated there, too, it was not always clear which IT application exactly was involved.

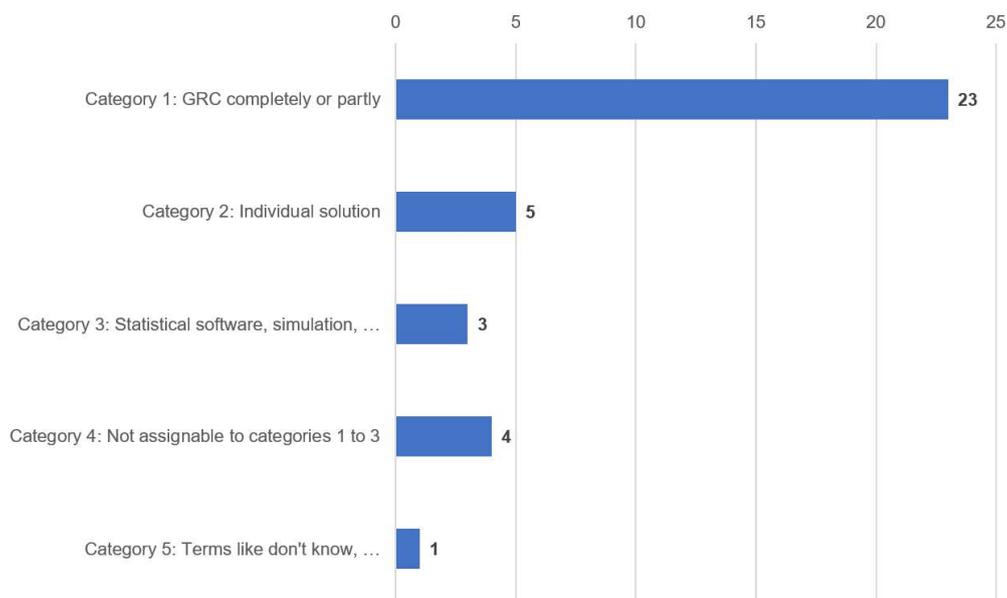

*Figure 13: IT applications in the internal control system*

Two of the three statements of Category 3 were given together with other information of IT applications. The last one was a single information.

## 12. Please specify the software that you use in the internal auditing.

**Question 12**: Please specify the software that you use in the internal auditing.

This question was shown only if internal auditing was chosen in Question 8 and therefore the information was given, that IT applications exist for internal auditing. This was a free text input.

From n=36 participants who stated in Question 8 that IT applications exist for internal auditing, n=42 persons answered Question 12, too. Because it was possible to enter free text and multiple inputs were possible, there are a total of 41 entries of software applications. The respondents gave a maximum of 6 mentions of IT applications.

| Number of software mentioned | Number of questionnaires |
|---|---|
| 1 | 25 |
| 2 | 3 |
| 3 | 0 |
| 4 | 1 |
| 5 | 0 |
| 6 | 1 |
| Total | 41 |

*Table 5: Number of respondents, who mentioned one to six IT applications in Question 12*

Table 5 shows the number of participants who mentioned one to six IT applications.

The classification is analogous to Table 2 of Question 9. For the reasons stated there, too, it was not always clear which IT application exactly was involved.

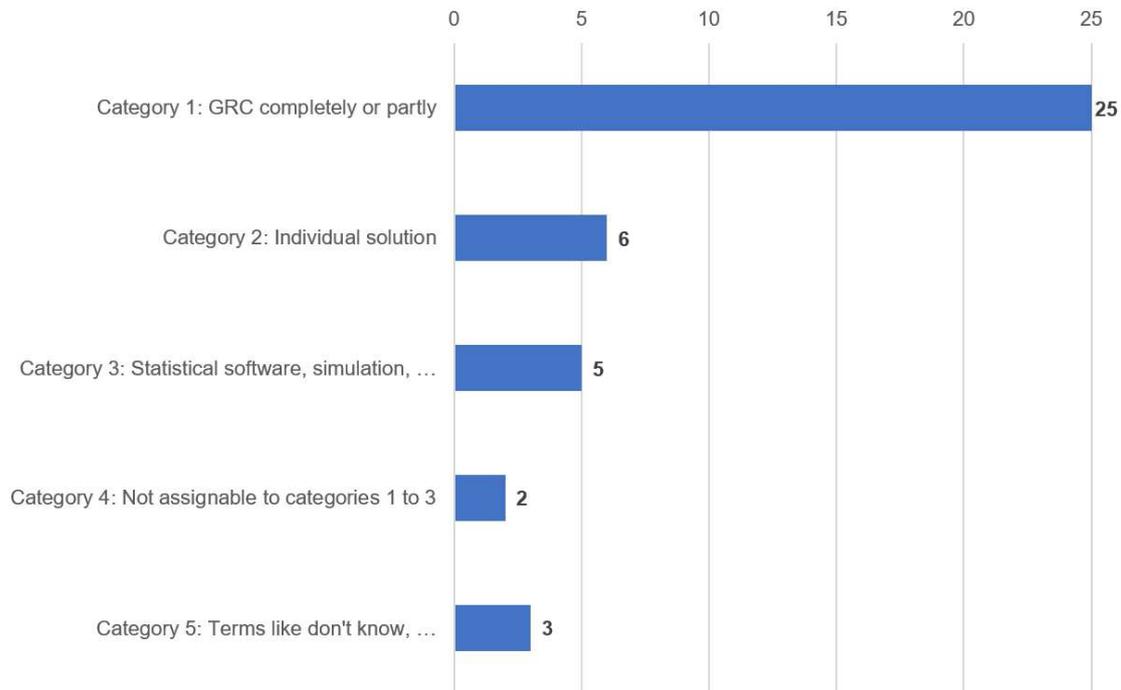

*Figure 14: IT applications in the internal auditing*

All information in Category 3 in Figure 14 are based on the two responses with four and five statements to IT applications. Looking at Category 2 there were only single data appointments.

## 9. to 12. Overall overview

In order to better illustrate the relationship of the contents of Question 9 to Question 12, this section provides a brief overview of the coherent response to the questions by the participants.

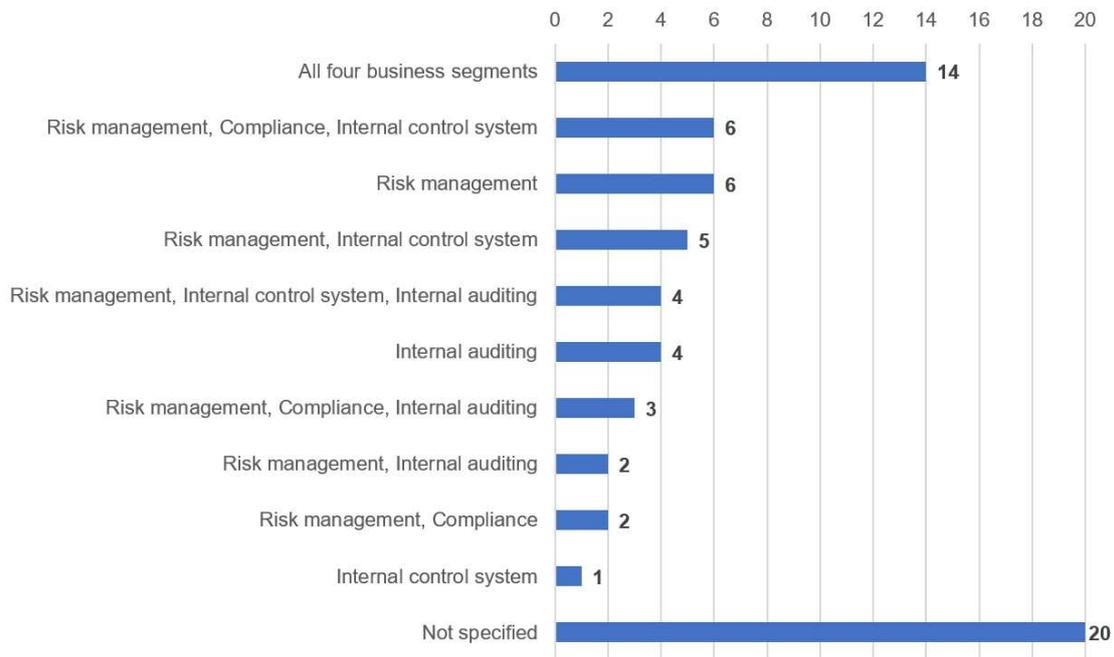

*Figure 15: Combination of the information of IT applications (n=67)*

As can be seen in Figure 15 a total of 47 participants stated information about IT applications. As shown in Figure 9, Question 8 this question was answered from a total of 57 persons. 10 participants were not shown the question at all because of their answers to Question 8, and 10 participants did not provide information about IT applications for any of the GRC business segments.

In Figure 15 only those business segments and combinations of business segments are shown, for which at least one IT application was mentioned. In most cases (30 %) looking at the 47 questionnaires providing information about IT applications there were information for all four GRC business segments. Terms like "don't know" or "not specified" were not taken into account as information for an IT application.

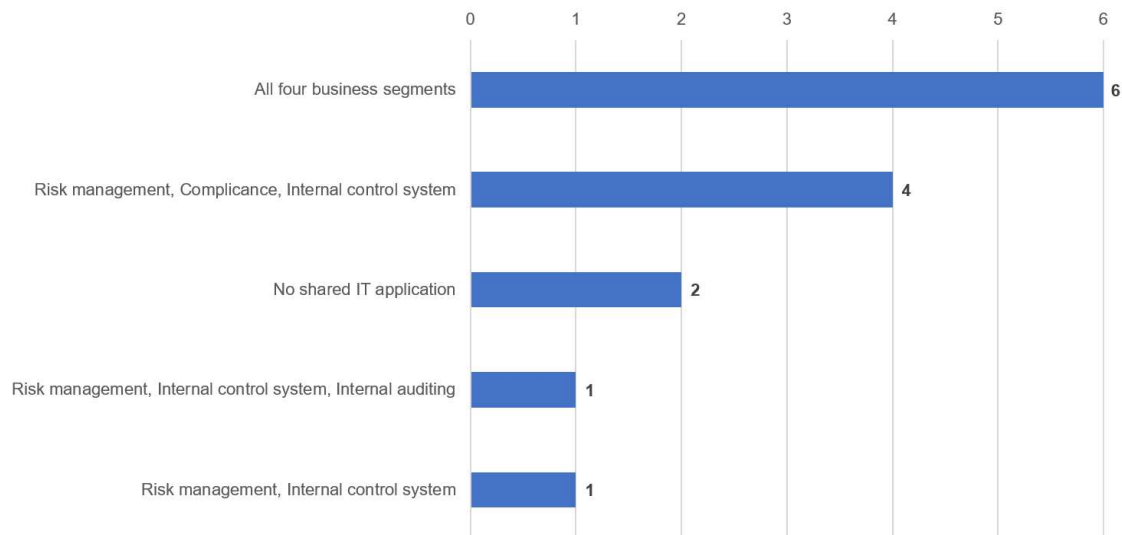

*Figure 16: Combination of identical IT applications from all four GRC business segments (n=14)*

The evaluation from n=14 questionnaires, that provide information of IT applications for all four GRC business segments is shown in Figure 16. Listed are the data for which at least one of the mentioned IT applications matches for each business segment. From these 6 persons list identical IT applications for all four business segments. However, four of these cases were applications of Category 2 (IT applications labeled as own-developed or individual industry solution).

## 13. How do you assess the level of automation of the IT applications in GRC in relation to data exchange?

**Question 13**: How do you assess the level of automation of the IT applications in GRC in relation to data exchange?

This question was shown only if at least one of the business segments risk management, compliance, internal control system and internal auditing was chosen in Question 8 and therefore the information was given, that IT applications exist for at least one of these segments.

A scale was displayed with the extreme values completely manually (1) and fully automated (11). The level of automation of data exchange could be selected between these two extreme values. In the evaluation the selected numerical values were presented in steps of 10 %. A specification of 1 means 0 % automation, 2 means 10 % automation and if 11 was given, this means an automation of 100 %.

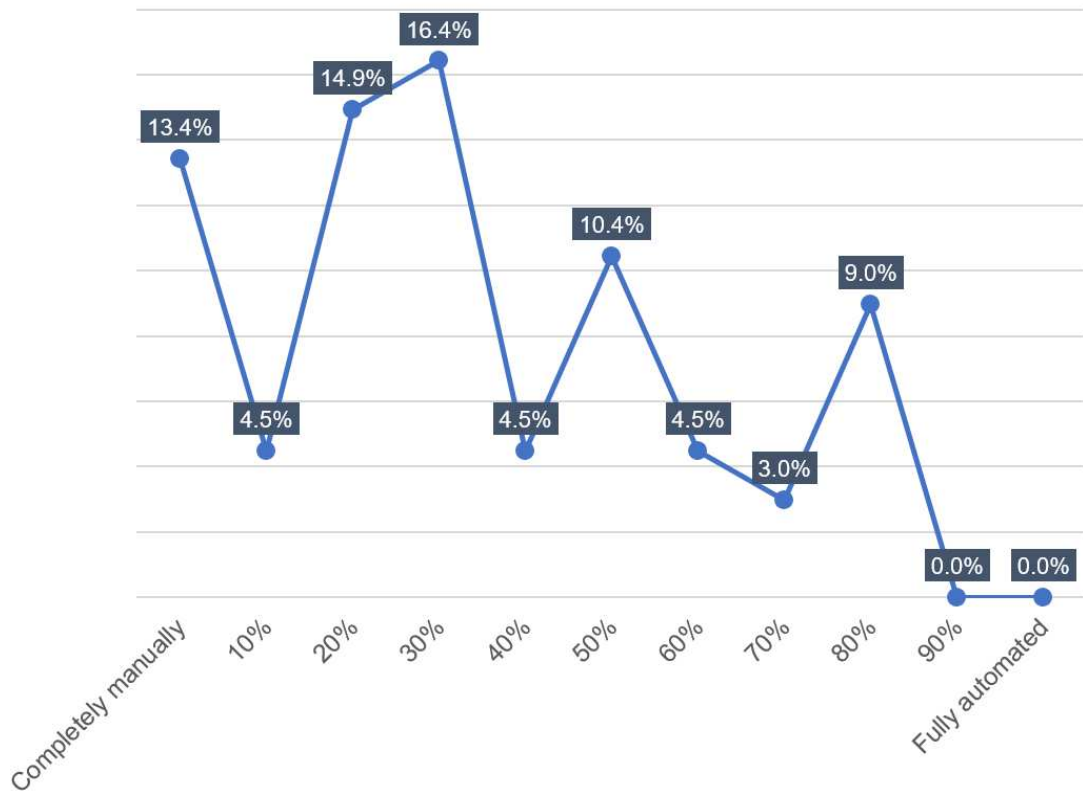

*Figure 17: Level of automation of data exchange (n=67)*

Figure 17 shows the level of automation of data exchange for all n=67 participants. Information was provided by n=54 persons and therefore the numerical values in the figure sum up to 80,6 %. Most data show a level of automation in the lower range less than 40 %. No one selected a level of automation of 90 % or 100 %. After all, 13,4 % of the respondents stated a completely manual data exchange. It would be interesting here to understand how completely manually was defined by the participants.

## 2.2 Part 2: AI Potentials in GRC

After clarifying the general framework conditions in part 1, part 2 should go into more detail with regard to potentials of artificial intelligence in governance, risk and compliance. For this purpose, the questions connect those two parts with each other.

Of course, all questions and answers provided for selection, listed in this chapter, were translated from German, too.

## 14. What applications do you see for using AI methods in GRC?

**Question 14**: What applications do you see for using AI methods in GRC?

The following note had been added to the text:

"Hereafter potentials should be identified which exist in the single GRC business segments concretely in your company."

For this purpose one or more of the segments risk management, compliance and internal control system could be selected for the following potentials:[4]

- Systematic categorization of data in classes,
- Discovery of similar structures in data with clustering,
- Forecasting of development through regression,
- Learning rewarding strategies of action.

---

[4] The potentials are derived from the application areas of the different learning types. Classification and regression are derived from supervised learning, clustering from unsupervised learning and learning strategies of action from reinforcement learning. See e.g. Jo, 2021, chapter 1.2 and 1.3.

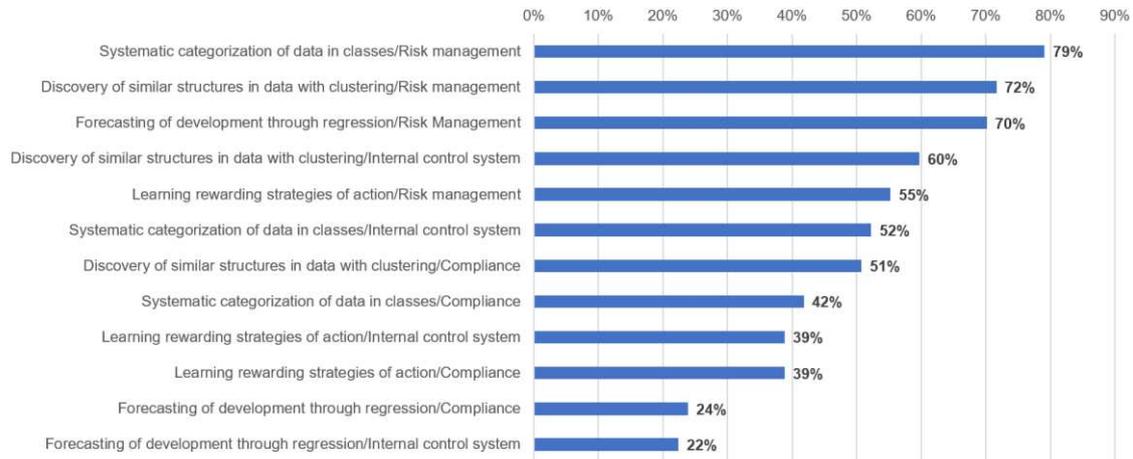

*Figure 18: Applications for the use of AI methods (n=67)*

Figure 18 shows all the results of n=67 questionnaires where potentials for the three GRC business segments risk management, compliance and internal control system were selected. All four potentials for risk management are among the five most frequently mentions.

## 15. If you see potentials or examples for risk management, please outline them.

**Question 15**: If you see potentials or examples for risk management, please outline them.

This was a free text input. Additional input fields were added where needed.

Altogether 26 participants answered this question. From these, 22 persons stated one potential, 3 persons stated two potentials and one person stated three potentials. Therefore, a total of 31 information to AI potentials or examples in risk management were given.

On the basis of the information of Question 15 to 17 a first classification of five categories was derived and the information was assigned accordingly. These categories are shown in Table 6. These categories are also used for Question 16 and 17.

| Category | Description |
|---|---|
| 1 | AI supported data analysis, for example for pattern detection, classification or clustering among others with the objective of risk identification and risk assessment, fraud detection, finding inconsistencies, quality assurance, control, independent assessment |
| 2 | Early warning system, forecasting, predictive analytics |
| 3 | Integration of systems, applications and segments |
| 4 | Modeling, simulation, scenario analysis, algorithms |
| 5 | Disclosures that cannot be directly linked to AI potentials and examples |

*Table 6: Categorization of the mentioned potentials*

Categorization of the results is made as best as possible. However, there is some room for interpretation in some cases, since there was, of course, no possibility to ask questions in the context of this survey.

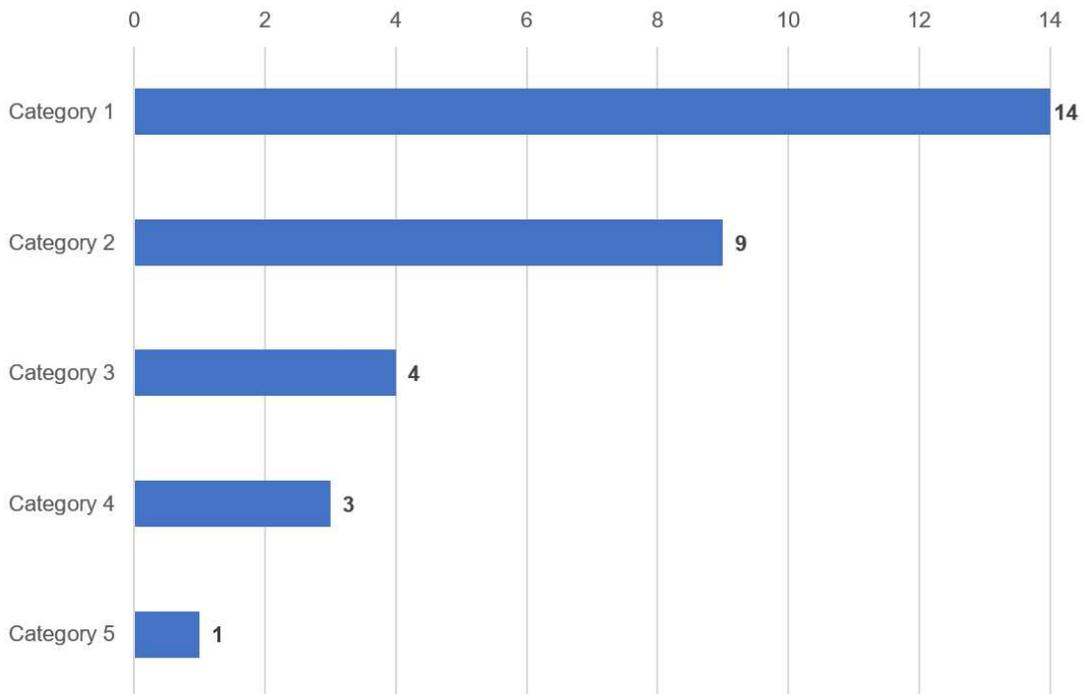

*Figure 19: Categorization of potentials or examples for risk management according to categories from Table 6*

As can be seen from Figure 19 the majority of participants (45 %) made disclosures that could be assigned to pattern detection, classification or clustering.

Answers can be seen as expansion of information from Question 14. Since Question 15 was asked subsequently of Question 14 which includes a list of possible potentials. Category 1 is also important for Question 16 (compliance) and Question 17 (internal control system).

Category 2 with information about forecasting and early warning systems can also be derived from Question 14. This category was not mentioned for compliance and the internal control system and thus represents a unique feature for application possibilities in risk management referred to this study. Categories 3 and 4 follows with distance.

### 16. If you see potentials or examples for compliance, please outline them.

**Question 16**: If you see potentials or examples for compliance, please outline them.

This was a free text input. Additional input fields were added where needed.

Altogether 18 participants answered this question. From these, all persons stated one potential. Therefore, a total of 18 information to AI potentials or examples in compliance were given.

The answers from the free text input were assigned to the categories from Table 6, too. This is shown in Figure 20.

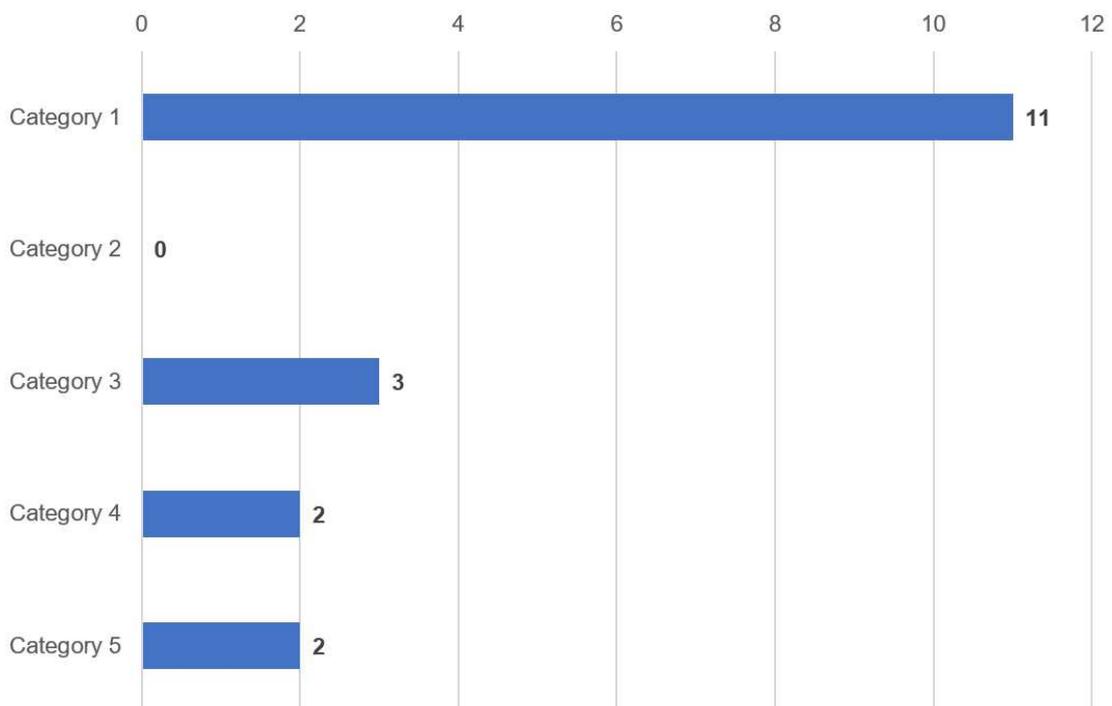

*Figure 20: Categorization of potentials or examples for compliance according to categories from Table 6*

Most information again could be assigned to Category 1 (61 %). Category 2 was not mentioned at all.

## 17. If you see potentials or examples for the internal control system, please outline them.

**Question 17**: If you see potentials or examples for the internal control system, please outline them.

This was a free text input. Additional input fields were added where needed.

Altogether 20 participants answered this question. From these, all persons stated one potential. Therefore, a total of 20 information to AI potentials or examples in the internal control system were given.

The answers from the free text input were assigned to the categories from Table 6, too. This is shown in Figure 21.

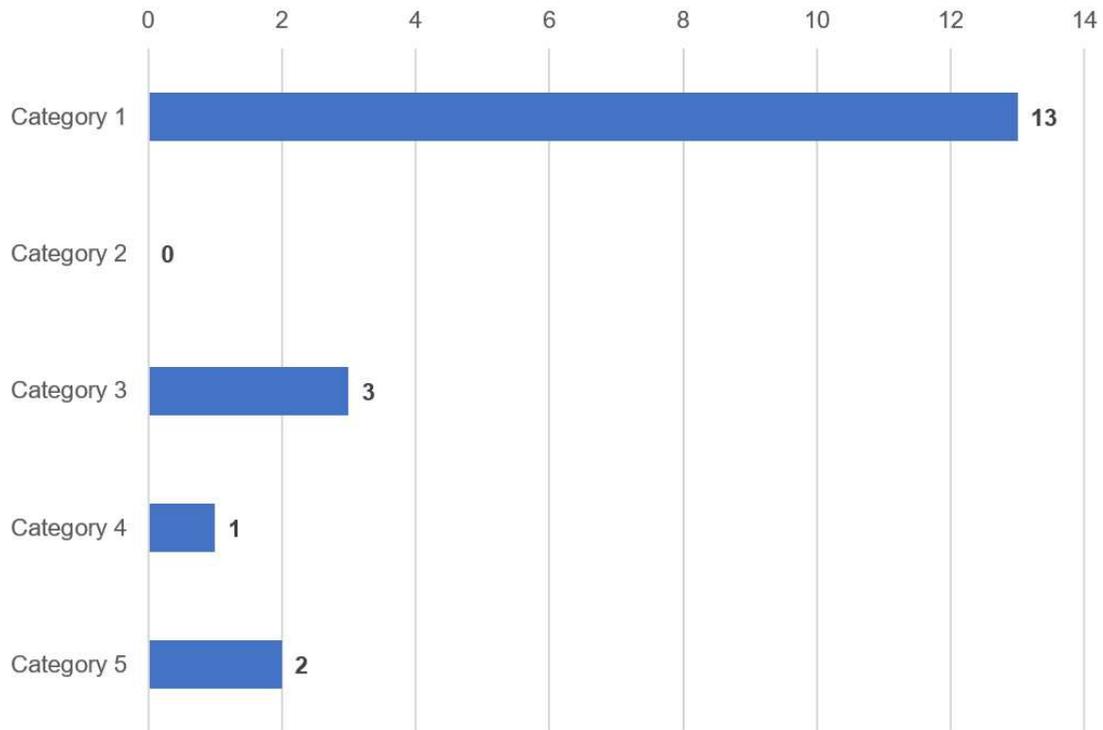

*Figure 21: Categorization of potentials or examples for the internal control system according to categories from Table 6*

Results are similar to Question 16. Information from Category 1 was mentioned from the majority as potential for the internal control system (65 %). Category 2 was not mentioned at all.

## 15. to 17. Overall Overview

Answers of Question 15 to 17 are closely related, because these questions are not independent but were answered directly one after the other from the participants.

Thereby 6 persons gave identical answers to Questions 15, 16 and 17 with regard to the different GRC business segments. Three answers were assigned to Category 3, two answers to Category 1 and one answer to Category 4.

10 persons used different categories answering Question 15, 16 and 17.

## 18. Is AI already in use in GRC?

**Question 18**: Is AI already in use in GRC?

Possible answers were: yes, no, don't know. (n=67)

This question differs from Question 1 insofar since it refers specifically to GRC and not to the entire company.

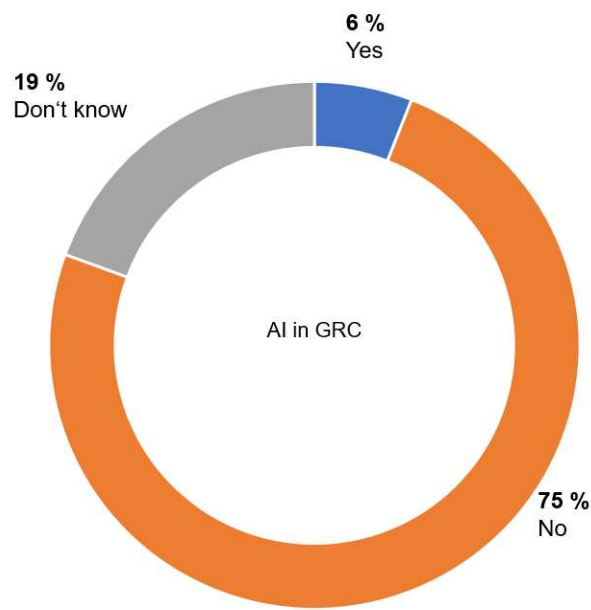

*Figure 22: Use of AI in GRC (n=67)*

As shown in Figure 22 a majority (75 %) of the n=67 participants stated, that AI is not used in GRC. Only 6 % answered that AI is used in GRC and 19 % answered that they do not know if AI is used in GRC in the company.

## 19. For which segments or systems with GRC reference do you see potentials for use of AI?

**Question 19**: For which segments or systems with GRC reference do you see potentials for use of AI?

Possible answers were:

- Internal auditing
- IT management
- Information security management system
- Additional segments or systems (with text input)

Multiple choice was possible.

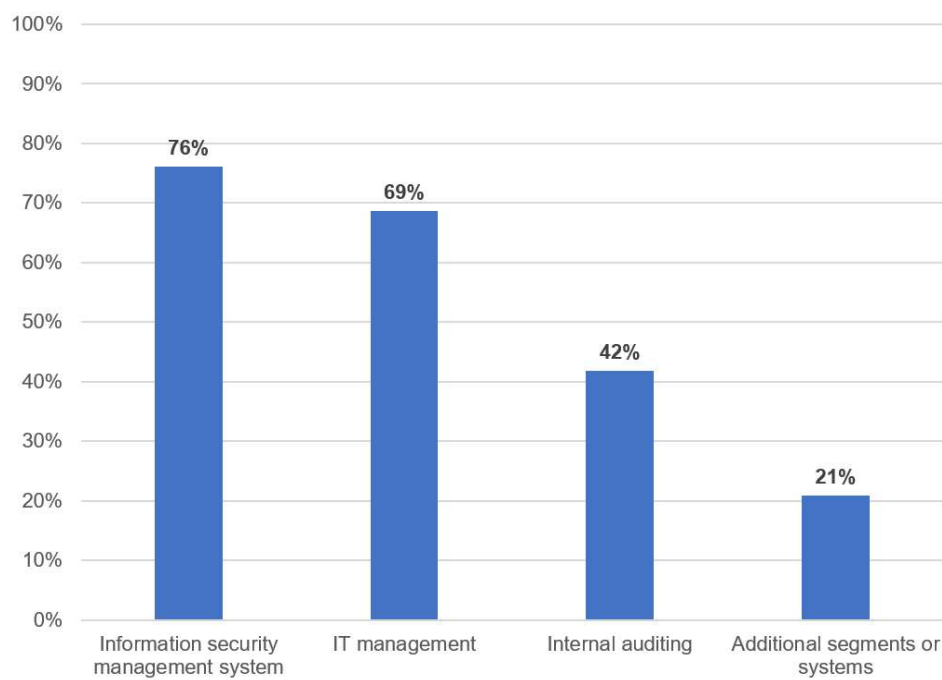

*Figure 23: Additional potentials using AI (n=67)*

Information security management system was selected in most cases (76 %), followed by IT management (69 %) and internal auditing (42 %). Additional segments or systems were selected by 21 %.

The following answers were given in the text input for additional segments or system (translated from German): Controlling (2), Quality management (2), Environmental, IT and Energy management system, SOX compliance, Fraud management, Data security, Money laundering, Business continuity management, Supply chain, Production control, Machine networking, Inventory management, Supervised learning processes, "all business functions".

## 20. What challenges do you see in your company for a successful use of AI, especially in GRC?

**Question 20**: What challenges do you see in your company for a successful use of AI, especially in GRC?

Possible answers were:

- Recruitment and staff training related to expertise in AI
- Quality and quantity of data
- Homogenous system environment
- Financial feasibility
- Tone of the top
- Transparency of legal requirements
- Ruggedness of possible regulatory requirements
- Identification of potentials for application
- Integration of AI technology in GRC functions and processes
- Extension of control function in GRC to include AI technology used
- Additional challenges (with text input)

Multiple choice was possible.

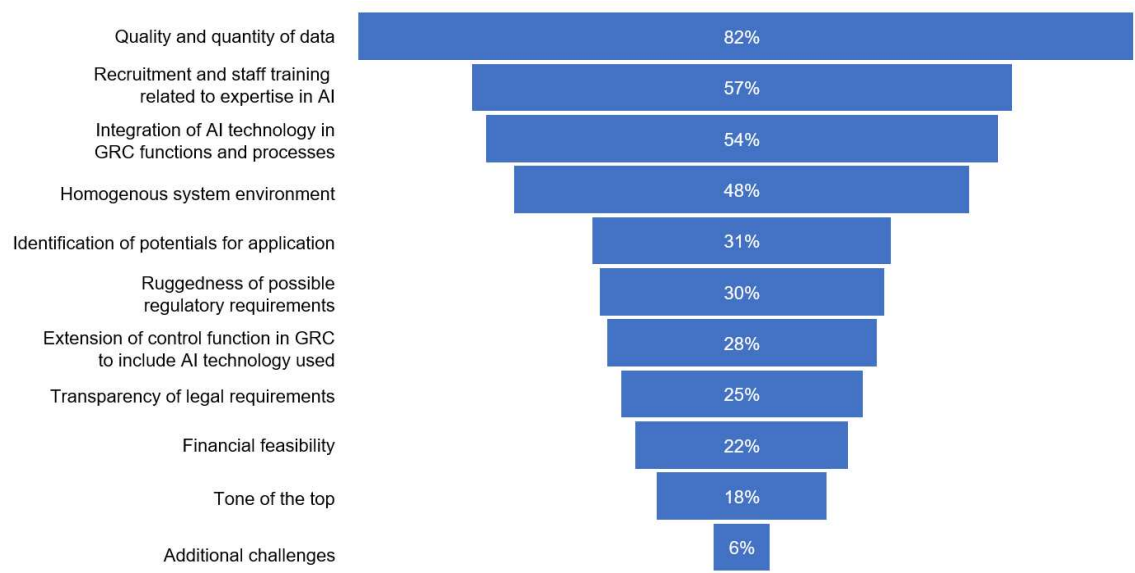

*Figure 24: Challenges for a successful use of AI in GRC (n=67)*

All except one person of the n=67 participants mentioned at least one challenge für a successful use of AI in GRC. Figure 24 lists the selected answers. Quality and quantity of data, which is an essential basis for a successful use of AI, was selected most often (82 %). Interestingly financial feasibility plays with 22 % only a role in the rear area.

Some additional challenges were specifically mentioned (translated from German): it usually fails because of hierarchies and silo thinking, lack of acceptance among risk owner, not much practical use, more important issues, clear return of investment for management.

## 2.3  Part 3: General Information about the Company

In order to better understand the answers, it was of interest to get information about who filled out the survey and to which framework conditions the answers refer. Therefore, in part 3 of the survey general information were collected.

Analogously to part 1 and 2, all questions and answers provided for selection listed in this chapter were translated from German, too.

### 21. – 23. Classification of very small, small and medium-sized firms

From Question 21, 22 and 23 it was possible to classify the companies along the definition of the EU Commission for very small, small and medium-sized firms (SME).[5] According to this the participants should answer the following three questions.

**Question 21: How many employees did your company havein the last financial year before COVID-19 pandemic?**

Possible answers were:

- Less than 10 employees

---

[5] See Eurostat.

- 10 to 49 employees
- 50 to 249 employees
- 250 or more employees
- Not specified

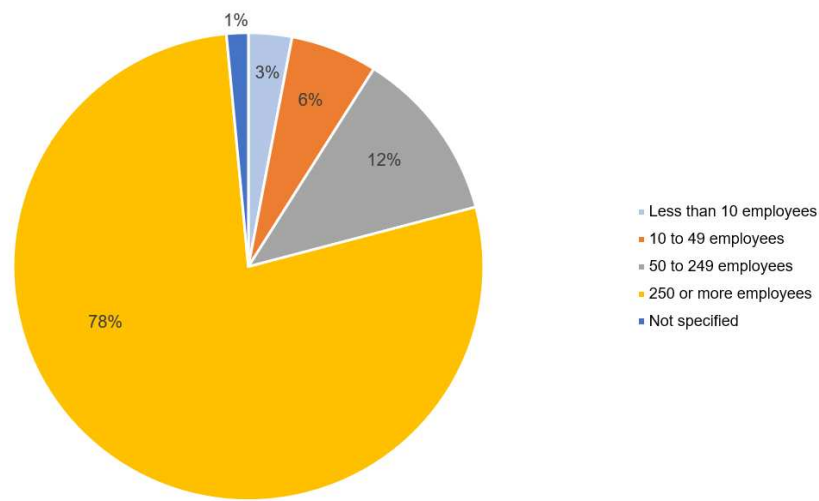

*Figure 25: Number of employees in the last financial year before COVID-19 pandemic (n=67)*

As can be seen in Figure 25 the majority (78 %) of the participants (n=67) stated, that the company has 250 or more employees. These companies are therefore no very small, small or medium-sized firms.

**Question 22: What was the annual turnover of the last financial year before COVID-19 pandemic?**

Possible answers were:

- Up to 2 Mio. €
- More than 2 Mio. € up to 10 Mio. €
- More than 10 Mio. € up to 50 Mio. €
- More than 50 Mio. €
- Not specified

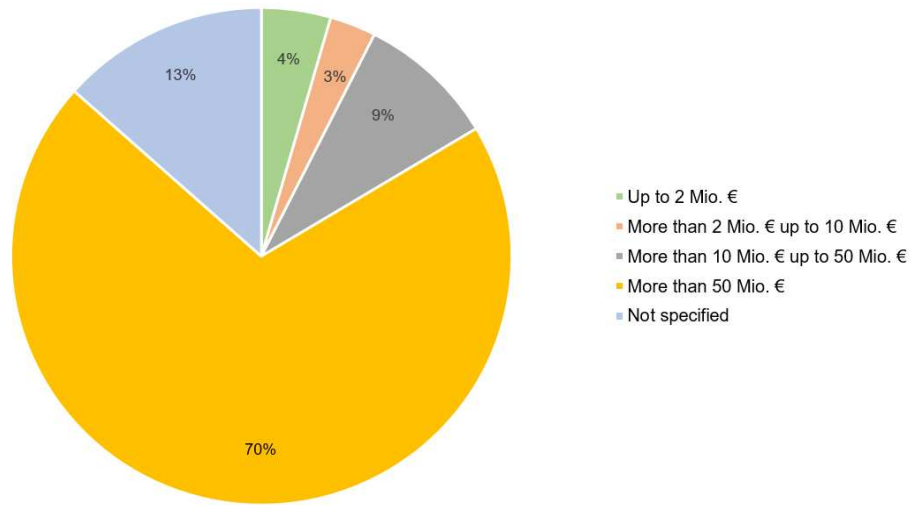

*Figure 26: Annual turnover of the last financial year before COVID-19 pandemic (n=67)*

Figure 26 shows, that the majority of the participants (70 %) are working in a company with an annual turnover of the last financial year before COVID-19 pandemic of more than 50 Mio. €. This specification describes a threshold value for classification of very small, small and medium-sized firms.

**Question 23: What was the balance sheet total of the last financial year before COVID-19 pandemic?**

Possible answers were:

- Up to 2 Mio. €
- More than 2 Mio. € up to 10 Mio. €
- More than 10 Mio. € up to 43 Mio. €
- More than 43 Mio. €
- Not specified

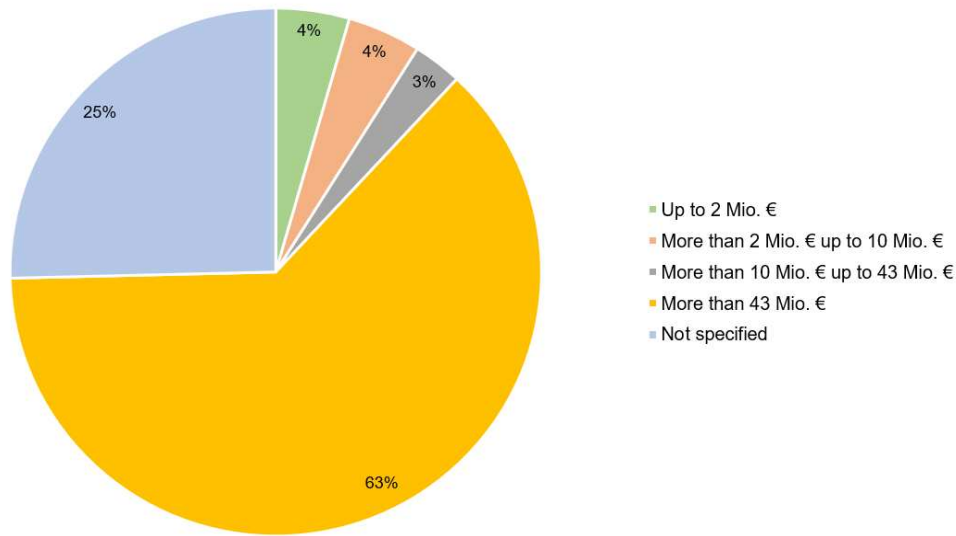

*Figure 27: Balance sheet total of the last financial year before COVID-19 pandemic (n=67)*

The threshold value for very small, small and medium-sized firms is 43 Mio. €. The majority of the answers was above (63%).

**Classification SME**

Evaluating the information from Question 21 to Question 23 using the definition of the EU Commission of SME[6] altogether the distribution of the classification can be seen in Figure 28. A clear majority of the companies (88 %) are not classified as SME.

---

[6] See Eurostat.

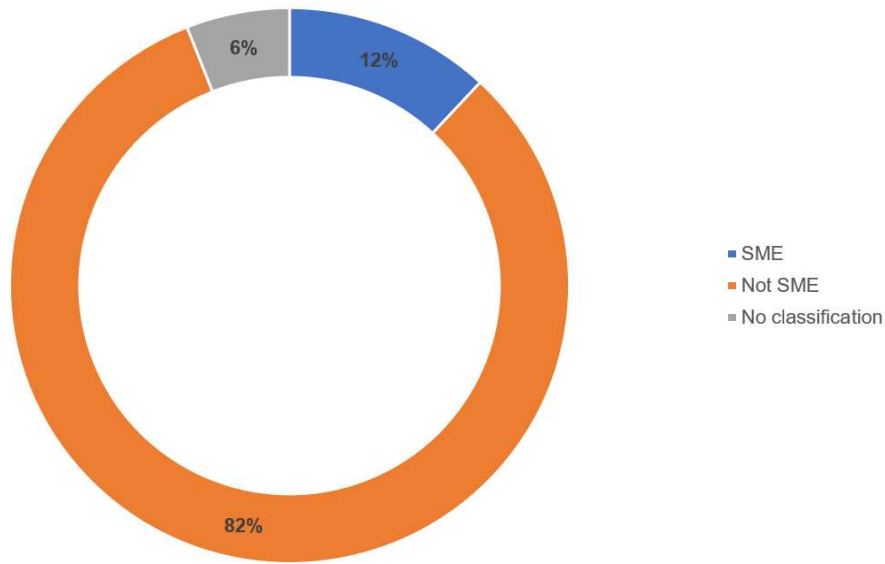

*Figure 28: Classification SME (n=67)*

## 24. What is the legal form of you company?

**Question 24**: What is the legal form of your company?

Possible answers were: [7]

- AG (listed)
- AG (not listed)
- Individual Entrepreneur
- GmbH
- KG
- KGaA
- OHG
- Other legal form (with text input)

---

[7] AG corresponds to stock company, GmbH to Ltd, KG to limited partnership, KGaA to limited partnership on shares and OHG to general partnership.

This question is relevant with regard to possible legal requirements depending on the legal form of the company.

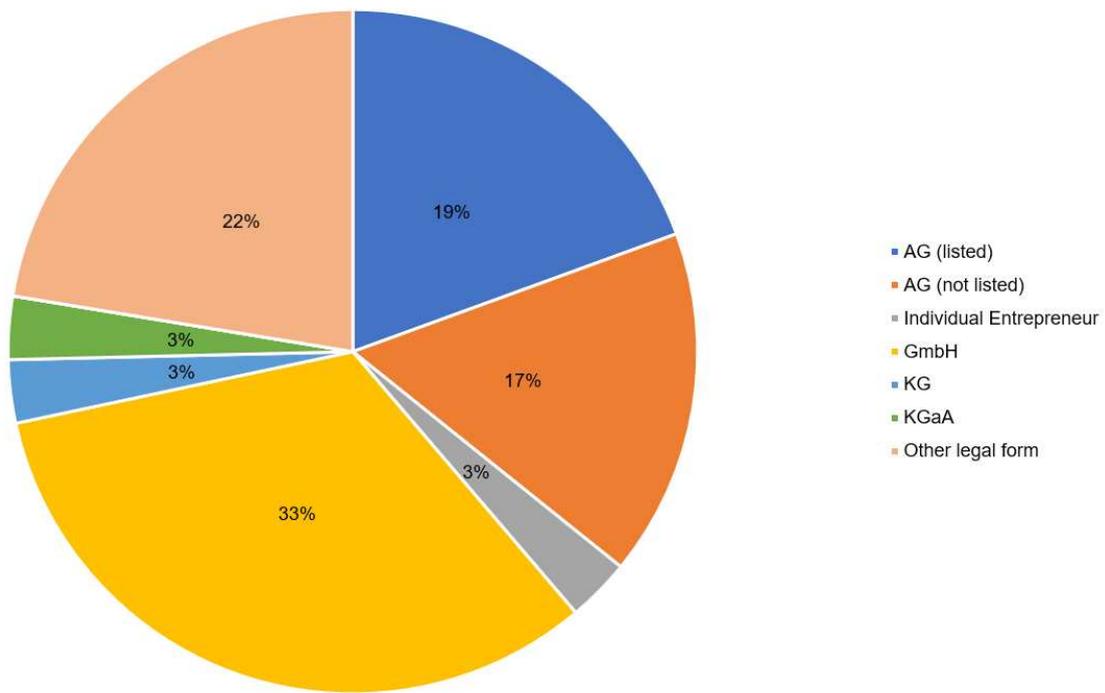

*Figure 29: Legal forms (n=67)*

The following input was made for "other legal form": joint institution, VVaG (3), K.ö.R., GbR, eG, AöR, fund, SE & Co. KG, educational facility, public legal form.

Figure 29 shows that the majority of the companies in this survey are listed and not listed stock companies (in total 36 %) and Ltd (33 %).

## 25. To which industry does your company belong?

**Question 25**: To which industry does your company belong?

The following economic sectors according to "NACE Rev. 2: Statistical classification of economic activities in the European Community" could be selected[8].

- Agriculture, forestry and fishing
- Manufacturing, mining and quarrying and other industry
    - Mining and quarrying
    - Manufacturing
    - Electricity, gas, steam and air conditioning supply
    - Water supply; sewage, waste management and remediation activities
- Construction
- Wholesale and retail trade, transportation and storage, accommodation and food service activities
    - Wholesale and retail trade; repair of motor vehicles and motorcycles
    - Transportation and storage
    - Accommodation and food service activities
- Information and communication
- Financial and insurance activities
- Real estate activities
- Professional, scientific, technical, administration and support service activities
    - Professional, scientific and technical activities
    - Administrative and support service activities
- Public administration, defense, education, human health and social work activities
    - Public administration and defense; compulsory social security
    - Education
    - Human health and social work activities
- Other services

---

[8] See Eurostat, 2017. This is a combination of the table on page 43 (high-level SNA/ISIC-aggregation A*10/11) and of the table on page 59 (broad structure of NACE Rev. 2).

- o Arts, entertainment and recreation
- o Other service activities
- o Activities of households as employers; undifferentiated goods- and services-producing activities of households for own use
- o Activities of extraterritorial organisations and bodies

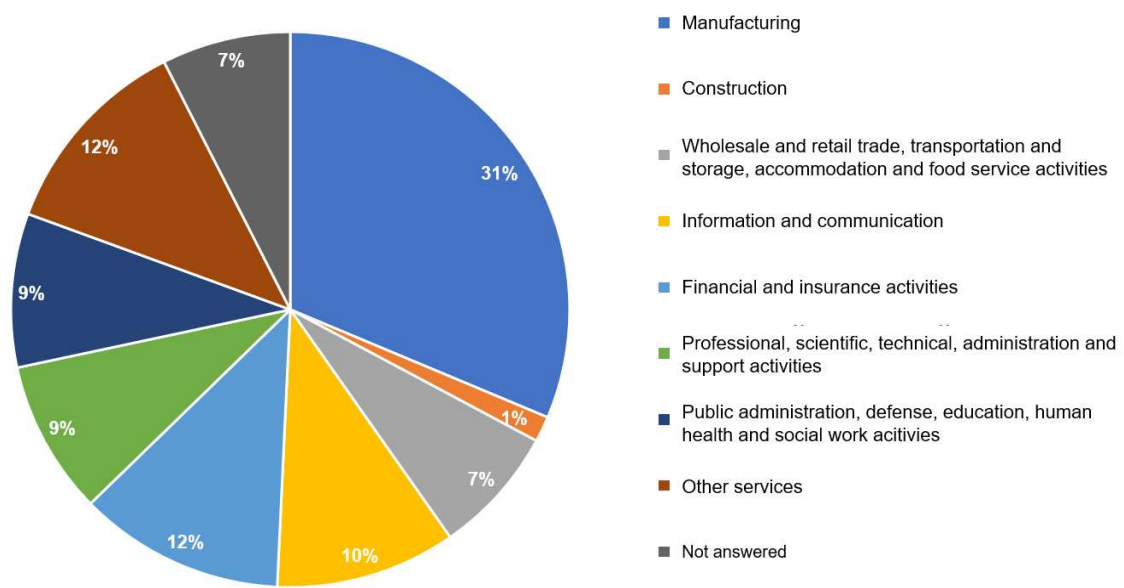

*Figure 30: Industry according to NACE Rev.2 (n=67)*

The answers aggregated according to top-level terms are shown in Figure 30. The information refers to all n=67 persons. Of these 5 persons (7 %) did not answer Question 25 at all. A large proportion of the respondents (31 %) belongs to manufacturing. Second important industry were financial and insurance activities as well as other services (both 12 %).

## 26. The company's location, where your workplace is located

**Question 26**: The company's location, where your workplace is located is in:

- Germany,
- Austria,
- Switzerland?

If Germany was answered in Question 26, the first two digits of zip code should be provided in Question 27.

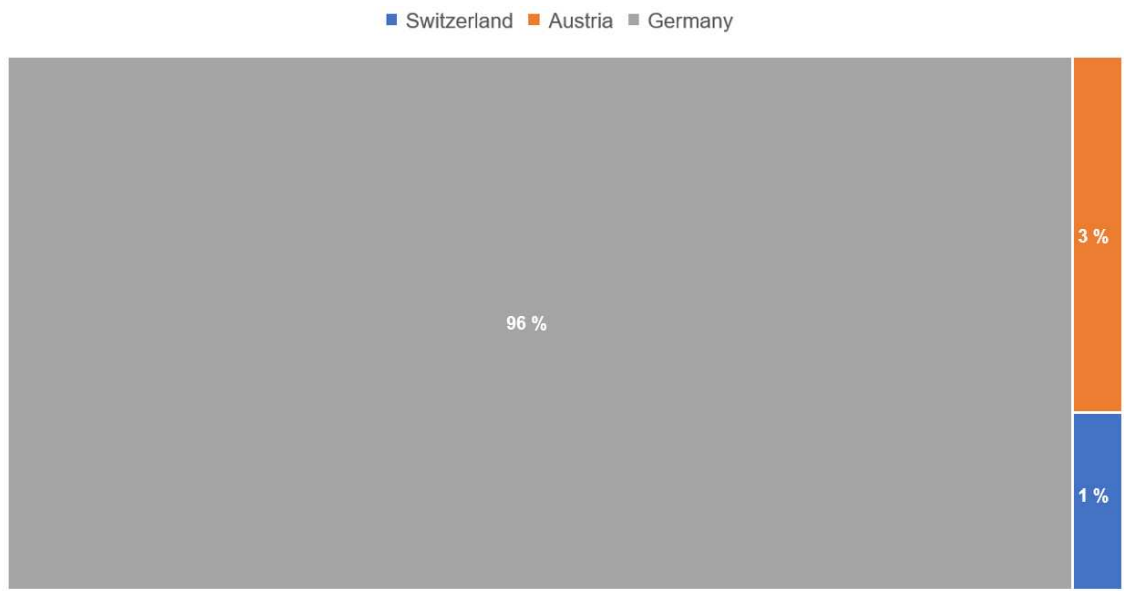

*Figure 31: The company's location (n=67)*

Question 26 was answered from all participants (n=67). The overwhelming majority (96 %) stated Germany as the company's location, as Figure 31 illustrates.

## 27. Please provide the first two digits of zip code of your company's location.

**Question 27**: Please provide the first two digits of zip code of your company's location.

This question was only shown if Question 26 was answered with "Germany".

The idea behind this question was to work out the regions within Germany from which the respondents came without removing the anonymity of the survey. From the n=64 participants, that stated in Question 26 that their company's location is Germany, 62 provided information about the zip code.

Considering a broad division in north, south and middle, as shown in Table 7, a distribution is created, which can also be taken from this table.

| Region | Zip code starting with | Number |
|--------|------------------------|--------|
| North  | 1, 2                   | 12     |
| Middle | 0, 3, 4, 5             | 33     |
| South  | 6, 7, 8, 9             | 17     |

*Table 7: Regions within Germany according to zip code (n=67), 2 persons made no specification*

## 28. What position do you, the interviewee, have in the company?

**Question 28**: What position do you, the interviewee, have in the company?

Possible answers were:

- Management board/Management
- Department risk management
- Department compliance
- Department internal control system
- Department internal auditing
- Department GRC
- Department risk controlling
- Department controlling
- None of the above but position with GRC reference
- None of the above but position without GRC reference

Multiple choice was possible. If a selection was made from the points department risk management to department controlling, a submenu opened asking whether it was an executive position or an employee's position.

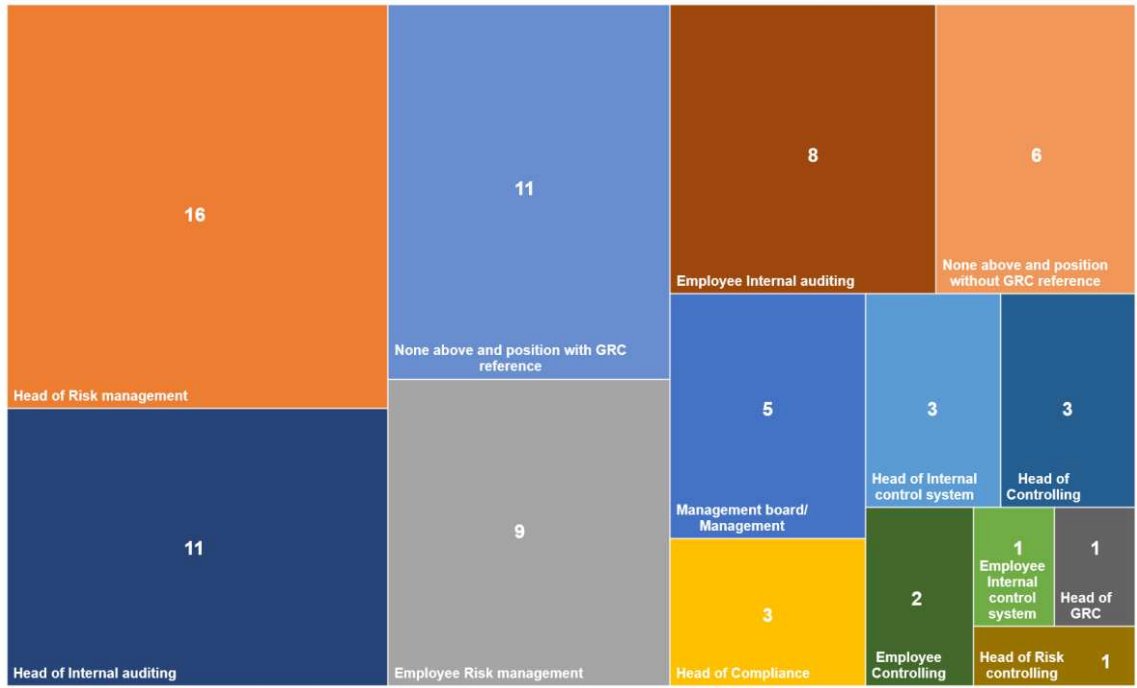

*Figure 32: Position of the respondents in the company (n=67, multiple choice possible)*

*It was mandatory to answer this question and all participants did answer this question. Because multiple choice for the position was possible, in*

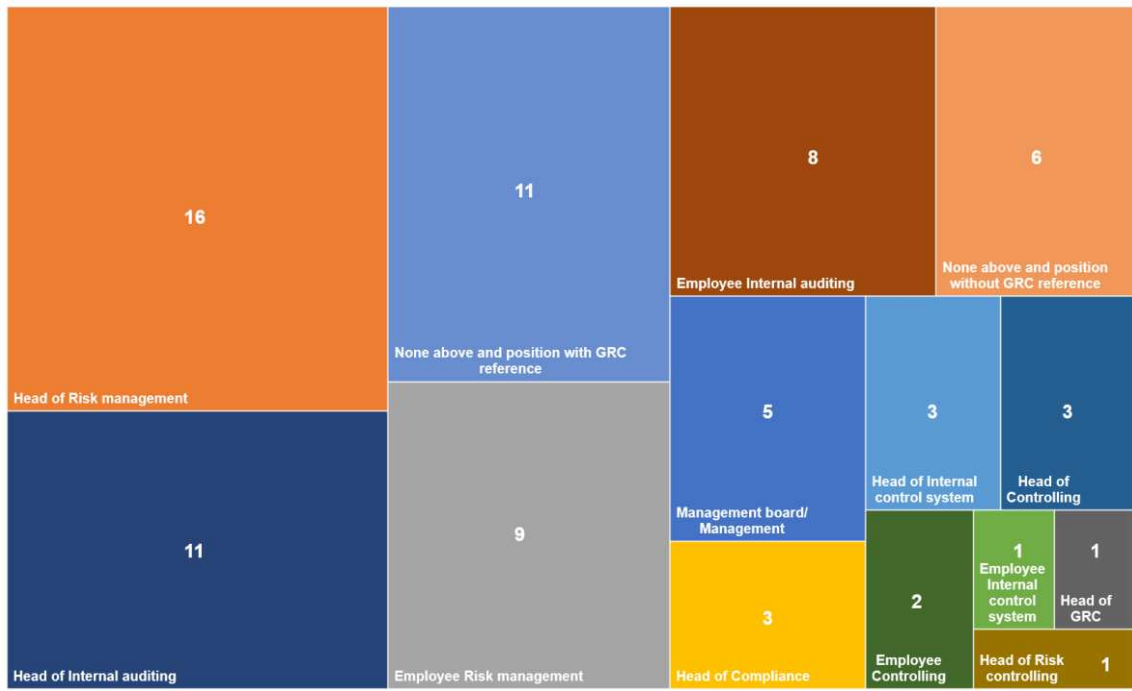

Figure 32, a total of 80 information, and therefore more than the number of 67 participants, are illustrated. As can be seen from the figure, risk management was strongly represented.

### *29. Do you have an expertise in IT auditing, IT security management or IT governance?*

**Question 29**: Do you have an expertise in IT auditing, IT security management or IT governance?

Possible answers were yes or no. A total of 46 % of the respondents (n=67) stated "Yes" whereas 54 % of the respondents have no expertise in IT auditing, IT security management or IT governance.

### *30. You have completed the part of the questionnaire for the survey. Would you like to make any final comments with regard to the topic or the survey?*

**Question 30**: You have completed the part of the questionnaire for the survey. Would you like to make any final comments with regard to the topic or the survey?

Answers could be filled in a free text form.

Directly to the study there were three remarks. In two cases there were information provided to clarify the information on the industry from Question 24. In one case there was a notice that the study in the actual topic is "thought too academic, too far away from operational practice".

## 3   Results

As results from the study several statements can be noted. Based on the results from Question 1 and Question 18, it can be determined that AI is indeed applied in companies (52 %), but only to a small extend in GRC (6 %). At the same time answers of Question 14 show, that the respondents see potentials for use of AI in GRC in different fields.

Therefore, there is a potential to analyze and optimize AI methods für use in GRC and to implement and supervise corresponding projects.

GRC business segments are definitely integrated to each other, as can be seen from the results of Question 4. There 78 % stated, that an integration exists. Further research would be relevant here to find out how this integration can provide a basis for data exchange, in terms of the business segments involved, and the type and structure of integration (Question 5 to 7), so that methods of artificial intelligence can be successful applied.

The fundamental statement that IT applications are used in GRC can be derived from the results of Question 8. The feedback to Question 9 to 12 provides an initial insight into shared use of IT applications in GRC. To derive clear instructions more information about the kind of data storage and the specific data exchange between systems are necessary. In this context, it would be also of interest to know which components the personal assessment of the level of automation in companies is determined. The respondents indicated the level of automation of IT applications in GRC in Question 13 as not very high.

Always bearing in mind that this is not a representative study, the answers given in Question 14 indicate a direction in which the use of AI in GRC might move. Risk management is seen as the biggest field of application for AI methods listed there. This impression is continued in the answers to Question 15 to 17. Further research on the basis of the information to potentials and examples mentioned there could lead to operational objectives and their prioritizations.

The four most frequently mentioned answers in Question 20 to challenges for successful use of AI in GRC are quality and quantity of data (82 %), personnel recruiting and training related to expertise in AI (57 %), integration of AI technology in GRC functions and processes (54 %) and homogenous system environment (48 %). These four corner pillars generally stand for challenges in the field of digitalization. It is precisely AI processes that require high-quality data in large quantities. In addition, the current scarcity of qualified specialists is noticeable with regard to necessary restructuring of processes to support the use of AI.

In the end, the results of the study are based on 82 % on people working in companies that are not classified as SME added with the given answers of the legal forms in Question 24, what is closely related with the framework of regulatory as well as reporting requirements. A closer look in framework conditions and needs of SMEs would be a further subject of research to evaluate the use of AI in GRC.

## 4    Conclusions

This paper provides an initial overview of the results of a study on potentials for the application of AI in GRC, that was conducted within the framework of an anonymous survey within the period from September 2021 to December 2021.

All questions were explained that were provided within the study. The results for every single question were presented and discussed. The study serves as a basis for more and complex analysis of the data obtained as well as deriving further research topics and discussions.

Even though the presented study is not representative, it does show some interesting approaches for further research. How exactly does the integration of the GRC business segments affect data exchange? How are data basis and data exchange developing in GRC? What are the challenges using AI in GRC in detail and how to counter them? What kind of AI methods can be usefully employed in GRC? When is a company ready for using AI? Which framework conditions must be recognized using AI in GRC and beyond this if when AI technology used in the company must be controlled in GRC? How can the problem of finding application potentials of AI in GRC be transferred to SME?

# List of Figures

## List of Tables

## List of Sources

Armbrüster, 2022: Armbrüster, C. "Digitalisierung und Nachhaltigkeit – rechtliche Herausforderungen für den Versicherungssektor, insbesondere beim Einsatz von Künstlicher Intelligenz", *ZVersWiss* **111**, 19–31 (2022), https://doi.org/10.1007/s12297-022-00518-3.

Eurostat: o.V. "Kleine und mittlere Unternehmen (KMU)", online available at https://ec.europa.eu/eurostat/de/web/structural-business-statistics/small-and-medium-sized-enterprises, last accessed 05.09.2022.

Eurostat, 2017: European Commission "NACE Rev. 2 – Statistical classification of economic activities in the European Community", Publications Office, 2017.

Jo, 2021: Jo, T. "Machine Learning Foundations – Supervised, Unsupervised, and Advanced Learning", Springer Nature, 2021.

Leiner 2019: Leiner, D. J. "Sosci Survey (Version 3.3.06)" [computer software], online available at https://www.soscisurvey.de.

Ransbotham et al., 2017: Ransbotham, S., Kiron, D., Gerbert, P., Reeves, M. "Reshaping Business with Artificial Intelligence – Closing the Gap Between Ambition and Action", MIT Sloan Management Review and The Boston consulting Group, September 2017.

Tarafdar et al., 2019: Tarafdar, M, Beath, C., Ross, J. "Using AI to Enhance Business Operations". MIT Sloan Management Review 60(4), 37-44 (2019).